\documentclass[aps,prx,twocolumn,english,showpacs,superscriptaddress,amssymb,amsfonts,nofootinbib]{revtex4-2}
\usepackage{hyperref}
\usepackage{amsmath}
\usepackage{braket}
\usepackage{tikz}
\usepackage{xypic}
\usetikzlibrary{positioning}
\usepackage{soul,xcolor}

\usepackage{placeins}
\usepackage[scr=boondox]{mathalpha}
\usepackage{adjustbox}

\newcommand{\Rom}[1]{\uppercase\expandafter{\romannumeral #1\relax}}
\usepackage{amsthm}

\usepackage{rotating}

\newtheorem*{assumption*}{\assumptionnumber}
\providecommand{\assumptionnumber}{}


\theoremstyle{definition}

\theoremstyle{theorem}
\newtheorem{theorem}{Theorem}[section]

\theoremstyle{corollary}

\theoremstyle{lemma}

\theoremstyle{Proposition}

\theoremstyle{definition}

\usepackage{amssymb}

\newcommand{\bst}{{\boldsymbol{T}}}

\makeatletter

\newcommand{\Rmnum}[1]{\expandafter\@slowromancap\romannumeral #1@}
\makeatother
\newcommand{\imth}{\hspace{1pt}\mathrm{i}\hspace{1pt}}

\newcommand{\mbz}{{\mathbb{Z}}}
\newcommand{\bea}{\begin{eqnarray}}
\newcommand{\eea}{\end{eqnarray}}
\newcommand{\bct}{\begin{center}}
\newcommand{\ect}{\end{center}}
\newcommand{\bpm}{\begin{pmatrix}}
\newcommand{\epm}{\end{pmatrix}}
\newcommand{\bal}{\begin{aligned}}
\newcommand{\eal}{\end{aligned}}
\newcommand{\bfr}{\begin{framed}}
\newcommand{\efr}{\end{framed}}

\newcommand{\expval}[1]{\langle{#1}\rangle}

\usepackage{tikz-cd}
\usetikzlibrary{matrix,arrows,backgrounds}

\makeatletter
\def\l@subsubsection#1#2{}
\makeatother

\begin{document}
	\title{Theory of topological defects and textures in two-dimensional quantum orders with spontaneous symmetry breaking}
	\author{Yan-Qi Wang}
	\affiliation{Department of Physics, University of California, Berkeley, Berkeley CA 94720, USA}
	\affiliation{Materials Sciences Division, Lawrence Berkeley National Laboratory, Berkeley, California 94720, USA}
	\author{Chunxiao Liu}
	\affiliation{Department of Physics, University of California, Berkeley, Berkeley CA 94720, USA}
	\author{Yuan-Ming Lu}
	\affiliation{Department of Physics, The Ohio State University, Columbus, OH 43210, USA}
	
	\date{\today}
	
	\begin{abstract}

We consider two-dimensional (2d) quantum many-body systems with long-range orders, where the only gapless excitations in the spectrum are Goldstone modes of spontaneously broken continuous symmetries. To understand the interplay between classical long-range order of local order parameters and quantum order of long-range entanglement in the ground states, we study the topological point defects and textures of order parameters in such systems. We show that the universal properties of point defects and textures are determined by the remnant symmetry enriched topological order in the symmetry-breaking ground states with a non-fluctuating order parameter, and provide a classification for their properties based on the inflation-restriction exact sequence. We highlight a few phenomena revealed by our theory framework. First, in the absence of intrinsic topological orders, we show a connection between the symmetry properties of point defects and textures to deconfined quantum criticality. Second, when the symmetry-breaking ground state have intrinsic topological orders, we show that the point defects can permute different anyons when braided around. They can also obey projective fusion rules in the sense that multiple vortices can fuse into an Abelian anyon, a phenomena for which we coin ``defect fractionalization''. Finally, we provide a formula to compute the fractional statistics and fractional quantum numbers carried by textures (skyrmions) in Abelian topological orders. 
\end{abstract}

	\maketitle

\tableofcontents

\clearpage{}

\section{Introduction}

One of the most successful theories in condensed matter physics is the Landau's theory of phases and phase transitions~\cite{Landau1937,Ginzburg2009,landau2013fluid,landau2013statistical}: phases are distinguished by symmetries, and phase transitions  are described by symmetry breaking. An ordered phase with broken symmetry is identified through the formation of off-diagonal long-range order and is characterized by a local order parameter. While the spontaneous breaking of a continuous symmetry leads to Goldstone modes~\cite{Naumbu1960,Goldstone1961} which are gapless excitations in the system,  a class of gapped excitations \--- defects and textures \--- may also be present in an ordered phase as a consequence of the nontrivial topology of the order parameter space~\cite{Mermin1979}. This \emph{classical topology} can lead to very rich physics. For example, the topological defects have their own dynamics and may also lead to phase transitions at finite temperatures~\cite{berezinskii1971destruction,berezinskii1972destruction,Kosterlitz1973}. Topological defects and textures also commonly appear in soft matter physics~\cite{kleman2003soft,vafa2022defect} and ultracold atom physics~\cite{lewenstein2012ultracold,stamper2013spinor}.

Since the discover of the quantum Hall effects in the 1980s~\cite{Klitzing1980,Tsui1982,Laughlin1983}, the notion of phases of matter have been extended beyond Landau's theory. Let us focus on gapped phases of matter, which, by definition, are phases with gapped excitations that are robust against local perturbations without closing the gap. In absence of symmetry, these different phases are  determined by the long-range entanglement structure of the ground state wave functions~\cite{Wen2017}, with the trivial one being an ``atomic insulator'' whose ground state shares the same phase as a collection of isolated atoms. A topological order, on the other hand, has a nontrivial entanglement structure in their wave function, which manifest itself, e.g. through ground state degeneracy when placed on a topologically nontrivial manifold~\cite{Wen1990}. Here, while the word ``topological'' in topological order still refers to the robustness of the low energy excitations against local perturbations without closing the gap, this feature is a direct consequence of the long-range entanglement of the wave function~\cite{Wen2017}. In presence of symmetry, either a topologically trivial state or a topologically ordered state may be further separated into different phases. The result is either a symmetry protected topological (SPT) phase~\cite{Gu2009, Chen2010,Chen2011a,Chen2011b,Chen2011c,Chen2012,Chen2013,Chen2014,Pollmann2012,Lu2012,Wen2013,Levin2012,Wen2014} or a symmetry enriched topological (SET) phase~\cite{Essin2013,Mesaros2013,Lu2016,Tarantino2016,Barkeshli2019}, and in both cases different gapped ground states are characterized by certain topological invariants. 
As the quantum state counterpart of the order parameter in a broken phase in Landau's theory, the topological invariants reflect the robustness of the state under small perturbations and are a manifestation of the \emph{quantum topology} that arises from many-body quantum entanglement in the wave function.

So far, most studies on topological phases --  SPT and SET -- preserve all symmetries of the system, in contrast to the Landau paradigm where spontaneously broken symmetries give rise to long-range orders. In other words, topological (SPT and SET) phases are usually discussed in a context that excludes long-range orders from spontaneous symmetry breaking. In nature, nevertheless, the coexistence of long-range order in a topological phase is not a rare phenomenon: nematic quantum Hall states in higher Landau levels~\cite{Feldman2016}, topological superconductors that spontaneously break charge conservation~\cite{Liu2019}, and magnetic fragmentation for spin ice~\cite{Petit2016}, to name a few. On the theoretical side, while most previous works studied examples with an emphasis on non-interacting fermion systems~\cite{Sondhi1993,Volovik1999,Ivanov2001,Read2000,Zhang2009,Ran2009,Teo2010,Barkeshli2013,Barkeshili2014,Balram2015,Wang2022}, a general theory for interacting topological phases with coexisting long-range orders is still lacking \cite{Else2021}.

This motivates us to establish a theoretical framework for topological phases in the presence of long-range orders~\cite{Else2021}, which is the main focus of the present work. We consider the ``gapped'' topological phases, where the Goldstone modes that arise from spontaneous breaking of continuous symmetries are the only gapless excitations in the system. Our approach is to study the universal properties of topological defects and textures of the spontaneously broken symmetries, as a first step towards a classification of topological phases in presence of long-range orders.

One theme of the present work is to establish a concrete connection between classical topology and quantum topology. We will be mainly focusing on topological point defects and smooth textures (i.e. skyrmions) in two-spatial dimensions and their interplay with an SPT or an SET phase. More precisely, when the full symmetry group $G$ spontaneously breaks down to a subgroup $H$, we consider a symmetry-breaking ground state where the order parameters are not fluctuating and fixed in a classical minimum of the free energy. Since the only gapless excitations in our systems are the Goldstone modes, these symmetry-breaking states must be the ground state of a gapped Hamiltonian that preserves $H$. In two spatial dimensions, they are either $H$-SPT phases in the absence of intrinsic topological orders, or more generally $H$-SET phases. We intend to understand how these $H$-SPT or $H$-SET ground states (``quantum topology'') affect universal properties of topological defects and textures of the order parameters (``classical topology'') in the associated long-range order. 

It turns out the crucial connection between classical and quantum topology can be established generally by a map (a ``connecting homomorphism''~\cite{Else2021}) from topological defects and textures of the order parameters to (extrinsic) symmetry defects~\cite{Levin2012,Barkeshli2019,Tarantino2016,Teo2016} in an $H$-SPT or $H$-SET phase. We use this map, and the classification of $H$-SPT and $H$-SET phases, to obtain a classification and characterization of the universal properties belonging to topological defects and textures in a long-range ordered quantum system. 

We first consider the (conceptually simpler) situation in the absence of intrinsic topological orders, where each ground state with fixed non-fluctuating order parameters is an $H$-SPT phase. We identify two phenomena out of interplay between classical topology and quantum topology: owing to the $H$-SPT ground state of the long-range order, the point defects of order parameters can carry a projective representation of the remnant symmetry $H$, while topological textures of the order parameters (i.e. skyrmions) can carry a nontrivial quantum number of the remnant symmetry $H$. This provides a new angle into a large family of Landau-forbidden quantum phase transitions: i.e. the deconfined quantum critical points (DQCPs)~\cite{Senthil20041,Senthil2004,Wang2017}.

Next we consider a more general situation, where each ground state with non-fluctuating order parameters is an $H$-SET phase with bulk anyon excitations~\cite{Essin2013,Barkeshli2019,Tarantino2016,Mesaros2013,Lu2016}. First, we reveal two exotic phenomena associated with point defects: (1) different types of anyons can be permuted after they are braided around a point defect, (2) multiple point defects, when combined together to form a trivial point defect, can instead fuse into an Abelian anyon, a phenomenon for which we coin the term ``defect fractionalization''. Then, in the case of smooth textures of order parameters, i.e. skyrmions in 2d, we develop a general field theory that couples a topological ordered system to a ferromagnetic order parameter via a topological term in the Lagrangian. Applying this to Abelian topological orders, we obtain the formula for the fractional statistics and fractional quantum numbers of skyrmions in the system.

Another interest of this work comes from the technical side. It has been known for long (and fairly familiar among condensed matter physicists) that classical topological defects are mathematically described by homotopy groups in algebraic topology~\cite{hatcher2002algebraic,rotman2009introduction}. This mathematical object is rather intuitive as it admits a real space picture: it models the topological defects as maps from real space (or spacetime) to the space of parameters (such as the order parameter space which is of interest to this work) and classifies them up to continuous deformation. On the other hand, the theory of symmetry defects and symmetry fractionalization are rather new in condensed matter physics~\cite{Wen2002,Levin2012,Essin2013,Tarantino2016,Barkeshli2019}, and the main mathematical tool employed are various homology (and cohomology) theories. While homology theory also stems from homotopy theory in algebraic topology, it has developed into an independent subject, whose application in physics is far more rich and profound. Even from this technical point of view, it would be a great pleasure -- and would bring great mathematical insight to the physical problem under consideration -- to see how these mathematical objects can be united in the treatment of classical topology and quantum topology.

This paper is organized as follows. In section \ref{sec:theory framework}, we describe a theoretical framework, which crucially connects the point defects and textures of the order parameters of the broken symmetries to the symmetry defects of the preserved symmetries. This connection allows us to classify and characterize the universal properties of point defects and textures using the topological properties of the ground states. In section ~\ref{sec:group cohomology}, we continue our theoretical framework by exploring the connection between classical topology and quantum topology, where we provide more mathematical details on group cohomology classification for point defects and textures in SPT and SET phases. The key word there are the so-called ``inflation map'' that appears in a five-term exact sequence for group cohomology, whose physical meaning will be investigated in great detail. Next we apply this framework to demonstrate universal properties of point defects and textures in 2d quantum orders. In section \ref{sec:spt+dqcp}, we focus on the simplest cases in the absence of intrinsic topological orders, where all ground states with a fixed order parameter configuration are SPT phases. We show that the exotic phenomena of DQCP can be captured in a concise manner within our framework. Next, we proceed with general cases where the ground states are SET phases with intrinsic topological order. In section \ref{sec:point defect}, we classify topological properties of point defects, highlighting two distinct phenomena: non-Abelian point defects that permute anyons when braided around, and a new phenomenon for which we coin ``defect fractionalization'' where multiple point defects fuse into Abelian anyons. In section \ref{sec:texture}, we study topological textures (i.e. skyrmions) in 2d SET phases, in particular, we compute the fractional statistics and quantum numbers of skyrmions. Finally we summarizes our main results and look into future directions in section \ref{sec:conclusion}. Clarification on the notations and a self-contained introduction to the mathematical tools can be found in the Appendices.

\section{General framework}\label{sec:theory framework}

We consider the ground state of a 2d quantum many-body system, which exhibits a long-range order associated with spontaneous symmetry breaking. To be precise, the symmetry group $G$ of the Hamiltonian spontaneously breaks down to a subgroup $H$ that is preserved in an ordered ground state. Moreover, we assume that the possible Goldstone modes, from spontaneously broken continuous symmetries, are the only gapless excitations in the bulk of the system. In other words, the ground state with a fixed nonzero order parameter $O(\vec r)=\expval{\hat O({\vec r})}\neq0$, is a gapped symmetric phase that preserves the remnant symmetry $H$. In the rest part of the paper, for simplicity, we assume the classical gapless Goldstone modes do not affect the topological data of the remnant gapped quantum phases that we are interested in. In two spatial dimensions, this means the ground state is an $H$-symmetry enriched topological order. The question we are answering is, what are the universal properties of topological defects and textures in the order parameters therein, when the symmetry-breaking ground state is a topological state enriched by the remnant symmetry $H$?

To address this question, we need to consider topological defects and textures as excitations in a symmetry-breaking ground state. They turn out to be connected to a special type of excitations known as extrinsic symmetry defects (or twist defects) in symmetry enriched topological (SET) phases. This correspondence allows us classify universal properties of topological defects and textures in ordered media with a nontrivial ground state topology. The key mathematical tool to establish this connection is the long exact sequence of homotopy groups for topological defects and textures. In this section, we outline this connection between classical topology of the order parameters and quantum topology of the entangled ground states, and then utilize this connection to classify point defects and textures in following sections.

\begin{table}[!h]
    \centering
    \begin{tabular}{|c|ccc|}
    \hline
         &$d=1$ & $d=2$ & $d=3$ \\
         \hline
         $\pi_0$& point defect & line defect & surface defect \\
         $\pi_1$& texture & point defect & line defect \\
         $\pi_2$& & texture & point defect \\
         $\pi_3$& && texture \\
         \hline
    \end{tabular}
    \caption{Homotopy group for topological defect/texture in $d=1,2,3$ dimension.}
    \label{Homotopy_Table}
\end{table}

\subsection{Homotopy theory of topological defects and textures: a brief review}

First we briefly review the homotopy theory of topological defects and textures in the order parameters~\cite{Mermin1979}. Mathematically, the long-range order of spontaneous symmetry breaking is described by a local order parameter %living on the manifold $\mathcal{M}$:
\bea
O(\vec r)\in\mathcal{M}=G/H,
\eea
i.e. the order parameter is valued on the (left) coset space of $G$ modulo $H$, where the full symmetry $G$ of the Hamiltonian is spontaneously broken down to a subgroup $H$ in a ground state with a fixed order parameter configuration. In particular, the remnant symmetry $H$ is the subgroup of $G$ which keeps the order parameter $\{\hat O(\vec r)\}$ invariant:
\bea
H\equiv\{h\in G|h\hat O(\vec r) h^{-1}=\hat O(\hat h\vec r)\}.
\eea
Given the order parameter manifold $\mathcal{M}=G/H$, in $d$ spatial dimensions, one can consider an order parameter configuration with point (line, surface etc.) defects, where the order parameter $O(\vec r)$ is a smooth function of spatial coordinate $\vec r$ except for singularities on isolated points (lines, surfaces etc.). Most generally, a $(d-D-1)$-dimensional defect (i.e. a defect of codimension $D+1$) is described by a continuous map:
\bea
\vec r\in S^D\rightarrow O(\vec r)\in\mathcal{M}=G/H
\eea
of order parameters on a submanifold enclosing the defect. The inequivalent classes of $(d-D-1)$-dimensional defects in $d$ spatial dimensions is hence classified by the homotopy group $\pi_{D}(G/H)$~\cite{Mermin1979} for $d\geq D+1$. Below we list a few defects in low dimensions:

(i) 0-dimensional point defects are classified by $\pi_{d-1}(G/H)$ for $d\geq1$;

(ii) 1-dimensional line defects are classified by $\pi_{d-2}(G/H)$ for $d\geq2$;

(iii) 2-dimensional defects are classified by $\pi_{d-3}(G/H)$ for $d\geq3$.

In addition to defect configurations where the order parameter $O(\vec r)$ becomes singular somewhere in space, homotopy theory also classifies textures of the order parameter configurations which are smooth everywhere. They are classified by the following continuous map:
\bea
\vec r\in S^{d}\rightarrow O(\vec r)\in\mathcal{M}=G/H
\eea
where we compactify the $d$-dimensional real space to $S^{d}$. As a result, topologically inequivalent textures in $d$ dimensions are classified by the homotopy group $\pi_d(G/H)$. Similarly one can also consider spacetime textures, classified by homotopy group $\pi_{d+1}(G/H)$ where the $(d+1)$-dimensional spacetime is compactified to $S^{d+1}$. Homotopy group for topological defect/texture in $d=1,2,3$ dimension can be found in Table.~\ref{Homotopy_Table}.

In this work, we shall restrict ourselves to two spatial dimensions ($d=2$), where different types of topological defects and textures are classified by the following homotopy groups:

(i) Domain walls with codimension 1, where order parameters are smooth everywhere except for along a line, are classified by $\pi_0(G/H)$;

(ii) Point defects (i.e. vortices) with codimension 2, where order parameters are smooth everywhere except for one point, are classified by $\pi_1(G/H)$;

(iii) Textures where order parameters are smooth everywhere, are classified by $\pi_2(G/H)$. A well-known example is a skyrmion in a $2+1$D O(3) nonlinear sigma model (NLSM), as will be discussed in detail later.

The main goal of this work is to establish a connection between topological defects and textures of the symmetry-breaking order parameters, and the underlying topological ground states. The main mathematical tool that reveals this connection is the long exact sequence of homotopy groups~\cite{Mermin1979}:
\begin{equation}\label{eq:long sequence}
    \begin{aligned}
        \cdots&\longrightarrow\pi_n(H)\longrightarrow\pi_n(G)\longrightarrow\pi_n(G/H)\\
        &\longrightarrow\pi_{n-1}(H)\longrightarrow\pi_{n-1}(G)\longrightarrow\cdots.
    \end{aligned}
\end{equation}
Here, ``exact'' means that for each term $G$ in the sequence $N\xrightarrow{~p~}G\xrightarrow{~q~}Q$, the kernel of the outgoing map $q$, $\mathrm{ker}(q) = \{m\in G|q(m)=0\}$ is equal to the image of the incoming map $p$, $\mathrm{im}(p) = f(N)$: $ {\rm im}(p) = {\rm ker}(q)$. We noticed that the general idea of mapping topological defects and textures to symmetry defects through ``connecting homomorphism'' $\pi_k(G/H)\rightarrow\pi_{k-1}(H)$ have been pointed out in Refs.~\cite{propitius1995topological,Else2021}.

\subsection{Domain walls}
 \begin{figure}[!h]
\centering
\includegraphics[width=1\columnwidth]{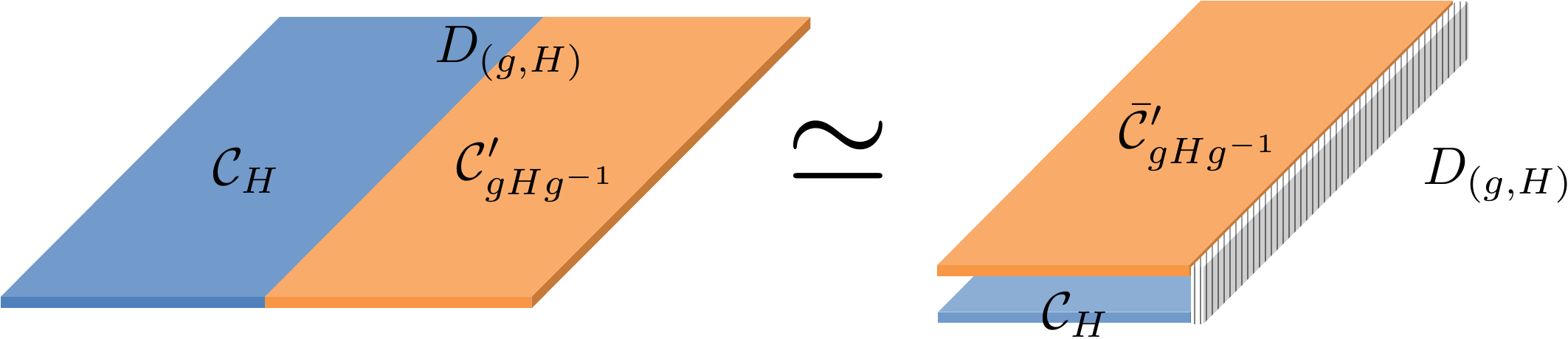}
\caption{\label{fig:Domain Wall} The equivalence between a domain wall between two phases $\mathcal{C}_H$ and $\mathcal{C}^\prime_{gHg^{-1}}$, and the boundary of a 2d phase (Eq.~\eqref{eq:folding trick}).}
\end{figure}

In 2d, a gapped phase that preserves remnant symmetry $H$ is generally an $H$-SET phase, whose anyon excitations are described by a unitary modular tensor category $\mathcal{C}$ \cite{Kitaev2006,Barkeshli2019,Tarantino2016}. We use $\mathcal{C}_H$ to label such an $H$-SET phase. The domain wall in two spatial dimension is a line defect characterized by $\pi_0(G/H)$. In a gapped system whose symmetry group $G$ of the Hamiltonian is spontaneously broken, a generic domain wall $D_{(g,H)}$ is labeled by a remnant subgroup $H<G$ and a group element $g\notin H$, such that it separates a left domain that preserves symmetry subgroup $H$, and a right domain that preserves subgroup $gHg^{-1}$, as shown in Fig. \ref{fig:Domain Wall}. We can label the $H$-SET phase on the left domain as $\mathcal{C}_H$, and the $(gHg^{-1})$-SET phase on the right domain would be
\bea
U_g\mathcal{C}_HU_{g}^{-1}\simeq\mathcal{C}^\prime_{gHg^{-1}}
\eea
where we use $U_g$ to label the action of broken symmetry element $g$ on the $H$-SET phase on the left domain. Note that the right domain and the left domain may not share the same topological order $\mathcal{C}$, e.g. in the case of a time reversal domain wall with $g=\bst$, the left and right domains have opposite chiralities and hence $\mathcal{C}^\prime=\bar{\mathcal{C}}$, where $\bar{\mathcal{C}}$ is defined as the time reversal counterpart of topological order $\mathcal{C}$. In this setup, using the folding trick, it is clear that the domain wall $D_{(g,H)}$ between left domain $\mathcal{C}_H$ and right domain $\mathcal{C}^\prime_{gHg^{-1}}$ can be mapped to the boundary of a 2d topological phase:
\bea\label{eq:folding trick}
(\mathcal{C}\boxtimes\bar{\mathcal{C}^\prime})_{H_g}\equiv\mathcal{C}_H\boxtimes\bar{\mathcal{C}}^\prime_{gHg^{-1}}
\eea
where we denote the remnant unbroken symmetry of the domain wall configuration as
\bea\label{H_g}
H_g\equiv H\cap(gHg^{-1})=\{h\in H|g^{-1}hg\in H\}
\eea
Therefore the universal properties of domain walls $D_{(g,H)}$ is captured by boundary excitations of a 2d topological phase described by (\ref{eq:folding trick}), as illustrated in Fig. \ref{fig:Domain Wall}.

The physics of boundary excitations in a 2d topological order is in fact a subject extensively studied in the literature \cite{Kitaev2012,Bridgeman2020,Hung2015,Lan2015,Hu2017,Bischoff2019,Lan2020,KONG2018,KONG2021}. In light of the above physical picture that maps a domain wall to a boundary, we will not attempt to classify topological properties of domain walls in SET phases in this manuscript. From now on we shall discuss only the point defects and textures in two spatial dimensions.

\subsection{Point defects}\label{subsec:vortex}

In two spatial dimensions, point defects are classified by the fundamental group $\pi_1(G/H)$. Two representative examples of point defects in 2d are the following:

(1) In a system of interacting spinless bosons whose ground state is an $m$-boson condensate, the boson number conservation symmetry $G=U(1)$ is spontaneously broken down to a $H=Z_n$ subgroup. The vortices of such an $m$-boson condensate are classified by $\pi_1(G/H)=\pi_1(U(1))=\mbz$. In particular, the fundamental vortex with unit winding number $\nu=1$ is equivalent to a $2\pi/m$ flux by a gauge transformation.

(2) For interacting ions that form a crystalline lattice in two dimensions, the continuous translation symmetry $G=\mathbb{R}^2$ is spontaneously broken down to a discrete subgroup $H=\mbz^2$. The associated point defects, i.e. dislocations, are classified by $\pi_1(G/H)=\mbz^2$, characterized by a Burgers vector $\vec b=b_1\vec a_1+b_2\vec a_2$, where $(b_1,b_2)\in\mbz^2$ and $\vec a_{1,2}$ are the two primitive lattice vectors. Physically, we can consider a close loop on a translation-invariant lattice. If a particle follows exactly the same path of the loop, which now encloses a dislocation, the particle will not return to the starting point after finishing the path. Instead, the final position differs from the initial position by the Burgers vector $\vec b$ of the dislocation enclosed by the path.

\begin{figure}[!h]
\centering
\includegraphics[width=1\columnwidth]{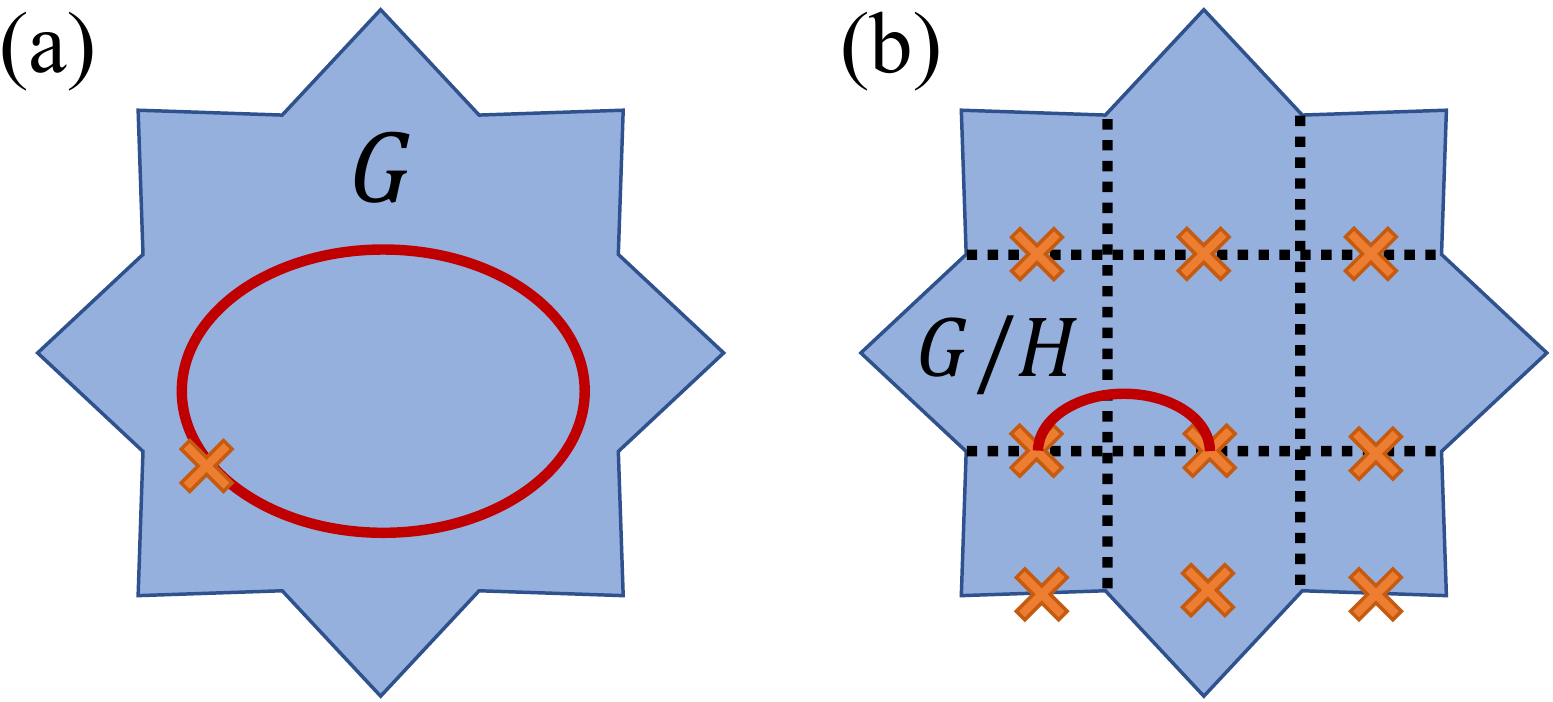}
\caption{\label{Symmetry_Breaking} Illustration of the fundamental group $\pi_1$ of the group manifold. (a) $\pi_1(G)$ captures the winding in $G$-space. (b) $\pi_1(G/H)$ captures the winding in $G/H$ space. Note that the open line in (a) may be considered as closed loop in (b) as those points are identified in $G/H$ space. }
\end{figure}

Both examples belong to the general case where a connected topological (continuous) group $G$ is broken down to a discrete subgroup $H$. We assume $H$ to be a normal subgroup of $G$ (i.e. $H\triangleleft G$), so that the whole point defect configuration of order parameters preserves symmetry $H$. Applying the long exact sequence Eq.~\eqref{eq:long sequence} to the $n=1$ case, we obtain the following \emph{short exact sequence}:
\begin{equation}\label{eq:short sequence}
    \begin{aligned}
       \cdots \longrightarrow \pi_1(H) =0 \xrightarrow{~f~}\pi_1(G) \xrightarrow{~i~}\pi_1(G/H) \\
       \xrightarrow{~p~} \pi_0(H) = H \xrightarrow{~g~} \pi_0(G) =0 \longrightarrow \cdots
    \end{aligned}
\end{equation}
where $p$ is known as the connecting homomorphism between the topological point defects and symmetry defect. The group $\pi_1(G)$ and $\pi_1(G/H)$ are illustrated in Fig.~\ref{Symmetry_Breaking}.

This short exact sequence of groups can be understood as follows: using the exactness of the sequence, it is clear that $i$ is an injective map, hence $\pi_1(G)$ is a normal subgroup of $\pi_1(G/H)$. The connecting homomorphism $p$ is a surjective map so $H$ is isomorphic to the quotient group $\pi_1(G/H)/\pi_1(G)$. Such a short exact sequence defines a group extension problem, and here we say that $\pi_1(G/H)$ is a group extension of the group $\pi_1(G)$ by $H$.

Physically, the surjective map $g$ in the short exact sequence (\ref{eq:short sequence}) connects topological point defects classified by $\pi_1(G/H)$, to symmetry defects associated with elements of the remnant symmetry group $H$ in the symmetry-breaking ground state, which is generally an $H$-SET phase. Note that the classification and characterization of $H$-SET phases~\cite{Barkeshli2019,Tarantino2016,Teo2016} is in fact built upon an algebraic theory of $h$-defects for $h\in H$, which has been extensively studied previously. The above group extension allows us to map each point defect classified by $\pi_1(G/H)$ to an $h$-defect associated with the group element $h\in H$, therefore allowing us to characterize the topological properties of point defects. In the two examples mentioned above: (1) Since $\pi_1(G=U(1))\neq0$, a vorticity-$\nu$ vortex in an $m$-boson condensate is mapped to a $2\pi\nu/m$ flux of the remnant $Z_m$ symmetry, and hence the map $p$ is surjective but not injective. (2) Since $\pi_1(G=\mathbb{R}^d)=0$, the dislocations are in one to one correspondence with translation symmetry defects, and hence the map $p$ is bijective.

Below we shall apply this idea to two different families of symmetry-breaking phases:

(i) When a symmetry-breaking ground state has no intrinsic topological order, it is described by an $H$-SPT phase. The point defects can carry linear or projective representations of the remnant symmetry $H$, a phenomenon closely related to deconfined quantum critical points (DQCP). We discuss point defects of this family in section \ref{sec:spt+dqcp}.

(ii) When there are intrinsic topological orders in a symmetry-breaking ground state, it is generally an $H$-SET phase. In this case, the point defects can be non-Abelian defects that permute anyons, or they can exhibit exotic fusion rules. We classify and discuss point defects of this family in section \ref{sec:point defect}.

\subsection{Textures}\label{subsec:skyrmion}

The most familiar example of a topological texture in two spatial dimension is a skyrmion. When a ferromagnetic order breaks the $G=SO(3)$ spin rotational symmetry to a uniaxial spin rotation subgroup $H=U(1)$, the order parameter manifold is a 2-sphere $\mathcal{M}=G/H=S^2$ with nontrivial textures classified by $\pi_2(S^2)=\mbz$. In fact, as summarized in Table \ref{tab:realization}, most familiar realizations of topological textures in 2d are essentially classified by $\pi_2(S^2)=\mbz$, and we shall focus on skyrmions as 2d textures in the manuscript.

In the case of skyrmions, the long exact sequence (\ref{eq:long sequence}) reduces to the short exact sequence
\begin{equation}\label{eq:short exact sequence:texture}
    \begin{aligned}
       \cdots &
       %&\longrightarrow \pi_2(U(1)) =0
       \xrightarrow{~~~} \pi_2(SO(3)) \!=\!0
        \xrightarrow{~~~} \pi_2(S^2) \!=\! {\mathbb Z}\xrightarrow{~i~} \pi_1(U(1)) \!=\! {\mathbb Z} \\ & \xrightarrow{~~~}\pi_1(SO(3)) \!=\! {\mathbb Z}_2\xrightarrow{~p~} \pi_1(S^2) \!=\!0 \longrightarrow  \cdots
    \end{aligned}
\end{equation}
This short exact sequence indicates that the connecting homomorphism $i$ is an injective map from skyrmions labeled by $\pi_2(S^2)=\mbz$ to fluxes labeled by $\pi_1(U(1))=\mbz$. More precisely, the group of $U(1)$ fluxes $\pi_1(U(1))=\mbz$ is a central extension of the skyrmion group $\pi_2(S^2)=\mbz$, with the center being the group of point defects $\pi_1(SO(3))=\pi_1(SU(2)/Z_2)=\mbz_2$ \footnote{Note that, here we have used the fact $\pi_1(SU(m)/{\mathbb Z}_m) = {\mathbb Z}_m$ from the following four-term exact sequence%~\cite{Lu2012}:
\begin{equation*}
    \begin{aligned}
       &\pi_1(SU(m))=0  \xrightarrow{~f_1~} \pi_1(SU(m)/{\mathbb Z}_m)  \\
       &\xrightarrow{~f_2~} \pi_0({\mathbb Z}_m) = {\mathbb Z}_m \xrightarrow{~f_3~} \pi_0(SU(m)) = 0,
    \end{aligned}
\end{equation*}
Using the exactness of the terms, one can show that $f_2$ is an isomorphism map, therefore $\pi_1(SU(m)/{\mathbb Z}_m) \cong {\mathbb Z}_m$.}.

Physically, it is known that the $\mbz_2$ point defect is nothing but a $2\pi$ vortex for the $SO(3)$ spins~\cite{Kawamura1984,Dombre1989}. In other words, the spin of a particle is rotated around one (any) axis by $2\pi$ after circling around the nontrivial point defect in $\pi_1(SO(3))=\mbz_2$. Eq.~\eqref{eq:short exact sequence:texture} defines a map from the element $a \in \pi_2(S^2)$ to  $b \in \pi_1(U(1))$ and then to the trivial element in $\pi_1(SO(3)) = {\mathbb Z}_2$. Since the non-trivial element for that ${\mathbb Z}_2$ point defect is a $2\pi$ vortex for spins, the trivial element of ${\mathbb Z}_2$ should correspond to a $4\pi$ vortex. Therefore, in Eq.~\eqref{eq:short exact sequence:texture}, a skyrmion with winding number $\nu\in\mbz$ is mapped to a $4\pi\nu$ flux of the $U(1)$ spin rotations, i.e., $2\nu\in\pi_1(U(1))=\mbz$ \cite{Grover2008,Else2021}.

The map in \eqref{eq:short exact sequence:texture} points to the nature of skyrmions in a ferromagnetic topological order where $SO(3)$ spin rotational symmetry is spontaneously broken down to a $U(1)$ subgroup, as the topological properties of each $\nu=1$ skyrmion can be extracted from those of $4\pi\nu$ flux/defect of the unbroken $U(1)$ symmetry. We shall follow this approach to identify the fractional stastistics of skyrmions in section \ref{sec:texture}.

\section{Group cohomology for point defects and textures}\label{sec:group cohomology}

\subsection{Group cohomology for symmetry defects: a brief review}\label{subsec: group cohomology for symmetry defects}
Following Ref.~\cite{Barkeshili2014}, below we give a definition for symmetry defects in both SPT and SET orders.
When the physical system has a symmetry $H$ with a given symmetry action $\rho$ on the quasi-particles, one can consider modifying the system by introducing a point-like defect $\tau_h$ associated with a group element $h \in H$. When a quasi-particle is braided around an $h$-symmetry defect $\tau_h$, it is acted upon by the corresponding symmetry action ${\rho}_{h}$. Since $H$ is a global symmetry, the point-like defects $\tau_h$ are not finite-energy excitations and must be extrinsically imposed by threading the symmetry flux of $h$. 

As group cohomology is a crucial mathematical object for this work, we give a self-contained introduction to it from the mathematical side in the App.~\ref{App:B2}. More detailed characterization and intuition for it from the physics side will be given here and in later sections.

\subsubsection{Symmetry protected topological (SPT) phases}\label{sec:recap_SPT}

We start by recalling the definition and classification of bosonic SPT phases. An $H$-SPT phase is a short-range entangled phase, which, in the presence of symmetry $H$, cannot be continuously connected to a trivial product state without closing the energy gap. In a system of interacting bosons in $d$ spatial dimensions, different $H$-SPT phases are classified by the $(d+1)$-th group cohomology $\mathcal{H}^{d+1}(H,U(1))$ \cite{Chen2013}. The cohomology group $\mathcal{H}^{d+1}(H,U(1))$ is an Abelian group, whose identity element labels the topologically trivial phase (a featureless product state), and the addition of group elements is implemented by stacking different SPT phases.

When symmetry $G$ is spontaneously broken down to $H$ in a given ground state, a long-range ordered ground state with fixed order parameters is an $H$-preserving short-range entangled phase\footnote{We use the definition of Ref.~\cite{Chen2013} for short-range entangled phase, which is different from Kitaev's definition~\cite{Kitaev2006}. Therefore we do not consider invertible phases such as topological superconductor in class D~\cite{Kitaev2009}, and $E_8$ states of $d=2$ interacting bosons~\cite{Kitaev2006,Lu2012}.} which are $H$-SPT phases classified by the group cohomology $\mathcal{H}^3(H,U(1))$\cite{Chen2013}.

We consider a 2d SPT phase protected by a symmetry group $H=B\times K$, which is a direct product of two groups $B$ and $K$. The K\"unneth formula indicates that the classification of $H$-SPT phases can be written as a direct product~\cite{Chen2013}:
\bea
\mathcal{H}^{d\!+\!1}(B\!\times \!K,U(1))=\prod_{i=0}^{d\!+\!1}\mathcal{H}^i\big(B,\mathcal{H}^{d\!+\!1\!-\!i}(K,U(1))\big)~~~
\eea
In the $d=2$ case which is the interest of this manuscript, we have
\bea\notag
&\mathcal{H}^3(B\!\times \!K,U(1))=\mathcal{H}^3(B,U(1))\times\mathcal{H}^2(B,\mathcal{H}^1(K,U(1)))\\
&\times\mathcal{H}^1(B,\mathcal{H}^2(K,U(1)))\times\mathcal{H}^3(K,U(1))\label{kunneth}
\eea
where each term has its own physical meaning. Clearly the first term $\mathcal{H}^3(B,U(1))$ classifies the SPT phases protected only by the subgroup $B$, while the last term $\mathcal{H}^3(K,U(1))$ classifies SPT phases protected only by the subgroup $K$.

In this work, we are particularly interested in the mixed anomaly of symmetry $B$ and $K$, which assigns quantum numbers or projective representations of subgroup $K$ to $B$ symmetry defects. The mixed anomaly is captured by the 2nd and 3rd terms in the K\"unneth expansion (\ref{kunneth}), as suggested by the ``decorated domain wall'' picture of the SPT phases \cite{Chen2014}.

(i) The 2nd term $\mathcal{H}^1(B,\mathcal{H}^2(K,U(1)))$ assigns a projective representation $[\omega]\in\mathcal{H}^2(K,U(1))$ to any symmetry defect $\alpha_g$ associated with $g\in B$. As discussed in section \ref{subsec:vortex}, in the exact sequence (\ref{eq:short sequence}), the connecting homomorphism $p$ maps a symmetry breaking point defect (i.e. vortex) of order parameters classified by $\pi_1(G/H)$ to a symmetry defect labeled by elements of $H=B\times K$. If the associated symmetry defect corresponds to an element $g\in B$, this suggests that the point defect of order parameters can carry a projective representation of $K$. This map will be discussed in detail in section \ref{subsec:cohomology:vortex}.

(ii) For topological textures of the order parameters in $d=2$, we shall focus on the case of skyrmions, where the remnant symmetry is $H=U(1)\times K$ with $B=U(1)$. In other words, we consider the full symmetry $G=SO(3)\times K$ is broken down to $H=U(1)\times K$ in the collinear long-range order. In this case, we utilize an important result from the theory of cohomology: for any finite Abelian group $M$ on which $U(1)$ and $\mathbb{Z}$ both act trivially\footnote{Here it is important to specify that by $\mathcal{H}^2(U(1),M)$ we are computing the Borel cohomology, which is the standard cohomology for bosonic SPT \cite{Chen2013}. The calculation uses the relation between Borel cohomology and simplicial cohomology $\mathcal{H}^2(U(1),M) = H^2(BU(1),M) = H^2(\mathbb{CP}^\infty,M)$ and the procedure of regarding $H^2(CP^\infty,M)$ as the limiting case of $H^2(CP^n,M)\cong M$ for $n\rightarrow \infty$ following Ref.~\cite{Chen2013}. On the other hand, the part $\mathcal{H}^1(\mbz,M)\cong M$ can be obtained in either standard group cohomology or Borel cohomology.},
%\ref{subsec:inflation-restriction}:
\bea\label{U(1) to Z}
\mathcal{H}^2(U(1),M)\cong M\cong \mathcal{H}^1(\mbz,M).
\eea

Mathematically, both cohomology groups classify different linear representations $\{R_\nu\in M|\nu\in\mbz\}$ of the integer ($\mbz$) group formed by the $2\pi\nu,~\nu\in\mbz$ fluxes of the remnant $U(1)$ symmetry with coefficients $R(\nu)\in M$.

As a result, with $B=U(1)$, we can rewrite the 3rd term in K\"unneth expansion (\ref{kunneth}) as
\bea\label{mixed anomaly:U(1)}
\mathcal{H}^2\big(U(1),\mathcal{H}^1(K,U(1))\big)=\mathcal{H}^1\big(\mbz,\mathcal{H}^1(K,U(1))\big)~~~
\eea
Physically, this can be interpreted as assigning quantum numbers (or charges) of symmetry group $K$, i.e. linear representation $\mathcal{H}^1(K,U(1))$, to the integer fluxes of the unbroken $U(1)$ subgroup.

As discussed in section \ref{subsec:skyrmion}, the connecting homomorphism $i$ in (\ref{eq:short exact sequence:texture}) maps the skyrmions labeled by $\nu_s\in\pi_2(G/H)=\mbz$ to integer fluxes labeled by $\nu_f\in\pi_1(U(1))=\mbz$. As a result, we can use the SPT classification $\mathcal{H}^2\big(U(1),\mathcal{H}^1(K,U(1))\big)$ for each symmetry-breaking ground state to determine the $K$ symmetry quantum numbers assigned to each skyrmion. This map will be discussed in detail in section \ref{subsec:cohomology:texture}.

\subsubsection{Symmetry enriched topological (SET) order}\label{sec:recap_SET}
Consider a topological order described by a unitary modular tensor category $\mathcal{C}$, enriched by symmetry $H$ that is a discrete group. Since the topological order preserves the unbroken symmetry $H$, an SET phase can arise which in addition to the topological order displays the phenomenon of fractionalization of symmetry~\cite{Essin2013,Tarantino2016,Barkeshli2019}. Here we briefly review the physical picture of symmetry fractionalization and the classification of SET.

For each group element $h \in H$, one can associate an extrinsic defect/flux $\tau_h$~\cite{Tarantino2016,Barkeshli2019} in the $H$-symmetry enriched topological order. A symmetry defect can be further \emph{labeled} by an anyon $\mathsf{a}\in\mathcal{C}$ of the topological order, with the identity symmetry defect (i.e. the symmetry defect associated with the identity element of the group) identified with the anyon $\mathsf{a}$ itself. The symmetry defects in this sense are generalized anyons and can have nontrivial fusion and braiding rules. In particular, anyons may be permuted when braided around a symmetry defect $\tau_h$. This braiding is one way of detecting the symmetry action $\rho \colon H\rightarrow \mathrm{Aut}(\mathcal{C})$, where $\mathrm{Aut}(\mathcal{C})$ denotes the automorphism group of the anyons. The first group cohomology group, $\mathcal{H}^1_\rho(H,\mathcal{A})$, classifies the (possibly crossed, when $\rho$ nontrivial) homomorphisms $[\omega]$ where $\omega\colon H\rightarrow \mathcal{A}$ describes anyon labeling that is consistent with the symmetry action on the anyons, $\rho$. Two anyon-labeling homomorphisms $\omega', \omega''$ belong to the same class $[\omega]$
if they differ only by an anyon permutation described by $\rho$.

The additional data to specify an SET phase is symmetry fractionalization (whose physical meaning will now be specified), classified by the second group cohomology $\mathcal{H}^2_{\rho}(H,\mathcal{A})$. Here $\rho$ denotes the symmetry action on the anyons as introduced above, and $\mathcal{A}\subset \mathcal{C}$ is the set of Abelian anyons viewed as an Abelian group under fusion ~\cite{Tarantino2016,Barkeshli2019}\footnote{Note that in the case of SPT discussed in Sec.~\ref{sec:recap_SPT}, the action of the symmetry $H$ on the $U(1)$ coefficient is uniquely determined by the symmetry $H$ itself. Thus here and after we only specify the action of the group cohomology for SET.}. 
The elements $[\omega] \in \mathcal{H}^2_{\rho}(H,\mathcal{A})$ are cocycle elements $\omega$ under the equivalence relation of anyon labeling. The cocycle elements $\omega(g,h) \in \mathcal{A}$ takes as an input two group elements $g,h \in H$ and returns an anyon. Physically, the cocycle, roughly speaking, describes the outcome of fusing three defects $\tau_g$, $\tau_h$, and $\tau_{(gh)^{-1}}$: the result is a trivial defect associated to the identity element of $H$ as required by the ``conservation law''of the symmetry $H$, and the only possible outcome is an Abelian anyon. The symmetry fractionalization data precisely refer to the anyon outcomes after fusion. Crucially, it is inappropriate to think of the outcome anyon as fixed since different anyons may be identified under the equivalence relation; rather symmetry fractionalization refers to the inequivalent ways of assigning defect fusion rules up to anyon relabeling.

To summarize, in an $H$-SET phase, two important rules exist: the defect ``anyon-labeling rule'', classified by $\mathcal{H}^1_\rho(H,\mathcal{A})$, and the defect ``fusion rule'', which encode symmetry fractionalization data classified by $\mathcal{H}^2_\rho(H,\mathcal{A})$. As we show below, all these data have correspondence in a phase with coexisting symmetry breaking defect and topological order.

\subsection{The inflation-restriction exact sequence}\label{subsec:inflation-restriction}

In this subsection we introduce the main mathematical tool we will be relying on in the physical interpretation of group cohomology, the \emph{inflation-restriction  exact sequence}.

Recall that the input data for group cohomology $\mathcal{H}^n_\rho(G,M)$ is a group $G$ and an Abelian group $M$ equipped with a $G$-action $\rho\colon G\rightarrow \mathrm{Aut}(M)$, $g\mapsto \rho(g)$ (we use $\rho(g)$ and $\rho_g$ interchangeably). For convenience, we denote the $G$-action on $M$ by the symbol ``$.$'', that is, for  $g \in G$ and $\mathsf{a}\in M$, we define $\rho(g)(\mathsf{a}) \equiv g.\mathsf{a}$. Suppose $N$ is a normal subgroup of $G$, and $Q:=G/N$ the associated quotient group. Formally, we say that $G$ is an extension of the group $Q$ by $N$ that fits into the short exact sequence
\bea \label{NGGN}
\xymatrix{0 \ar[r] & N \ar[r]^i\ar[rrd]_{\rho_N} &G \ar[r]^{p\qquad} \ar[dr]^\rho & Q = G/N\ar[d] \ar[r] &0 \\
&&&\mathrm{Aut}(M)&}
 \eea
Note that here the action $\rho_N\colon N\rightarrow \mathrm{Aut}(M)$ is inherited from the action $\rho$. There is also an action $\rho_Q\colon Q\rightarrow \mathrm{Aut}(M^N)$, here $M^N$ denotes the subgroup of $M$ that are stabilized under the action of $N$: $M^N = \{\mathsf{a} \in M|n.\mathsf{a}=\mathsf{a}~\forall n \in N\}$. 

Given these data, a five-term exact sequence exists
\cite{rotman2009introduction}
\begin{equation}\label{five_term}
\begin{aligned}
0\rightarrow &\mathcal{H}^1(Q,M^N)
\rightarrow \mathcal{H}^1(G,M)
\xrightarrow{\text{res}} \mathcal{H}^1(N,M)^Q\\
&\xrightarrow{d_2}\mathcal{H}^2(Q,M^N)\xrightarrow{\text{inf}}
\mathcal{H}^2(G,M),
\end{aligned}
\end{equation}
this is the {inflation-restriction exact sequence}. Here $\mathcal{H}^1(N,M)^Q$ denotes the subgroup of $\mathcal{H}^1(N,M)$ that are stabilized under the action of $Q$ in the following sense: for $[\omega] \in \mathcal{H}^1(N,M)$ represented by the cocycle $\omega$,
\begin{equation}\label{actupperQ}
[\omega] \in \mathcal{H}^1(N,\mathcal{A})^Q
~~ \Longleftrightarrow ~~ q.(\omega(q^{-1}nq)) = \omega(n),
\end{equation}
for any $q \in Q$ and $n \in N$. The maps ``res'', ``$d_2$'' and ``inf'' are called the restriction, the transgression (or the differential), and the inflation maps, respectively, and can be defined explicitly. A sketch of the proof of the five-term exact sequence and further information about the maps can be found in App.~\ref{app:c}.

In the following, we will apply the five-term exact sequence to various short exact sequences of homotopy groups. These homotopy groups describe either point defects or textures in the broken symmetry phase, and the broken symmetry phase is either an SPT or an SET. Below we discuss each cases separately.

\subsection{Group cohomology for point defects}\label{subsec:cohomology:vortex}

\subsubsection{Point defects in SPT}\label{cohomology:vortex:spt}

Consider a symmetry breaking from $G = A \times K$ to $H = B \times K$, where $A$ is a continuous group and $B$ a discrete group. A symmetry-breaking ground state with a fixed order parameter is an $H$-SPT, and we focus on the mixed anomaly described by the  $\mathcal{H}^1(B,\mathcal{H}^2(K,U(1)))$ term in the K\"{u}nneth formula (\ref{kunneth}).

In this case, the topological point defects of the order parameter $\phi(\vec{r}) \in G/H$ is classified by the fundamental group  $\pi_1(G/H) =\pi_1(A/B)$ which fits into the short exact sequence \eqref{eq:short sequence}. Recall from the K\"unneth decomposition \eqref{kunneth} that the term $\mathcal{H}^1(B,\mathcal{H}^2(K,U(1)))$ assigns projective representations in $M:=\mathcal{H}^2(K,U(1))$ to the symmetry defects of $B$. Note that here we consider the case of $B$ consisting of only unitary symmetry, therefore $B$ acts trivially on $M$. By \eqref{NGGN}, now the point defects labeled by $\pi_1(A)$ and $\pi_1(A/B)$ all act trivially on $M$.

Then, applying the five-term exact sequence \eqref{five_term} gives
\begin{equation}\label{five_term:vortex:spt}
\begin{aligned}
&0\rightarrow\mathcal{H}^1(B,{\mathcal H}^2(K,U(1)))
\rightarrow \mathcal{H}^1(\pi_1(A/B),{\mathcal H}^2(K,U(1)))\\
&\xrightarrow{\text{res}} \mathcal{H}^1(\pi_1(A),{\mathcal H}^2(K,U(1)))\\
&\xrightarrow{d_2}\mathcal{H}^2(B,{\mathcal H}^2(K,U(1)))\xrightarrow{\text{inf}}
\mathcal{H}^2(\pi_1(A/B),{\mathcal H}^2(K,U(1))),
\end{aligned}
\end{equation}

The term $\mathcal{H}^1(B,\mathcal{H}^2(K,U(1)))$ is exactly the term in the K\"unneth decomposition \eqref{kunneth} discussed above, which gives classification for $H$-SPT phases with the mixed anomaly. The image of injective map $d_1$ in $\mathcal{H}^1(\pi_1(A/B),\mathcal{H}^2(K,U(1)))$ classifies the $K$ projective representation carried by point defects in $\pi_1(A/B)$.

Since $\mathcal{H}^2(B,{\mathcal H}^2(K,U(1)))$ is one piece out of the K\"unneth decomposition for $\mathcal{H}^4(B\times K,U(1))$, it physically corresponds to an anomalous symmetry implementation of group $H=B\times K$ on the surface of a three-dimensional $H$-SPT phase. It can only happen on the two-dimensional surface of a three-dimensional $H$-SPT phase, but not in any two-dimensional lattice models with onsite symmetry actions. Therefore the image of the restriction map is the same as the kernel of the transgression  map $d_2$ in the exact sequence (\ref{five_term:vortex:spt}).

More physics of the point defects in $H$-SPT phases will be discussed in section \ref{sec:dqcp:vortex}.

\subsubsection{Point defects in SET}\label{cohomology:vortex:set}

We consider a continuous symmetry $G$ spontaneously broken down to a discrete subgroup $H$, where each symmetry-breaking ground state is an $H$-SET phase, whose intrinsic topological order is specified by the anyons $\mathcal{C}$.

Recall from previous discussions that SET is equipped with a symmetry action on the anyons, $\rho\colon H\rightarrow \mathrm{Aut}(\mathcal{C})$ \cite{Barkeshli2019,Tarantino2016}. Suppose that the system started with a larger (continuous)
symmetry $G$ that spontaneously broke down to a discrete symmetry group $H$. The topological point defects of the order parameter is again classified by $\pi_1(G/H)$ %are present in the post-SSB system, whose classification, $\pi_1(G/H)$,
that fits into the short exact sequence \eqref{eq:short sequence}.
 Then, one can pull back the symmetry $H$-action $\rho$ to obtain actions of the topological point defects on the anyons:
 \bea \label{inducedaction:vortex:set}
\xymatrix{\pi_1(G) \ar[r]^i\ar[rrd]_{\tilde{\tilde{\rho}}} &\pi_1(G/H)\ar[r]^p \ar[dr]^{\tilde{\rho}} & H\ar[d]^\rho\\
&&\mathrm{Aut}(\mathcal{C}).}
 \eea
Applying the five-term exact sequence \eqref{five_term} to Eq.~\eqref{inducedaction:vortex:set} gives
\begin{equation}\label{five-term:vortex:set}
\begin{aligned}
    0 &\rightarrow
{\mathcal H}^1_\rho(H,\mathcal{A})\rightarrow
{\mathcal H}^1_{\tilde{\rho}}(\pi_1(G/H),\mathcal{A})\rightarrow {\mathcal H}^1(\pi_1(G),\mathcal{A})^H\\
&\xrightarrow{d_2}
{\mathcal H}^2_\rho(H,\mathcal{A})
\xrightarrow{{\rm inf}}
{\mathcal H}^2_{
\tilde{\rho}}(\pi_1(G/H),\mathcal{A}),
\end{aligned}
\end{equation}
where the second group cohomology ${\mathcal H}^2_\rho(H,{\mathcal A})$ classifies the symmetry fractionalization in an $H$-SET phase~\cite{Barkeshli2019}. Since the second cohomology group classifies defect fusion rules with anyon as outcome, Eqs.~\eqref{five-term:vortex:set} is a physical statement that the symmetry fractionalization class of symmetry defects may induce a nontrivial fusion rule of the topological point defects. We coin the term {\it defect fractionalization} for this phenomenon, in reference to the terminology ``symmetry fractionalization'' for symmetry defects~\cite{Essin2013,Barkeshli2019,Tarantino2016}. While different symmetry fractionalization classes of the $H$-SET phases are classified by $\mathcal{H}^2_\rho(H,\mathcal{A})$, different defect fractionalization classes are classified by the image of the map ${\rm inf}\colon \mathcal{H}^2_\rho(H,\mathcal{A})\rightarrow {\mathcal H}^2_{
\tilde{\rho}}(\pi_1(G/H),\mathcal{A})$, as describe by the above sequence (\ref{five-term:vortex:set}). 

We will discuss more about point defects in $H$-SET phases in section \ref{sec:point defect}.

\subsection{Group cohomology for textures}\label{subsec:cohomology:texture}

\subsubsection{Textures in SPT}\label{cohomology:skyrmion:spt}
Consider the symmetry breaking from $G = SO(3)\times K$ to $H = U(1)\times K$. In the absence of intrinsic topological orders, a symmetry-breaking ground state with fixed order parameters is an $H$-SPT phase. Here we focus on those $H$-SPT phases described by the mixed anomaly $\mathcal{H}^2(B=U(1),\mathcal{H}^1(K,U(1)))$ in the K\"{u}nneth decompostion (\ref{kunneth}). Due to relation (\ref{mixed anomaly:U(1)}), the mixed anomaly is also captured by $\mathcal{H}^1(\mbz,\mathcal{H}^2(K,U(1)))$.

The topological textures of the order parameter are skyrmions classified by $\pi_2(SO(3)/U(1))$. The point defects associated with $SO(3)$ and $U(1)$ are classified by $\pi_1(SO(3))$ and $\pi_1(U(1))$, respectively. Together they fit into the short exact sequence \eqref{eq:short exact sequence:texture}. Both defects can carry quantum number of the symmetry $K$, which is classified by the linear representations in  $M:=\mathcal{H}^1(K,U(1))$, which we assume to be a finite Abelian group. 
Note that $\pi_1(U(1)) = {\mathbb Z}$, $\pi_1(SO(3)) = {\mathbb Z}_2$ and $\pi_2(SO(3)/U(1)) = \pi_2(S^2)=\mathbb{Z}$ all act trivially on $M$. Applying the inflation-restriction exact sequence \eqref{five_term} to short exact sequence \eqref{eq:short exact sequence:texture} gives
\begin{equation}\label{five_term:texture:spt}
\begin{aligned}
0\rightarrow &\mathcal{H}^1(\pi_1(SO(3)),M) = \{m \in M|m\cdot m = 1_M \in M\}\\
&\rightarrow \mathcal{H}^1(\pi_1(U(1)),M)=M\\
&\xrightarrow{\text{res}} \mathcal{H}^1(\pi_2(S^2),M)=M\\
&\xrightarrow{d_2} \mathcal{H}^2(\pi_1(SO(3)),M)\xrightarrow{\text{inf}=0}
\mathcal{H}^2(\pi_1(U(1)),M)=0.
\end{aligned}
\end{equation}

Physically, as discussed in section \ref{subsec:skyrmion}, a point defect of an $SO(3)$ breaking noncollinear magnetic order classified by $\pi_1(SO(3))=\mbz$, is equivalent to a $2\pi$ flux of any $U(1)$ subgroup of $SO(3)$. A skyrmion with topological charge $\nu\in\pi_1(S^2)=\mbz$, is therefore equivalent to a $4\pi\nu$ flux of the remnant $U(1)$ symmetry. Note that a $2\pi$ flux is nothing but the symmetry defect of the remnant $U(1)$ symmetry, which can carry a linear representation (i.e. charges) of the unbroken subgroup $K$ due to the mixed anomaly (\ref{mixed anomaly:U(1)}) in the $H$-SPT phase. The restriction map (``res'') in exact sequence above \eqref{five_term:texture:spt} physically states the following: if the $K$-symmetry charge (or mathematically a linear representation of $K$) $[R]\in M=\mathcal{H}^1(K,U(1))$ is assigned to each $2\pi$ flux, the $K$-symmetry charge carried by a fundamental ($\nu=1$) skyrmion is given by that of a $4\pi$ flux: $[R\otimes R]\in M$.

The above sequence shows that, if $M$ does not contain an order-2 element, then $\{m \in M|m\cdot m = 1_M \in M\}=\emptyset$ and hence the restriction map (``res'') is injective: i.e. the quantum number of the symmetry $K$ carried by the skyrmions is fully determined by that carried by the $U(1)$ vortices. More interestingly, if $M$ contains an order-2 element, then the restriction map is the  ``multiply by 2'' map, while $d_2$ is the mod 2 map. In this case, skyrmion symmetry quantum number is twice that of the $U(1)$ flux. If we further allow the process of completely breaking the $SO(3)$ symmetry down and then restoring the $U(1)$ subgroup, a skyrmion may carry any quantum numbers allowed on a $U(1)$ flux, due to a possible projective representation carried by the $SO(3)$ defect via the transgression map $d_2$.

We will discuss in more details the physics of topological textures in $H$-SPT phases in section \ref{sec:dqcp:skyrmion}.

\subsubsection{Textures in SET}\label{cohomology:skyrmion:set}

Here we consider the spontaneous symmetry breaking (SSB) from $G = SO(3)$ to $H = U(1)$, resulting in an $H$-SET phase in each symmetry-breaking ground state with fixed order parameters. More generally, we could consider an extra symmetry group $K$ that survives the SSB as in the SPT case discussed above. The $K$-symmetry charges (i.e. linear representations) carried by skyrmions can be determined in parallel to the $H$-SPT case. Here we will not discuss this aspect, but focus on the fractional statistics of skyrmions in the $H=U(1)$-SET phase, with $G=SO(3)$ and $H=U(1)$.

Note that due to the relation (\ref{U(1) to Z}) we have
\bea\label{U(1) to Z:SET}
\mathcal{H}^2(U(1),\mathcal{A})=\mathcal{H}^1(\pi_1(U(1))=\mbz,\mathcal{A})
\eea
where $\mathcal{A}$ is the set of all Abelian anyons, and $\mbz=\pi_1(U(1))$ labels the integer flux quanta of the remnant $U(1)$ symmetry. Since distinct $U(1)$-SET phases are classified by $\mathcal{H}^2(U(1),\mathcal{A})$~\cite{Barkeshli2019,Tarantino2016}, the above identity (\ref{U(1) to Z:SET}) implies that $U(1)$-SETs are fully characterized by the Abelian anyon $a\in\mathcal{A}$ assigned to each $2\pi$ flux (or ``fluxon''\cite{LU2020}) in the SET phase. Based on discussions in section \ref{subsec:skyrmion}, a skyrmion with a topological charge $\nu\in\mbz=\pi_2(S^2)$ is equivalent to a $4\pi\nu$ flux of the remnant $H=U(1)$ symmetry, and we can assign Abelian anyon $a^{2\nu}$ to such a skyrmion accordingly.

Mathematically, we apply the five-term exact sequence \eqref{five_term} to the short exact sequence (\ref{eq:short exact sequence:texture}) to obtain
\begin{equation}\label{five_term:texture:set}
\begin{aligned}
0\rightarrow &\mathcal{H}^1_\rho(\mbz_2,\mathcal{A})
\rightarrow \mathcal{H}^1_{\tilde \rho}(\mbz,\mathcal{A})\xrightarrow{\text{res}} \mathcal{H}^1(\pi_2(S^2),\mathcal{A})\\
&\xrightarrow{d_2}\mathcal{H}^2_\rho(\pi_1(SO(3)),\mathcal{A})\xrightarrow{\text{inf}}
\mathcal{H}^2_{\tilde \rho}(\mbz,\mathcal{A})=0.
\end{aligned}
\end{equation}
The fractional statistics carried by a skyrmion of topological charge $\nu\in\mbz$ is determined by the image of the restriction map (``res'') in the exact sequence above. Physically, since $[R]\in\mathcal{H}^1(\mbz,\mathcal{A})$ assigns an Abelian anyon $a\in\mathcal{A}$ to each $2\pi$ flux, its image in the restriction map $\tilde R\in\mathcal{H}^1(\pi_2(S^2),\mathcal{A})$ implies that Abelian anyon $a^{2\nu}$ is assigned to each skyrmion of topological charge $\nu$. This will determine the fractional statistics of a skyrmion.

We will discuss the physics of topological textures in $H$-SET phases in more detail in section \ref{sec:texture}.

\section{Defects and textures in the absence of intrinsic topological orders}\label{sec:spt+dqcp}

\subsection{From SPT physics of ordered ground states to deconfined quantum critical points}

In this section, we first discuss the simpler cases where every long-range ordered ground state with fixed order parameters exhibits no intrinsic topological order. In other words, when symmetry $G$ is spontaneously broken down to $H$ in a given ground state, a long-range ordered ground state with fixed order parameters (not a cat state!) is a $H$-preserving short-range entangled phase\footnote{We use the definition of Ref.~\cite{Chen2013} for short-range entangled phase, different from Kitaev's definition~\cite{Kitaev2006}. Therefore we do not consider invertible phases such as topological superconductor in class D~\cite{Kitaev2009}, or $E_8$ states of $d=2$ interacting bosons~\cite{Kitaev2006,Lu2012}.}. In two spatial dimensions, they belong to the $H$-symmetry protected topological ($H$-SPT) phases classified by group cohomology $\mathcal{H}^3(H,U(1))$ \cite{Chen2013}. What are the physical consequences if each symmetry-breaking ground state is a nontrivial $H$-SPT? As will become clear soon, this SPT physics is closely related to the physics of deconfined quantum critical points (DQCP)~\cite{Senthil20041,Senthil2004,Wang2017}.

An $H$-SPT phase is a short-range entangled phase, which, in the presence of symmetry $H$, cannot be continuously connected to a trivial product state without closing the energy gap. In a system of interacting bosons in two spatial dimensions, different $H$-SPT phases are classified by the 3rd group cohomology $\mathcal{H}^{3}(H,U(1))$ \cite{Chen2013}. The cohomology group $\mathcal{H}^{3}(H,U(1))$ is an Abelian group, whose identity element labels the topologically trivial phase of product states, and the addition of group elements is implemented by stacking different SPT phases.

A deconfined quantum critical point (DQCP) describes the continuous phase transition between two different long-range orders, who are not related to each other by Landau-Ginzburg-Wilson paradigm of spontaneous symmetry breaking~\cite{Senthil20041,Senthil2004,Wang2017}. More precisely, the remnant symmetry groups $H_1,H_2\leq G$ of the two long-range ordered phases do not have a subgroup relation: in other words, $H_1$ is not a subgroup of $H_2$ and vice versa. Therefore, a DQCP is clearly beyond the Landau-Ginzburg-Wilson paradigm and provides new mechanism to understand direct quantum phase transitions between different long-range orders.

For example, DQCP is believed to describe the direct transition between a Neel order and a valance bond solid (VBS) on a square lattice \cite{Senthil2004,Senthil20041}. While the Neel order spontaneously breaks the time reversal and spin rotational symmetries, it preserves the 4-fold rotation symmetry $C_4$ around each lattice site. On the other hand, the columnar VBS phase spontaneously breaks $C_4$, but preserves both time reversal and spin rotational symmetries. Therefore a direct continuous phase transition between these two long-range orders are incompatible with the Landau-Ginzburg-Wilson paradigm. There are two alternative and complementary physical pictures to understand this DQCP.

(i) The first point of view starts with the following property of the VBS phase: each vortex of the VBS order parameter carries a spin-$1/2$~\cite{Levin2004}, which forms a projective representation $[\omega]\in\mathcal{H}^2\big(SO(3)\times Z_2^\bst,U(1)\big)$ of the $SO(3)$ spin rotation and time reversal ($\bst$) symmetries. Therefore, condensing the VBS vortex will necessarily breaks both time reversal and spin rotational symmetries, while restoring the crystalline $C_4$ symmetry. This leads to a direct transition from the columnar VBS to the Neel order via a DQCP.

(ii) The other viewpoint starts from the Neel order side, where each fundamental skyrmion (with unit topological charge $\nu=1\in\pi_2(S^2)=\mbz$) of the Neel order parameter carries a unit $C_4$ angular momentum \cite{Haldane1988}. As a result, condensing the skyrmion will necessarily breaks the $C_4$ crystalline rotation symmetry, while restoring time reversal and spin rotational symmetries. This corresponds to a direct transition from the Neel order to the columnar VBS phase via the DQCP~\cite{Senthil20041,Senthil2004}.

To unify the above two pictures (i) and (ii), and to treat VBS and Neel order parameters on an equal footing, one can introduce a 5-component order parameter $\vec n=(n^1,\cdots,n^5)$, where the first 3 components $(n^1,n^2,n^3)$ represent the Neel vector, while $(n^4,n^5)$ serves as the columnar VBS order parameter~\cite{Levin2004,Tanaka2005,Senthil2006}. The interplay of the VBS and Neel order parameters, i.e. the spin-$1/2$ VBS vortex and the $C_4$ angular momentum of a Neel skyrmion, is captured by a (2+1)-dimensional Wess-Zumino-Witten (WZW) term~\cite{Wess1974,Witten1983} of the 5-component order parameter $\vec n$~\cite{ABANOV2000,Tanaka2005,Senthil2006,Wang2017}:
\bea\label{WZW:O(5)}
\mathcal{S}_\text{WZW}\!=\!\frac{2\pi\epsilon_{abcde}}{\text{Area}(S^4)}\int_0^1\text{d}u\text{d}^3\vec q~n^a\partial_x n^b\partial_y n^c\partial_t n^d\partial_u n^e~~~
\eea
where we use $\vec q=(x,y,t)\in S^3$ to parametrize the spacetime manifold $S^3$, and $u$ is introduced to parametrize a smooth interpolation (extension) between $\vec n(\vec q,u=0)=(0,0,0,0,1)$ and $\vec n(\vec q,u=1)=\vec n(\vec q)$. While the physical system only has a microscopic symmetry of $G=C_4\times SO(3)$, an enlarged $SO(5)$ symmetry that rotates the 5 components of $\vec n$ was argued to emerge at the DQCP described by a NLSM with the above WZW term~\cite{Senthil2006,Wang2022}.

The aforementioned two pictures for the Neel-VBS transition are both captured by the above WZW term. On the VBS side, a classical vortex configuration for VBS order parameters $(n^4,n^5)$ reduces the topological term (\ref{WZW:O(5)}) to an $O(3)$ NLSM with a WZW term, physically corresponding to a spin-$1/2$ at the vortex core~\cite{ABANOV2000}. On the Neel side, a classical skyrmion configuration of Neel order parameters $(n^1,n^2,n^3)$ reduces (\ref{WZW:O(5)}) to a $U(1)$ rotor action of VBS order parameters $(n^1,n^2)$, which carries a unit angular momentum of crystalline rotation $C_4$.

In this work, we want to point out a connection between the two physical pictures of DQCP, and the SPT physics of the symmetry-breaking ground states. More precisely, in a long-range order which spontaneously breaks symmetry $G$ down to a subgroup $H$, if a symmetry-breaking ground state belongs to certain $H$-SPT phases, the condensation of topological point defects or textures of the order parameters will lead to a direct transition described by a DQCP. The other side of the direct transition must spontaneously breaks $H$: it is generally a long-range order with remnant symmetry $H^\prime$, which is neither a subgroup nor a supergroup of $H$. This observation can be summarized in two classes:

(i) If a symmetry-breaking ground state belongs to certain $H$-SPT phases, to be elaborated in section \ref{sec:dqcp:vortex}, a point defect (i.e. vortex) of the order parameter $\hat O(\vec r)\in G/H$ can carry a projective representation $[\omega]\in\mathcal{H}^2(H,U(1))$ of the remnant symmetry $H$. This is a generalization of viewpoint (i) for Neel-VBS transition from the VBS side. Condensing the point defects (vortices) will spontaneously breaks (a part of) symmetry $H$ across the DQCP.

(ii) If a symmetry breaking ground state belongs to certain $H$-SPT phases, to be elaborated in section \ref{sec:dqcp:skyrmion}, a topological texture (i.e. a skyrmion) of the order parameter can carry a charge (or a linear representation $[\omega]\in\mathcal{H}^1(H,U(1))$) of the remnant symmetry $H$. This is a generalization of viewpoint (ii) for the Neel-VBS transition from the Neel side. Condensing the skyrmions will spontaneously break the remnant symmetry $H$ across the DQCP.

This general connection allows us to determine whether a given long-range order is in proximity to a DQCP in the phase (parameter) space, and to systematically construct examples of DQCPs based on the $H$-SPT classification for the symmetry-breaking ground states. For the rest of this section, we shall discuss the two classes in detail: (i) point defects carrying a projective representation of $H$, in section \ref{sec:dqcp:vortex}; (ii) skyrmions carrying a linear representation (i.e. quantum numbers) of $H$, in section \ref{sec:dqcp:skyrmion}. We will use the mathematical classification based on group cohomology in section \ref{sec:group cohomology}, and its related ``decorated domain wall'' picture~\cite{Chen2014}, to elucidate the aforementioned connection between SPT phases in symmetry-breaking ground states and the DQCP.

\subsection{Point defects}\label{sec:dqcp:vortex}

Without loss of generality, let's consider the following situation:
\bea
G=A\times K,~~~H=B\times K,~~~H\triangleleft G.
\eea
where $\times$ is the direct product of two groups, and we assume that $H$ is a discrete normal subgroup of a continuous group $G$. In other words, the subgroup symmetry $A$ is broken down to $B\triangleleft A$ while the subgroup symmetry $K$ is preserved in the long-range order.

The question we plan to address is the following: does a point defect in $\pi_1(G/H=A/B)$ carry a projective representation of the remnant symmetry $H$? As discussed in section \ref{cohomology:vortex:spt}, given the mixed SPT anomaly in $\mathcal{H}^1(B,\mathcal{H}^2(K,U(1)))$ in a symmetry-breaking ground state with fixed order parameters, due to exact sequence (\ref{five_term:vortex:spt}), the projective representation carried by point defects is classified by the image of injective map $d_1$ in $\mathcal{H}^1(\pi_1(A/B),\mathcal{H}^2(K,U(1)))$. This gives the following criterion for point defect bound states:

\emph{(i) When a full symmetry $G=A\times K$ is spontaneously broken down to a subgroup $H=B\times K$, if a symmetry-breaking ground state is an $H$-SPT with a mixed anomaly described by $[R]\in\mathcal{H}^1(B,\mathcal{H}^2(K,U(1)))$, the projective representation of subgroup $K$ carried by the point defect in $\pi_1(G/H)$ is specified by the image $[\tilde R]\in\mathcal{H}^1(\pi_1(G/H),\mathcal{H}^2(K,U(1)))$ of map $d_1$ in exact sequence (\ref{five_term:vortex:spt}).}

Physically, a point defect of the order parameters can be mapped to a symmetry defect by the connecting homomorphism $p$ in exact sequence (\ref{eq:short sequence}). The mixed anomaly $[R]\in\mathcal{H}^1(B,\mathcal{H}^2(K,U(1)))$ assigns a projective representation $\mathcal{H}^2(K,U(1))$ to each symmetry defect $b\in B$. This in turn determines the projective representation assigned to a point defect in $\pi_1(G/H)$.

Below we discuss one example of such nature, where each fundamental vortex with a unit winding number carries a spin-$1/2$ in the pair superfluid phase.

\subsubsection{Symmetry protected pair superfluid with spin-$1/2$ vortices}

We consider a bosonic system on a two-dimensional lattice, which consists of hard-core bosons $\{b_i\}$ and a single spin-$1/2$ d.o.f. $\vec s_i$ per unit site. The full symmetry of the Hamiltonian is $G=U(1)\times SO(3)$, which is spontaneously broken down to $H=Z_2\times SO(3)$ in a pair superfluid phase. In the notation the general discussions above, we have $A=U(1),~K=B=Z_2$ and $K=SO(3)$\footnote{One can also replace $K=SO(3)$ symmetry by a discrete time reversal symmetry $K=Z_2^\bst$.}. Each hard-core boson $b_i$ carries a unit charge but no spin (or spin-0), while each $\vec s_i$ carries no charge but transforms as a spin-$1/2$ projective representation of $SO(3)$ symmetry. The filling number for bosons is $\bar\rho=2$ per unit cell.

In the long-range order, we consider a pair superfluid phase where two bosons form a condensate with $\expval{(b_i)^2}\neq0$ but $\expval{b_i}=0$. To achieve the desired SPT properties, we further require the system to preserve the following magnetic translation symmetry:
\bea
T_1T_2T_1^{-1}T_2^{-1}=(-1)^{\sum_ib^\dagger b_i}
\eea
where $T_{1,2}$ are magnetic translations along the two primitive vectors $\vec a_{1,2}$. In other words, the bosons experience a $\pi$ flux in each unit cell when traveling around the lattice. A pair of bosons, carrying charge 2, only experiences $2\pi$ flux per unit cell and can condense without breaking the translation symmetry $T_{1,2}$, driving the system into a translation invariant pair superfluid phase. Note that there is a $\pi$ flux and a single spin-$1/2$ in each unit cell. Due to the Lieb-Schultz-Mattis theorem for SPT phases~\cite{Lu2017,Yang2018}, in the presence of the magnetic translation symmetry, any short-range entangled ground state preserving the magnetic translation and $H=\mbz_2\times SO(3)$ symmetry must be a $H$-SPT phase, exactly described by a mixed anomaly $[R]\in\mathcal{H}^1(\mbz_2,\mathcal{H}^2(SO(3),U(1)))<\mathcal{H}^3(H,U(1))$. Physically, in a ground state of the pair superfluid phase, each symmetry defect of the remnant $\mbz_2$ (i.e. a $\pi$ flux) carries a projective representation of the remnant $SO(3)$ symmetry (i.e. a spin-$1/2$).

On the other hand, the point defect of the order parameter $\expval{(b_i)^2}$ are vortices in the pair superfluid, labeled by an integer-valued vorticity $\nu\in\mbz$. In particular, each vorticity-1 vortex corresponds to a $\pi$ flux, and hence carries a spin-$1/2$, exactly captured by the mixed anomaly $[\tilde R]\in\mathcal{H}^1(\mbz,\mathcal{H}^2(SO(3),U(1)))$, the image of map $d_1$ in exact sequence (\ref{five_term:vortex:spt}). As a result, condensing the elementary ($\nu=\pm1$) vortices will drive the system into a magnetic order that spontaneously breaks $K=SO(3)$ symmetry, via a DQCP. The Neel vector $(n^1,n^2,n^3)$ and the superfluid order parameters $\expval{(b_i)^2}\sim n^4+\imth n^5$ are described by a NLsM with an $O(5)$ WZW term.

\subsection{Textures}\label{sec:dqcp:skyrmion}
% {\color{blue} below restrict to one component spinon}
We consider the symmetry group $G=SO(3)\times K$ to be spontaneously broken down to a subgroup $H=U(1)_{S^z}\times K$, where $K$ is a subgroup of onsite unitary symmetries. In the absence of intrinsic topological orders, the symmetry breaking phase can be an $H$-SPT phase in two spatial dimensions, classified by $\mathcal{H}^3(H,U(1))$. As discussed in section \ref{subsec:skyrmion}, a skyrmion of topological charge $\nu\in\mbz$ is equivalent to a $4\pi\nu$ flux of the remnanat $U(1)$ symmetry.

To understand the universal properties of skyrmions in the $H$-SPT phases, again we use the K\"unneth decomposition in (\ref{kunneth}):
\begin{equation}\label{kunnethtexture}
\begin{aligned}
&\mathcal{H}^3(H,U(1))\\
&=\mathcal{H}^3(U(1)_{S^z},U(1))\times\mathcal{H}^2(U(1)_{S^z},\mathcal{H}^1(K,U(1)))\\
&\quad\times\mathcal{H}^1(U(1)_{S^z},\mathcal{H}^2(K,U(1)))\times\mathcal{H}^3(K,U(1)),
\end{aligned}
\end{equation}
Note that two of the four terms indicates different topological properties of the flux of the unbroken $U(1)_{S^z}$ symmetry, and hence of the skyrmions. The first term labels the $U(1)_{S^z}$ quantum number $S^z=q\in\mathcal{H}^3(U(1)_{S^z},U(1))\cong \mathcal{H}^1(U(1)_{S^z},U(1))=\mbz$ carried by each $U(1)_{S^z}$ flux quantum. The third term $\mathcal{H}^1(U(1),\mathcal{H}^2(K,U(1)))$ in \eqref{kunnethtexture} vanishes when $\mathcal{H}^2(K,U(1))$ is any finite Abelian group. Finally, the fourth term labels the $K$-SPT phases that do not require the protection of the $U(1)_{S^z}$ symmetry.

Due to the relation \eqref{mixed anomaly:U(1)}, the 2nd term is equivalent to $\mathcal{H}^1(\mbz,M)$ with $ M\equiv{\mathcal H}^1(K,U(1))$, and hence can be understood as the linear representation (i.e. the charge) of the unbroken subgroup $K$ carried by each flux quantum (or $2\pi$ flux) of the $U(1)_{S^z}$ symmetry. Now that each $\nu=1$ skyrmion can be viewed as a $4\pi$ flux, it carries the linear representation $[R \otimes R]\in M$ of unbroken subgroup $K$. Condensing skyrmions with nontrivial $K$-symmetry charges will inevitably break the $K$ symmetry while restoring the $SO(3)$ spin rotational symmetry, through a DQCP.

As we are interested in the interplay between the two subgroups $U(1)_{S^z}$ and $K$ of the remnant symmetry $H=U(1)\times K$, we will be focusing on the second term in the K\"{u}nneth decomposition \eqref{kunnethtexture}. According to the discussions in Sec.~\ref{cohomology:skyrmion:set}, the following statement describes the symmetry quantum numbers of topological textures:

\emph{(ii) When a full symmetry group $G=SO(3)\times K$ is spontaneously broken down to a subgroup $H=U(1)\times K$, if a symmetry-breaking ground state is an $H$-SPT described by $[R]\in\mathcal{H}^1(\mbz,M)\equiv\mathcal{H}^2(U(1),M)$ with $M=\mathcal{H}^1(K,U(1))$, the quantum number (or linear representation) of unbroken subgroup $K$ assigned to each skyrmion is specified by the image $[\tilde R]\in\mathcal{H}^1(\pi_2(S^2),M)$ of the restriction map (``res'') in the exact sequence (\ref{five_term:vortex:spt}).}

Below we discuss two familiar examples of this type in more details.

\subsubsection{Charge-$2e$ skyrmions in quantum spin Hall insulators}

The first example we consider is a fermion system with charge conservation $K=U(1)_c$ and spin rotational symmetry, hence a full symmetry group of $G=SO(3)\times U(1)_c$. A time-reversal-invariant collinear order parameter can spontaneously breaks the symmetry $G$ down to a $H=U(1)_s\times U(1)_c$ subgroup, where $U(1)_s$ is the subgroup of $U(1)$ spin rotational symmetry along e.g. $z$-axis:
\bea
f_\uparrow\xrightarrow{e^{\imth\theta S^z_{\rm tot}}} e^{-\imth\theta/2}f_\uparrow,~~~f_\downarrow\xrightarrow{e^{\imth\theta S^z_{\rm tot}}} e^{\imth\theta/2}f_\downarrow
\eea

We are interested in the case where each symmetry-breaking ground state is a quantum spin Hall (QSH) insulator~\cite{Kane2005,Bernevig2006,Konig2007}, with a pair of helical edge states protected by the remnant symmetry $H$.

It is well known that such a QSH insulator exhibits a mixed anomaly between the $U(1)_c$ and $U(1)_s$ subgroups, captured by group cohomology
\bea
\mathcal{H}^2(U(1)_s,M)=\mathcal{H}^1(\mbz,M)\subset\mathcal{H}^3(H,U(1))
\eea
where $M\equiv\mathcal{H}^1(U(1)_c,U(1))$ labels $U(1)_c$ quantum numbers i.e. electric charges. Specifically, the quantized spin Hall conductance $\sigma_{xy}^{c,s}=e/2\pi$~\cite{Bernevig2006a} indicates that each $2\pi$ flux of spin symmetry $U(1)_s$ would carry a unit electric charge. As a result, each skyrmion of topological charge $\nu\in\mbz=\pi_2(S^2)$, equivalent to a $4\pi\nu$ flux of spin rotational symmetry $U(1)_s$, carries an electron charge of $2\nu e$.

In particular, an elementary $\nu=1$ skyrmion carries charge $2e$, as pointed out in Ref.~\cite{Grover2008}. Condensing these elementary skyrmions hence induce a superconducting state that spontaneously breaks the $U(1)_c$ symmetry, which was recently proposed to be one mechanism for superconductivity in magic-angle twisted bilayer graphene~\cite{Khalaf2021}.

\subsubsection{Neel order in a spin-$1/2$ model on the square lattice}

Next we discuss a familiar example related to the DQCP, i.e. the Neel order in a spin-$1/2$ system on the square lattice. Before studying the full space group symmetry, let us first consider a simplified situation where we ignore the translations and mirror reflection: we only take into account the onsite $SO(3)$ spin rotational symmetry and the site-centered 4-fold crystalline rotation symmetry $K=C_4$. In a collinear Neel order, the full symmetry $G=SO(3)\times C_4$ is spontaneously broken down to the remnant symmetry $H=U(1)_{S^z}\times C_4$. In the latter $H$-preserving Neel ordered phase, there is a quantized topological term of the Wen-Zee type~\cite{Wen1992,Han2019}, which corresponds to an element in the cohomology class $\mathcal{H}^2(U(1)_{S^z},\mathcal{H}^1(K,U(1)))\subset{\mathcal H}^3(H,U(1))$ and characterizes the mixed anomaly between $C_4$ rotation and $U(1)_{S^z}$ spin rotational symmetries. Physically, this cohomology class and associated Wen-Zee term in the continuum field theory implies that each  $U(1)_{S^z}$ flux quantum (i.e. $2\pi$ flux) carries a $C_4$ eigenvalue of $\imth$ (i.e. a unit angular momentum). As a result, each skyrmion (equivalent to $2\pi$ flux of $U(1)_{S^z}$ for a spin-1/2 system, see section \ref{sec:SU(2)}) also carries a $C_4$ eigenvalue of $\imth$~\cite{Haldane1988}, and condensing the skyrmions (which restores the $SO(3)$ spin rotational symmetry) will necessarily break $C_4$ crystalline rotational symmetry, as is the case of the valence bond solids on the square lattice~\cite{Haldane1988,Senthil2004}.

Next, we include the translations and consider the full space group symmetry of the square lattice.  The full symmetry for the paramagnetic phase is $G=p4m\times SO(3)\times Z_2^\bst$, where $\bst$ is the time reversal operation and $p4m$ is the wallpaper group that describes the symmetry of the square lattice, generated by translations $T_1$, $T_2$, site-centered rotation $C_4$, and reflection $M$ (with respect to a site-crossing mirror plane). After the transition to Neel order, $G$ is broken down to $H= U(1)_{S^z}\rtimes p_p4m$, where $p_p4m$ is the magnetic space group for the Neel order~\cite{litvin2008tables}, generated by magnetic translations $\widetilde{T}_{x,y}=\bst\cdot T_{x,y}$ and point group symmetries $C_4$ and $M$. Note that spin rotation $U(1)_{S^z}$ and the magnetic space group do not commute with each other, hence the semidirect product. Due to this semidirect product structure, the K\"unneth decomposition \eqref{kunnethtexture} can no longer be used to calcualte the cohomology. Nevertheless one can show through a spectral sequence calculation that $\mathcal{H}^3(H,U(1))$ contains a summand
\begin{equation}
\mathcal{H}^2(p_p4m,\mathcal{H}^1(U(1)_{S^z},U(1))) \subset \mathcal{H}^3(H,U(1)).
\end{equation}
One can further show that $\mathcal{H}^2(p_p4m,\mathcal{H}^1(U(1)_{S^z},U(1))) = \mathcal{H}^1(p_p4m,U(1)) = \mathbb{Z}^2_2$, where the two $\mathbb{Z}_2$ summands label the eigenvalues of $C_4$ and $M$, respectively. Compared to the case of $K=C_4$ where the eigenvalues of $C_4$ are $\{\pm 1, \pm \mathrm{i}\}$, now considering the full magnetic space group $p_p4m$ reduces the eigenvalues of $C_4$ to $\{1,\mathrm{i}\}$ due to the magnetic translations (the $C_4$ eigenvalues $\pm 1$ are identified and so are the eigenvalues $\pm \mathrm{i}$). We see that the analysis in the case $K= C_4$ above still holds. This means that, taking into account the full lattice symmetry, condensing the skyrmions (hence restoring the $SO(3)$ internal symmetry) will indeed break the $C_4$ rotation symmetry spontaneously. Note that a similar analysis has been carried out in Refs.~\cite{PhysRevB.98.085140,dqcp2021topological}.

\section{Point defects in symmetry enriched topological orders}\label{sec:point defect}
When the ground state is a SPT phase protected by the unbroken symmetry $H$, we have shown previously that point defects (or vortices) of the symmetry-breaking order parameters can carry a projective representation of the unbroken symmetry. Below we discuss the more general situation, i.e. two-dimensional intrinsic topological orders with spontaneously broken symmetries, where each symmetry-breaking ground state is a $H$-symmetry enriched topological ($H$-SET) phase. As discussed previously in section \ref{subsec:cohomology:vortex}, due to the short exact sequence (\ref{eq:short sequence}) that maps a topological point defect (an element of $\pi_1(G/H)$) to a symmetry defect (an element of $H$), one can derive universal properties of point defects from those of symmetry defects, which were extensively studied in the context of SET phases~\cite{Essin2013,Tarantino2016,Barkeshli2019}. We found that when topological orders coexist with spontaneous symmetry breaking, due to the presence of anyons which obey fractional statistics, two classes of new phenomena can occur:

(1) Point defects can permute anyons in the topological order when braided around. In other words, after traveling around a point defect, one anyon of a certain type can be transmuted into an anyon of a different type. In these cases, the point defect (vortex) is mapped into a non-Abelian symmetry defect (or twist defect)~\cite{Barkeshli2012,Barkeshli2013,Barkeshli2019,Tarantino2016,Teo2014,Teo2016}.

(2) Multiple point defects can fuse into Abelian anyons, which we coin ``defect fractionalization'' for reasons that we describe below in details.

\subsection{Defect fractionalization phenomenon}\label{Sec: Defect fractionalization}

Importantly, the topological point defects may obey a nontrivial fusion rule: upon annihilating with each other, these topological defects may leave behind Abelian anyons, similar to the symmetry defects in SET phase.

We have named this phenomenon  {\it defect fractionalization} in Sec.~\ref{cohomology:vortex:set}. Here we stress that the understanding of defect fractionalization parallels that of symmetry fractionalization~\cite{Essin2013,Barkeshli2019}: for SSB $G\rightarrow H$, different defect fractionalization classes in the broken symmetry phase should correspond to equivalence classes $[\omega] \in \mathcal{H}^2_{\tilde{\rho}}(\pi_1(G/H),\mathcal{A})$, meaning that for $g,h \in \pi_1(G/H)$, the group element $\omega(g,h)$ in ${\mathcal H}^2_{\tilde \rho}(\pi_1(G/H),{\mathcal A})$ denotes the residual anyons after fusing the defects $g$, $h$ and $(gh)^{-1}$. 

In the simplest case of $\pi_1(G)=0$, we have $\pi_1(G/H)\cong H$ from \eqref{eq:short sequence}, namely there is a one-to-one correspondence between the topological point defects (or vortices) of the order parameter $\phi(\vec{r}) \in \pi_1(G/H)$ and the (extrinsic) symmetry defects $\tau_h,h\in H$. In this case $p$ is an isomorphism, hence $\tilde\rho=\rho$. As a result, there is a one-to-one correspondence between defect fractionalization described by $\mathcal{H}^2_{\tilde\rho}(\pi_1(G/H),\mathcal{A})$, and symmetry fractionalization described by $\mathcal{H}^2_{\rho}(H,\mathcal{A})$, i.e. $p^\ast \equiv {\rm inf}$ is a bijective map in exact sequence (\ref{five-term:vortex:set}). One such example is lattice dislocations, where $G=\mathbb{R}\times{\mathbb{R}}$ and $H=\mbz^2$, where dislocations with a Burgers vector $\vec b=b_1\vec a_1+b_2\vec a_2$ corresponds to symmetry defect of translation operation $T_1^{b_1}T_2^{b_2}$, where $\vec a_{1,2}$ are the two primitive translation vectors of the two-dimensional lattice. In section \ref{sec:exp:dislocation}, we compute and explicitly demonstrate the nontrivial fusion rules of dislocations in the toric code.

\subsection{Defect fractionalization vs symmetry fractionalization}\label{Defect fractionalizationvssymmetry fractionalization}
It is tempting to conclude that defect fractionalization in the general case of $\pi_1(G)\neq 0$ is classified by ${\mathcal H}^2_{\tilde{\rho}}(\pi_1(G/H),\mathcal{A})$. However, as we shall show in explicit model calculations, ${\mathcal H}^2_{\tilde{\rho}}(\pi_1(G/H),\mathcal{A})$ contains elements that are redundant, and the physical ones are classified by the subgroup which is the image of the inflation map $p^*={\rm inf}$ in the exact sequence (\ref{five-term:vortex:set})
\begin{equation}
\text{im}(p^*) \subset {\mathcal H}^2_{\tilde{\rho}}(\pi_1(G/H),\mathcal{A}),
\end{equation}
namely, the classes of symmetry fractionalization that survives the inflation map. When $\pi_1(G)=0$ is trivial, the inflation map becomes an isomorphism, and correctly reproduces the classification mentioned above.

We claim in the above that in the general case when $\pi_1(G)\neq 0$, the classification for defect fractionalization is fully determined by specifying the inflation map $p^*$. The inflation map $p^*$ sends the cohomology of the quotient group (in our case, $H={\pi_1(G/H)}/{\pi_1(G)}$) to that of the group extension (in our case, $\pi_1(G/H)$). Physically, it simply sends any symmetry fractionalization class $[\omega]$ \--- cocycle elements $\omega(h_1,h_2)=\mathsf{a}, \mathsf{a}\in \mathcal{A}, h_1,h_2 \in H$ \--- to defect fractionalization class $[p^*(\omega)]$ by sending symmetry $h\in H$ to any defect $\phi(\vec{r}) \in \pi_1(G/H)$ that maps to $h$ under $p$. Here, however, what is nontrivial is that such a class $[\omega]$ may be \emph{trivialized} under the map $p^*$: the defect fusion rule inherited from $\omega$ can be continuously deformed to the one that fuses to no anyons. This happens whenever the the map $p^*$ has a nontrivial kernel \--- that is, $\mathrm{ker}(p^*) := \{[\omega] |p^*([\omega])=0\} \neq \emptyset$.

At this point, the exact sequence \eqref{five-term:vortex:set} is introduced as a mathematical computational tool, whose physical meaning is yet to be specified. We now try to understand the physical meaning of each piece and the exactness among them.
To achieve this, recall from Sec.~\ref{sec:recap_SET} that elements of $ {\mathcal H}^1_\rho(H,{\mathcal A})$ describes the ``anyon-labeling rule'' for the extrinsic symmetry defects $\tau_h,h\in H$. The topological defects $\phi(\vec{r}) \in \pi_1(G/H)$ inherits a similar defect ``anyon-labeling rule'' consistent with their actions on the anyons, classified by $\mathcal{H}^1_{\tilde{\rho}}(\pi_1(G/H),\mathcal{A})$. Such a heritage is easy to understand since the symmetry $H$ is intact after the SSB $G\rightarrow H$. However, one can imagine the alternative, indirect, physical process in which the symmetry $G$ is completely broken down (to $\{1\}$), and then restored to $H$. We will call this the indirect SSB process $G\rightarrow \{1\}\rightarrow H$ from now on (note here the arrows are written in a physical sense, not in a mathematical sense). In this scenario, the relevant defect ``anyon-labeing rule'' is that for the topological defects $\psi(\vec{r})\in \pi_1(G)$, classified by $\mathcal{H}^1(\pi_1(G),\mathcal{A})$, but when restoring $H$, only those ``anyon-labeling rules'' for $\pi_1(G)$ that are \emph{invariant} under the action of $H$ makes sense after $H$ is restored. Here the invariance is defined in the sense of Eq.~\eqref{actupperQ}.
%, we mean any $[\omega] \in \mathcal{H}^1(\pi_1(G),\mathcal{A})$, represented by the cocycle $\omega(g) =\mathsf{a}$ for $\mathsf{a} \in \mathcal{A}$ and $g\in G$, such that
%\begin{equation}\label{eq:midgroup}
%h.\big(\omega(h^{-1}gh)\big) =\omega(g)=\mathsf{a},\quad \forall h \in H
%\end{equation}
%any such element $\omega$ defines a class $[\omega]$ in the object $\mathcal{H}^1(\pi_1(G),\mathcal{A})^H$ (note the superscript) that appears in \eqref{eq:midgroup}.
Intuitively, both the defect $g \in \pi_1(G)$ and the anyon $\mathsf{a}\in \mathcal{A}$ may transform nontrivially under $H$, but the defect--anyon composite must transform \emph{covariantly} under $H$, implying that the ``anyon-labeling rules'' for $\pi_1(G)$ is invariant under $H$.

Together with the physical meaning of $\mathcal{H}^2_\rho(H,\mathcal{A})$ and $\mathcal{H}^2_{\tilde{\rho}}(\pi_1(G/H),\mathcal{A})$ as symmetry fractionalization and defect fractionalization, respectively, that have been introduced before, we are now in a position to understand the exactness of the sequence \eqref{five-term:vortex:set}. As mentioned before, to know $\mathrm{im}(p^*)$ it suffices to known $\mathrm{ker}(p^*)$. The exactness $\mathrm{ker}(p^*)=\mathrm{im}(d_2)$ states that, after the indirect SSB process $G\rightarrow \{1\}\rightarrow H$,
for any resulting defect (now object in $\pi_1(G/H)$), the only possible ``defect fusion rule'' compatible with its ``anyon-labeling rule'' is the trivial one. The exactness at $\mathcal{H}^1(\pi_1(G),\mathcal{A})^H$ states that, if a ``defect fusion rule'' can be realized in both the direct SSB process $G\rightarrow H$ and the indirect process $G\rightarrow \{1\}\rightarrow H$, the only possible ``defect fusion rule'' compatible with its ``anyon-labeling rule'' is the trivial one. The exactness at $\mathcal{H}^1_{\tilde{\rho}}(\pi_1(G/H),\mathcal{A})$ states that, the topological defect ``anyon-labeling rule'' that originates from a symmetry defect ``anyon-labeling rule'' cannot be realized at the end of the indirect SSB process $G\rightarrow \{1\}\rightarrow H$. Finally, the exactness at $\mathcal{H}^1_\rho(H,\mathcal{A})$ states that every symmetry defect ``anyon-labeling rule'' can be realized in the topological defects of $\pi_1(G/H)$.

We note that the five-term exact sequence \eqref{five-term:vortex:set} is a corollary of spectral sequences. We will be using spectral sequences in the calculation of cohomology groups (especially those with nontrivial action on the anyons) and an elementary introduction can be found in App.~\ref{app:c}. We present one quite useful statement about this map:
\begin{theorem}\label{thm1}
Given Eq.~\eqref{five-term:vortex:set}. If $\pi_1(G/H)$ is of a semi-direct product form $\pi_1(G/H)\cong \pi_1(G)\rtimes H$, then $d_2=0$, hence $p*$ is injective. Consequently, the symmetry fractionalization classes are in one-to-one correspondence with the defect fractionalization classes.
\end{theorem}
{ This statement is powerful in that it works regardless of the action $\rho$ (trivial or nontrivial alike).

In Table.~\ref{cases1to4} we give examples of the $\mathbb{Z}_2$ topological order (toric code) enriched by two different symmetries: (1) $G=U(1)$, $H=Z_2$, and (2) $G=SO(3)$ and $H=D_2 = \{ 1, X,Y,Z\}$,  with $\mathcal{A} = \mathbb{Z}_2^{\mathsf{e}}\times \mathbb{Z}_2^{\mathsf{m}}$. For each example we consider the cases of trivial and nontrivial $H$-actions. All these examples lie outside the application of Theorem \ref{thm1}\footnote{$G=N\rtimes Q$ if and only if the short exact sequence $0\rightarrow N\rightarrow G\xrightarrow{p} Q\rightarrow 0$ \emph{splits}, i.e. there is a homomorphism $j\colon Q\rightarrow G$ s.t. the composed map $p\circ j$ is the identity map on $Q$. This is not the case for either $0\rightarrow Z\xrightarrow{2}Z\rightarrow Z_2\rightarrow 0$ or $0\rightarrow D_2\rightarrow Q_8\rightarrow Z_2$. In fact, $Q_8$ does not admit a semi-direct product structure.}

\begin{table*}[t]
\begin{tabular}{c|c|c|c|ccccccccc}
\hline
No.&$G$ & $H$& $H$-Action & ${\mathcal H}^1_\rho(H,{\mathcal A})$ & $\rightarrow$ & $ {\mathcal H}^1_{\tilde{\rho}}(\pi_1(G/H),\mathcal{A})$ & $\rightarrow$  & $  {\mathcal H}^1_{\mathrm{id}}(\pi_1(G),\mathcal{A})^H$ & $\rightarrow$ & ${\mathcal H}^2_\rho(H,{\mathcal A})$ & $\rightarrow$ & ${\mathcal H}^2_{\tilde{\rho}}(\pi_1(G/H),\mathcal{A})$ \\ \hline
1 & $U(1)$ & $Z_2$ & Trivial & ${\mathbb Z}_2^{\mathsf e} \times {\mathbb Z}_2^{\mathsf m}$ & $\xrightarrow{\cong}$ & ${\mathbb Z}_2^{\mathsf e} \times {\mathbb Z}_2^{\mathsf m}$ &  $\xrightarrow{0}$ & ${\mathbb Z}_2^{\mathsf e} \times {\mathbb Z}_2^{\mathsf m}$ &  $\xrightarrow{\cong}$ & ${\mathbb Z}_2^{\mathsf e} \times {\mathbb Z}_2^{\mathsf m}$ & $\rightarrow$ &0 \\ %\hline
2 & $U(1)$ &  $Z_2$ & Nontrivial & 0 & $\rightarrow$ & ${\mathbb Z}_2^{\mathsf e} \times {\mathbb Z}_2^{\mathsf m}/ {\mathbb Z}^{\varepsilon}_2$ & $\xrightarrow{\cong}$ & ${\mathbb Z}_2^\varepsilon$ & $\rightarrow$ & 0 & $\rightarrow$ & 0  \\ %\hline
3 & $SO(3)$ & $D_2$ & Trivial & $({\mathbb Z}_2^2)^{\mathsf e} \times ({\mathbb Z}_2^2)^{\mathsf m}$ & $\xrightarrow{\cong}$ & $({\mathbb Z}_2^2)^{\mathsf e} \times ({\mathbb Z}_2^2)^{\mathsf m}$ & $\xrightarrow{0}$ & ${\mathbb Z}_2^{\mathsf e} \times {\mathbb Z}_2^{\mathsf m}$ &  $\xrightarrow{\text{inj.}}$ & $({\mathbb Z}_2^3)^{\mathsf e} \times ({\mathbb Z}_2^3)^{\mathsf m}$ & $\xrightarrow{\text{surj.}}$ & $({\mathbb Z}_2^2)^{\mathsf e} \times ({\mathbb Z}_2^2)^{\mathsf m}$ \\ %\hline
4 & $SO(3)$ & $D_2$ & Nontrivial & ${\mathbb Z}_2^\varepsilon$  & $\xrightarrow{\cong}$ & ${\mathbb Z}_2^\varepsilon$ & $\xrightarrow{0}$ & ${\mathbb Z}_2^\varepsilon$ & $\xrightarrow{\cong}$ & ${\mathbb Z}_2^\varepsilon$ & $\xrightarrow{0}$ & ${\mathbb Z}_2^\varepsilon$\\ \hline
\end{tabular}
\caption{\label{cases1to4} Defect fractionalization classes from Eq.~\eqref{five_term}, for the toric code with coexisting long-range orders, where the fully symmetry $G$ is spontaneously broken down to a subgroup $H$. Different $H$ symmetry actions on the anyons are also considered.}\label{fivetermESS}
\end{table*}

\subsection{Two examples}

Below we will use two primary examples to demonstrate the aforementioned properties of point defects in SET phases: i.e. (1) Point defects can permute different anyons when braided around. (2) Point defects can fuse into Abelian anyons, coined defect fractionalization. The two representative examples we consider are

(i) Toric codes with a coexisting pair superfluid order, where the $G=U(1)$ symmetry is spontaneously broken down to a $H=Z_2$ subgroup. In this example, the point defects are vortices classified by fundamental group $\pi_1(G/H)=\mbz$, labeled by integer-valued vorticity $\nu\in\mbz$. A microscopic model of such will be constructed in section \ref{sec:exp:pair superfluid}.

(ii) Toric codes with a coexisting biaxial nematic order, where the spin rotational symmetry $SO(3)$ is spontaneously broken down to a $H=D_2\simeq Z_2^2$ subgroup. In this example, the point defects are classified by a non-Abelian fundamental group $\pi_1(G/H)\simeq Q_8$, the quarternion group~\cite{Mermin1979}. Two microscopic models of such will be constructed later: for anyon-permuting vortices in section \ref{sec:exp:biaxial nematic}, and for defect fractionalization in section \ref{sec:exp:defect fractionalization}.

\subsubsection{Non-Abelian point defects that permute anyons}

The simplest case of such nature is example (1), i.e. toric code with the pair superfluid order, where $G=U(1)$ spin rotational symmetry is broken down to an $H=Z_2$ Ising symmetry. The resulting symmetry enriched topological phases can be obtained by gauging the fermion parity in fermionic non-chiral topological superconductors with an Ising symmetry \cite{Qi2013,Gu2014}. In this case, there is an $m\in \mathbb{Z}_8$ classification, where each Ising symmetry defect can permute ${\mathsf e}$ and ${\mathsf m}$ anyons in the toric code for $m=1\mod2$, while while the Ising symmetry fractionalization happens for $m=2\mod4$. Now that any $\nu=$~odd vortex is mapped into the Ising symmetry defect by (\ref{eq:short sequence}), we conclude that each $\nu=$~odd vortex can permute $e$ and $m$ anyons, if the pair superfluid ground state is a $Z_2$-SET phase with $m=1\mod2$. As shown in Table \ref{cases1to4}.2, in this case with a nontrivial $H$-action on anyons, both the symmetry fractionalization class and the defect fractionalization class are trivial. A microscopic model of this example is constructed in section \ref{sec:exp:pair superfluid}.

Next we consider example (2), i.e. a biaxial nematic toric code phase with $G = SO(3)$ and $H = D_2  = \{1,X,Y,Z\} \cong Z_2^2$. The associated homotopy groups are $\pi_1(G) \cong Z_2$ and $\pi_1(G/H) \cong Q_8$; the latter is the non-Abelian quarternion group. The $H$-action can be nontrivial in a $H$-enriched toric code: e.g. the generator $X$ in $H = Z_2^X\times Z_2^Y$  permutes the $\mathsf{e}$ and $\mathsf{m}$ particles while $Y$ does not. This is the only nontrivial $H$-action possible up to isomorphism. This also means that the last group element, $Z = XY$, permutes $\mathsf{e}$ and $\mathsf{m}$ particles as well. We construct a microscopic lattice model for this phase in section \ref{sec:exp:biaxial nematic}.

The five-term exact sequence (\ref{five-term:vortex:set}) for this case is given in Table. \ref{cases1to4}.4. Interestingly, we find that in this case the defect fractionalization is not inherited from the symmetry defect (see the last map in Table. \ref{cases1to4}.3, which is a zero map). Thus according to our classification, although the symmetry defect fractionalization class is nontrivial, $\mathcal{H}_\rho^2(H,\mathcal{A})=\mbz_2^\epsilon$ in this case, there is no nontrivial fractionalization for the topological point defect.

\subsubsection{Nontrivial fusion rules of fractionalized defects}

Previously we have shown that a nontrivial $H$-action on anyons leads to a trivial defect fractionalization class in both examples. In fact, in a toric code with a pair superfluid order, when the $H$-action is trivial, even with a nontrivial symmetry fractionalization class $\mathcal{H}_{\text{id}}^2(H=Z_2,\mathcal{A})=\mbz_2^{\mathsf e}\times\mbz_2^{\mathsf m}$, the defect fractionalization class is still trivial, as shown in Table \ref{cases1to4}.1. 

As we show below, example (2), i.e. the toric code with a biaxial nematic order, can realize a nontrivial defect fractionalization class if the $H$-action is trivial on anyons. Again consider the case $G=SO(3)$ and $H = D_2\cong Z_2^2$, this time with trivial $H$-action $\rho=\text{id}$ on the anyons. The five-term exact sequence is given in Table \ref{fivetermESS},
where the surjective map on the right allows us to extend the sequence to a six-term exact sequence by appending ``$\rightarrow 0$'' to the right-hand side.
We see that defect fractionalizations are \emph{fully determined} by symmetry fractionalizations. This is the only case with a nontrivial topological defect fractionalization among those considered in Eq.~\eqref{cases1to4}. Only a subgroup $(\mathbb{Z}_2^2)^{\mathsf{e}}\subset H^2(D_2,\mathbb{Z}_2^\mathsf{e})$ survives the inflation map $p^*$. Denote the three nontrivial cocycles of this $(\mathbb{Z}_2^2)^{\mathsf{e}}$ as $\omega_{1,2,3}$. They can be distinguished by the following values
\begin{equation}
\begin{aligned}
\omega_1(X,X) &= \mathsf{1},\quad \omega_1(Y,Y)=\mathsf{1},\quad \omega_1(XY,XY)=\mathsf{e},\\
\omega_2(X,X)&=\mathsf{1},\quad\omega_2(Y,Y)=\mathsf{e},\quad \omega_2(XY,XY)=\mathsf{1},\\
\omega_3(X,X)&=\mathsf{e},\quad\omega_3(Y,Y)=\mathsf{1},\quad \omega_3(XY,XY)=\mathsf{1}.
\end{aligned}
\end{equation}
the cocycle $[\omega_4]\in {\mathcal H}^2(D_2,\mathbb{Z}_2^{\mathsf{e}})$ with $\omega_4(X,X)=\omega_4(Y,Y)=\omega_4(XY,XY) = \mathsf{e}$ does not survive the $p^*$ map and becomes a coboundary in ${\mathcal H}^2(\pi_1(G/H)\!=\!Q_8,\mathbb{Z}_2^{\mathsf{e}})$. The other half with the coefficient $\mathbb{Z}_2^{\mathsf{m}}$ can be analyzed in a similar manner.

A microscopic model of this case is constructed in section \ref{sec:exp:defect fractionalization}.

\subsection{Lattice models}\label{sec4:c:lattice_models}
\begin{figure}[!h]
\centering
\includegraphics[width=1\columnwidth]{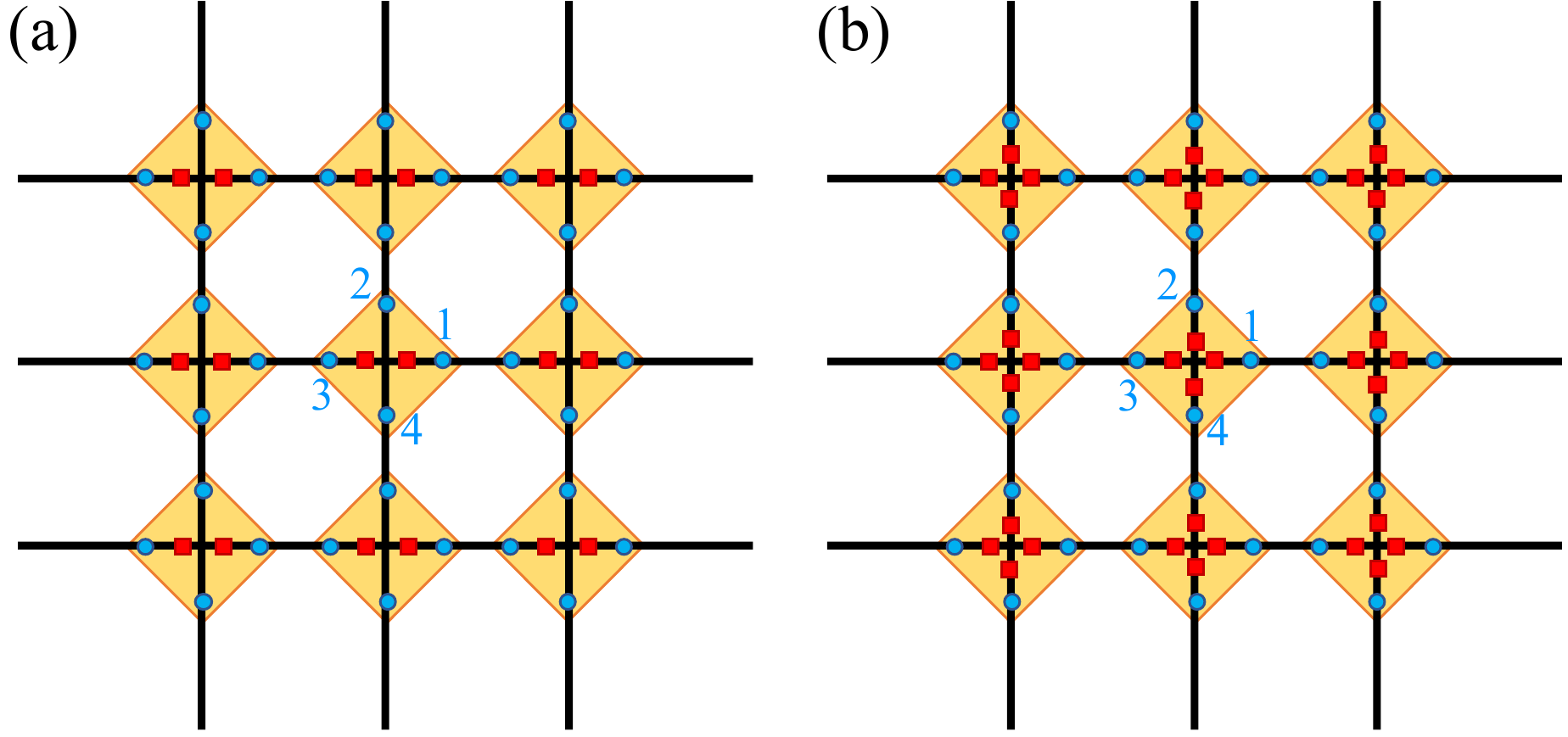}
\caption{\label{Picture02} Illustration for lattice model on square lattice. Each orange kite stands for a unit cell, the blue circle stands for a Majorana, and the red square stands for a complex fermion. (a) A toric code with pair superfluid order, featuring anyon-permuting vortices. (b) A toric code with biaxial nematic order.}\label{fig:lattice model}
\end{figure}

Our general construction of lattice models has the following form:
\bea\label{eq:full ham}
\hat H=\hat H_\text{TO}+\hat H_\text{SSB}+\hat H_\text{int},
\eea
where $\hat H_\text{TO}$ is a Hamiltonian for the symmetry enriched topological order, $\hat H_\text{SSB}$ is a Hamiltonian for the (classical) long-range order associated with spontaneous symmetry breaking, and $\hat H_{\rm int}$ describes the coupling between the topological order and the order parameters of the long-range order.

\subsubsection{Pair superfluids with anyon-permuting vortices}\label{sec:exp:pair superfluid}

We first construct a model for the toric code enriched by a $H=Z_2$ symmetry, which is spontaneously broken down from a $G=U(1)$ group. The model \eqref{eq:full ham} consists of three parts: the $\hat H_\text{TO}$ is responsible for the $Z_2$ topological order in the toric code. $\hat H_\text{SSB}$ is an XY model describing a superfluid that spontaneously breaks $G=U(1)$ symmetry. Meanwhile $\hat H_\text{int}$ describes the interaction/coupling between the superfluid and the topological order.

Built on the square lattice, the Hilbert space of the model consists of a $2^3=8$ dimensional qudit (or 3 qubits) on each site, and an extra spin-$1$ on each site. The site qudit can be represented by a pair of spin-$1/2$ complex fermions $f_{\uparrow,\downarrow}$ and four Majorana fermions $\{\chi_l|1\leq l\leq4\}$ (see Fig.~\ref{Picture02}(a)), satisfying the following constraint of an even fermion parity on each site $i$ in the lattice $\Lambda$ (see Fig. \ref{fig:lattice model}(a)):
\bea
(-1)^{f^\dagger_\uparrow f_\uparrow+f^\dagger_\downarrow f_\downarrow}\chi_1\chi_2\chi_3\chi_4=1,~~~\forall~i\in\Lambda.
\eea
Similar to the Kitaev honeycomb model, this can be viewed as a $Z_2$ gauge constraint (Gauss's law) on each vertex/site. In terms of these fermions, $\hat H_\text{TO}$ writes
\begin{equation}
    \begin{aligned}
       \hat H_\text{TO} &\!=\!\sum_{\expval{i,j}}\imth\chi_{i,l(ij)}\chi_{j,l(ji)}\sum_{\sigma=\uparrow,\downarrow}t_\sigma f^\dagger_{i,\sigma}f_{j,\sigma}\\
       &\!-\!\sum_{i,\sigma}\mu_\sigma f^\dagger_{i,\sigma}f_{i,\sigma}-E_m\sum_{\expval{i,j,k,p}}\hat P_{\expval{i,j,k,p}}.
    \end{aligned}
\end{equation}
We define $\hat P_{\expval{i,j,k,p}}$ as the $Z_2$ flux on each square plaquette with vertices $i,j,k,p$: $\hat P_{\expval{i,j,k,p}}\equiv\chi_{i,l(ij)}\chi_{j,l(ji)}\chi_{j,l(jk)}\chi_{k,l(kj)}\chi_{k,l(kp)}\chi_{p,l(pk)}\chi_{p,l(pi)}\chi_{i,l(ip)}$, where the Majorana label $l(i,j)$ on NN link $\langle{i,j}\rangle$ is defined as: $l(i,i+\hat x) \!=\! 1,~l(i,i-\hat x)\!=\!3,~l(i,i+\hat y)\!=\!2,~l(i,i-\hat y)\!=\!4$. If $E_m\gg|t_\sigma|,|\mu_\sigma|$, the plaquette term $\hat P_{\expval{ijkp}}$ favors zero flux in each square plaquette in the ground state, rather than the $\pi$ flux state favored in the fermion hopping model at half filling~\cite{Lieb1994}.

The Hamiltonian for the spin-$1$'s $\vec S_i$ takes the form of the Bose Hubbard model~\cite{Zhang2013}:
\begin{equation}
  \hat H_\text{SSB}\!=\! \!-\!\sum_{\expval{i,j}}[tS^+_i S^-_j+t_p(S_i^+)^2(S_j^-)^2+{\text h.c.}]\!-\!\mu\sum_iS_i^z.
\end{equation}
Quantum Monte Carlo simulations revealed that it favors a pair superfluid ground state with an order parameter $\expval{(S^+_i)^2}\neq0$ in a finite parameter range, e.g. for $t_p/t\geq g_c$ when $\mu=0$ ($g_c\approx2.5$ on the triangular lattice)~\cite{Zhang2013}.

Finally the coupling term between the $Z_2$ gauge theory and the link spins have the following form:
\begin{widetext}
\bea\notag
&\hat H_{\rm int}=-\sum_{\expval{i,j}}\imth\chi_{i,l(ij)}\chi_{j,l(ji)}\big(S^+_iS^+_j\Delta_\uparrow e^{\imth\arg(j-i)} f_{i,\uparrow}f_{j,\uparrow}+\Delta_\downarrow e^{-\imth\arg(j-i)}f_{i,\downarrow}f_{j,\downarrow}+h.c.\big)-\Delta_0\sum_i(S^-_if^\dagger_{i,\uparrow}f_{j,\downarrow}+{\text{h.c.}}).
\eea
\end{widetext}
Clearly the full Hamiltonian (\ref{eq:full ham}) preserves a $G=U(1)_{S^z}$ spin rotational symmetry:
$S^+_i\rightarrow e^{\imth\theta}S^+_i,~~~f_{i,\uparrow}\rightarrow e^{-\imth\theta}f_{i,\uparrow},~f_{i,\downarrow}\rightarrow f_{i,\downarrow},~\chi_{i,l}\rightarrow\chi_{i,l}$.
In the pair superfluid phase with $\expval{(S^+_i)^2}\neq0$, the $G=U(1)_{S^z}$ spin rotational symmetry is broken down to a $H=Z_2$ subgroup generated by $S^+_i\rightarrow-S^+_i,~f_{i,\uparrow}\rightarrow-f_{i,\uparrow}$. Note that the parity $Z_2^F$ of spin-$1/2$ fermions is always preserved:
\bea
(-1)^{\hat F}\equiv(-1)^{\sum_if^\dagger_{i,\uparrow} f_{i,\uparrow}+f^\dagger_{i,\downarrow}f_{i,\downarrow}}
\eea
In the pair superfluid phase with $\expval{S_i^+S_j^+}\neq0$ and $\expval{S_i^+}=0$, the $f^\dagger_\uparrow$ fermions enter a $p+\imth p$ topological superconducting phase, while the $f_\downarrow$ fermions form a $p-\imth p$ topological superconductor. A fundamental $\nu=1$ vortex of the pair superfluid will translate into a vorticity-1 vortex in the $p+\imth p$ superconductor of $f_\uparrow$'s, hence trapping a single Majorana zero mode at the vortex core. Therefore an odd-vorticity vortex of the pair superfluid permutes $e$ and $m$ sectors in the $Z_2$ toric code.

\subsubsection{Biaxial nematics with anyon-permuting vortices}\label{sec:exp:biaxial nematic}

Another Hamiltonian of the form (\ref{eq:full ham}) can also give rise to a biaxial nematic phase with $Z_2$ toric code topological order, which spontaneously breaks the $SO(3)$ spin rotational symmetry down to a $D_2=(Z_2)^2$ subgroup. Similarly we build our Hilbert space out of fermionic partons: four complex fermions of one $s$ ($f_s$) and three $p$ ($f_{x,y,z}$) orbitals, and four Majoranas $\{\chi_i|1\leq i\leq4\}$. The Majoranas and $f_s$ fermions are spinless, while $f_{x,y,z}$ form a vector (spin-1) representation of the $G=SO(3)$ spin rotational symmetry. Again there is a $Z_2$ gauge constraint for fermion parity on each site of the square lattice (see Fig. \ref{fig:lattice model}(b)):
\bea
(-1)^{f^\dagger_sf_s+\sum_{a=x,y,z}f_a^\dagger f_a}\chi_1\chi_2\chi_3\chi_4=1,~~~\forall~i\in\Lambda.
\eea

The topologically ordered Hamiltonian writes
\begin{widetext}
\bea\notag
&\hat H_\text{TO}=-E_m\sum_{\expval{i,j,k,p}}\hat P_{\expval{i,j,k,p}}-\sum_i(\mu_sf^\dagger_{i,s}f_{i,s}+\mu_p\sum_{a=x,y,z}f^\dagger_{i,a}f_{i,a})+\\
&\sum_{\expval{i,j}}\imth\chi_{i,l(i,j)}\chi_{j,l(j,i)}\big[\Delta_s e^{-\imth\arg(j-i)}f_{i,s}f_{j,s}-t_sf^\dagger_{i,s}f_{j,s}+\sum_{a=x,y,z}(\Delta_p e^{\imth\arg(j-i)}f_{i,a}f_{j,a}-t_pf^\dagger_{i,a}f_{j,a})+~{\text h.c.}\big]
\eea
\end{widetext}

It preserves $SO(3)$ symmetry with a ground state with zero flux in each square plaquette, where the $f_s$ fermions form a $p-\imth p$ superconductor and each flavor of $f_{x,y,z}$ fermions forms a $p+\imth p$ superconductor.

In addition to the $2^5=32$-dimensional qudit described by partons, the physical Hilbert space contains another spin-1 $\vec S_i$ on each site. The nematic order parameter is given by the following $3\times 3$ matrix
\bea
Q_{a,b}=\expval{S^a_{i}S^b_i},~~~a,b=x,y,z.
\eea

The topological order couples with the spin-1's in the following way:
\bea
\hat H_{\rm int}=-J\sum_{i,a,b}f^\dagger_{i,a}f_{i,b}S_i^aS_i^b
\eea

Once the spin-1 Hamiltonian $\hat H_\text{SSB}[\vec S_i]$ \cite{Diener2006,Bernier2006,Song2007} favors a biaxial nematic ground state with
\bea
\hat Q=\bpm q_2-q_1/2&&\\&-q_2-q_1/2&\\&&q_1\epm
\eea
the $G=SO(3)$ spin rotational symmetry is spontaneously broken down to a $H=D_2$ group, generated by $\pi$ rotation along the $\hat x$ and $\hat y$ axis.

In the limit of $q_1=0$ and $J|q_2|\gg|t_p|,|\mu_p|$, the $f_x$ and $f_y$ fermions are driven into a strong-pairing atomic superconductor, giving rise to a $Z_2$ toric code ground state, with a $p+\imth p$ superconductor of $f_z$'s and a $p-\imth p$ superconductor of $f_s$'s. Since $f_z$ is odd under a $\pi$ rotation along either $x$ or $y$ axis, both the $\pm\imth\sigma_x$ and $\pm\imth\sigma_y$ vortices can trap a single Majorana zero mode of $f_z$ and hence permute $e$ and $m$ anyons.

We note that the $SO(3)$ symmetric phase with $Q_{a,b}\equiv0$ in this example is an Abelian $Z_2$ topological order with the following ${\bf K}$ matrix~\cite{Wen1995}: ${\bf K}=4$. It describes the $\nu=2$ state in Kitaev's 16-fold way \cite{Kitaev2006}, where each elementary anyon of statistical angle $\Theta=\imth$ carries spin-$1/2$ (hence a ``spinon''), and each fermion $\{f_a|a=s,x,y,z\}$ is a bound state of two such spinons. 

\subsubsection{Biaxial nematics with defect fractionalization}\label{sec:exp:defect fractionalization}

As discussed previously, when the point defects (or vortices) do not permute anyons in a biaxial nematic order with $H=D_2$ symmetry that is broken down from $G=SO(3)$, they can exhibit defect fractionalization phenomenon captured by the group cohomology $\mathcal{H}^2(\pi_1(G/H)=Q_8,\mathcal{A})$. A model for this phenomenon in the $Z_2$
 topological order (toric code) can be constructed in a similar way as the biaxial nematic order with anyon-permuting vortices in the previous section.

Again we consider an $s$ orbital ($f_s$) and three $p$ orbitals ($f_{x,y,z}$) of complex fermions (see Fig. \ref{fig:lattice model}(b)), coupled to a $Z_2$ gauge field implemented by spinless Majorana fermions. The $p$ orbitals $f_{x,y,z}$ transform as a spin-1 representation of the $G=SO(3)$ symmetry. Now we require $f_x$ and $f_y$ fermions to each form a $p+\imth p$ superconductor, while $f_z$ and $f_s$ fermions each forms a $p-\imth p$ superconductor. In this case, since $f_{x,y}$ fermions are both odd under the $Z\equiv e^{\imth\pi S^z}$ spin rotation, the $\pm\imth\sigma_z$ vortices will each trap a $e^{\imth\pi S^z}$ symmetry defect. In such a $D_2$ symmetry enriched topological order, the symmetry fractionalization class \cite{Gu2014} $[\omega]\in\mathcal{H}^2(D_2,\mathcal{A})$ is characterized by
\bea
\omega(X,X)=\omega(Y,Y)=1,~~~\omega(Z,Z)=\epsilon.
\eea
where $X=e^{\imth\pi S^x},~Y=e^{\imth\pi S^y},~Z=e^{\imth\pi S^z}$ are $\pi$ rotations along $x,y,z$ axis.
As a result, the associated fractionalization class $[\Omega]\in\mathcal{H}^2(Q_8,\mathcal{A})$ for the vortices is characterized by
\begin{equation}
    \begin{aligned}
       \frac{\Omega(\imth\sigma_z,\imth\sigma_z)}{\Omega(\imth\sigma_x,\imth\sigma_x)}\!=\!\frac{\omega(Z,Z)}{\omega(X,X)}\!=\!1,~~\frac{\Omega(\imth\sigma_z,\imth\sigma_z)}{\Omega(\imth\sigma_y,\imth\sigma_y)}\!=\!\frac{\omega(Z,Z)}{\omega(Y,Y)}\!=\!\epsilon.
    \end{aligned}
\end{equation}
Physically this means fusing two $\imth\sigma_x$ vortices differ from fusing two $\imth\sigma_y$ vortices by a fermion $\epsilon$ in the toric code.

One can also arrange $f_{y,z}$ fermions each in a $p+\imth p$ superconductor and $f_{x,s}$ fermions each in a $p-\imth p$ superconductor, to achieve
\bea
\omega(Z,Z)=\omega(Y,Y)=1,~~~\omega(X,X)=\epsilon.
\eea
or arrange $f_{x,z}$ fermions each in a $p+\imth p$ superconductor and $f_{y,s}$ fermions each in a $p-\imth p$ superconductor, to achieve
\bea
\omega(Z,Z)=\omega(X,X)=1,~~~\omega(Y,Y)=\epsilon.
\eea

\begin{figure}[!t]
\centering
\includegraphics[width=1\columnwidth]{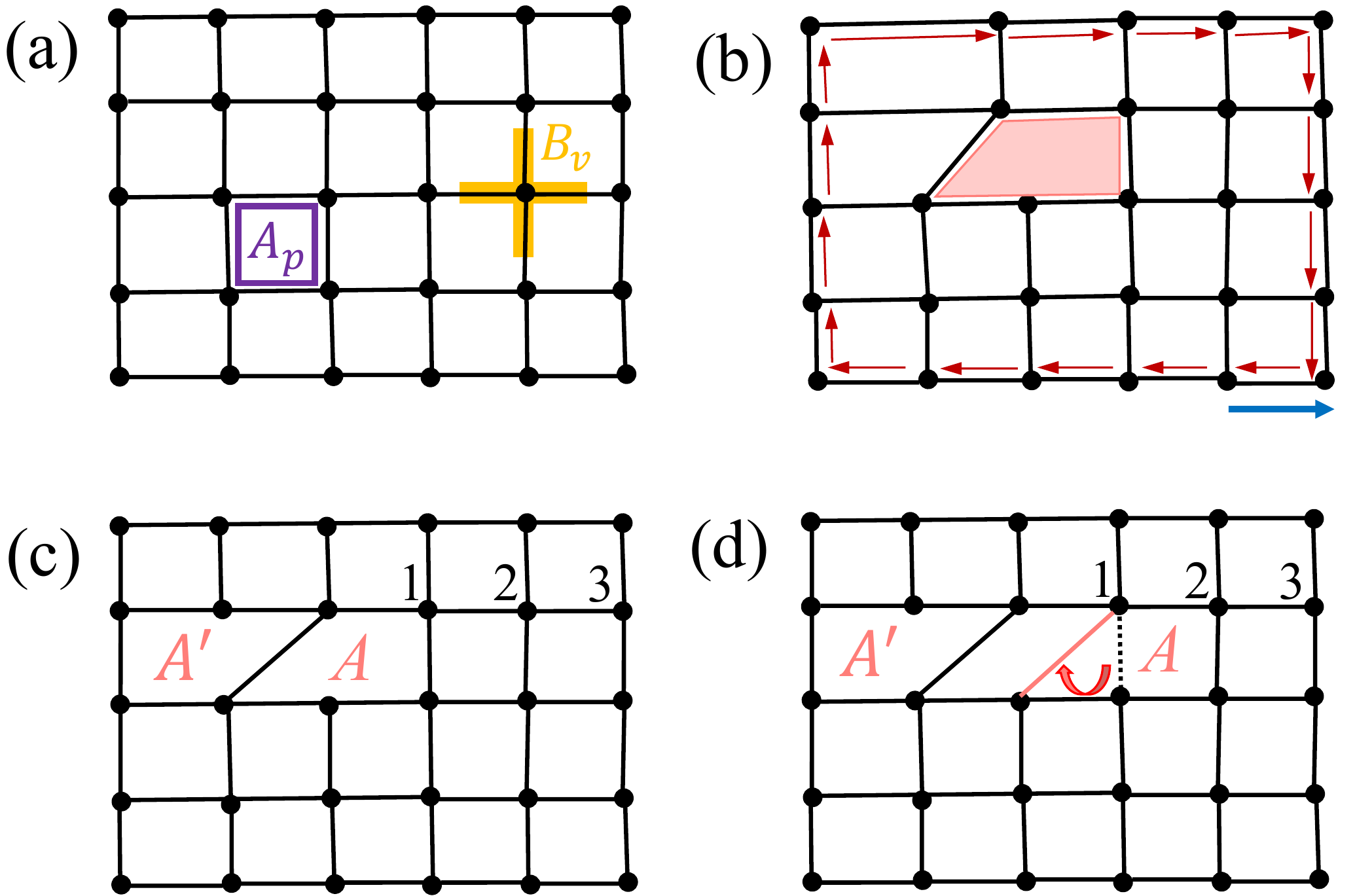}
\caption{\label{Dislocation_Illustration}(a) Illustration of the model Hamiltonian Eq.~\eqref{Toric_Code_Model} without any defects. (b) A single dislocation, with the blue arrow denote the Burger vector. (c) A pair of dislocations $A$ and $A^\prime$. (d) Moving the dislocation $A$ rightward for one unit cell.}
\end{figure}

\subsubsection{Fusion rules of dislocations in the toric code}\label{sec:exp:dislocation}
\begin{figure*}[!t]
\centering
\includegraphics[width=500.0pt]{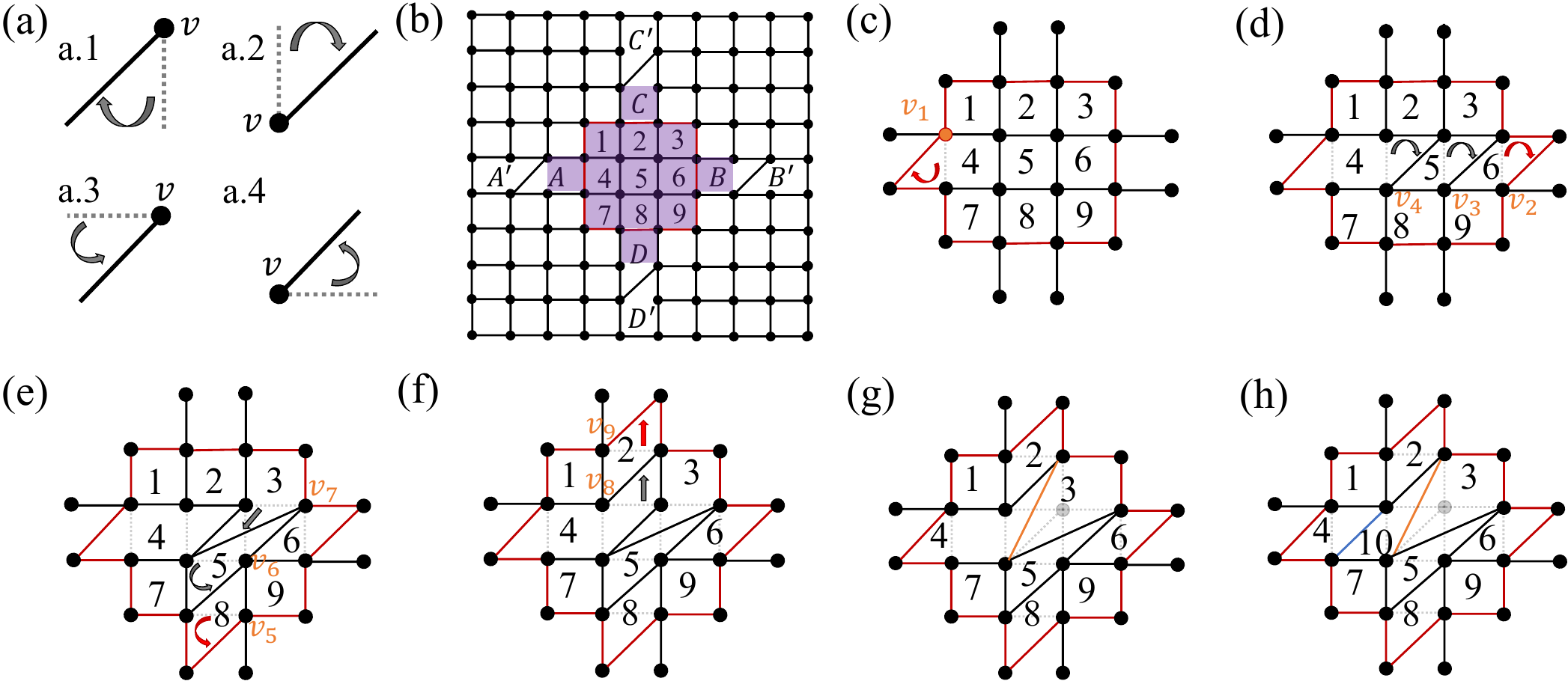}
\caption{\label{Defect_Fusion} Fusion of four dislocation defects. (a) Fundamental operations for defect movements. (b) Toric code with eight dislocation defects. The shaded part is the area that we are interested in. (c) Right move the dislocation $A$ for one lattice constants. (d) Left move dislocation $B$ for three lattice constants. (e) Move dislocation $D$ upwards for three lattice constants. (f) Move dislocation $C$ downward for two lattice constants. (g) Connect two bonds, see Fig.~\ref{Defect_Operation}(b) (h) Add one bond, see Fig.~\ref{Defect_Operation}(c)}
\end{figure*}

The defect fractionalization is captured by ${\mathcal H}^2(\pi_1(G/H),{\mathcal A})$, as we discussed in Sec.~\ref{Sec: Defect fractionalization}. Consider the fusion of dislocations with trivial action in toric code. Here we choose $G = {\mathbb R} \times {\mathbb R}$ and $H = {\mathbb Z} \times {\mathbb Z}$, such that $\pi_1(G/H) = {\mathbb Z}\times {\mathbb Z}$ captures the dislocation defects in two spatial dimension. The toric code topological order has (Abelian) anyon content ${\mathcal A} = {\mathbb Z}_2^{\mathsf e} \times {\mathbb Z}_2^{\mathsf m}$. Since the symmetry fractionalization is of a product form ${\mathcal H}^2_{\rm id}(\pi_1(G/H),{\mathbb Z}_2^{\mathsf e} \times {\mathbb Z}_2^{\mathsf m}) = {\mathcal H}^2_{\rm id}(\pi_1(G/H), {\mathbb Z}_2^{\mathsf e}) \times {\mathcal H}^2_{\rm id}(\pi_1(G/H), {\mathbb Z}_2^{\mathsf m})$, without loss of generality below we only consider the magnetic anyon sector which has non-trivial defect fractionalization  labeled by ${\mathcal H}^2_{\rm id}(\pi_1(G/H), {\mathbb Z}_2^{\mathsf m}) = {\mathcal H}^2_{\rm id}({\mathbb Z}^2,{\mathbb Z}_2^{\mathsf m}) = {\mathbb Z}_2$.  In the following, we will show such a non-trivial defect fractionalization via fusing four dislocations in a lattice model.

The Hamiltonian for toric code on a square lattice is given by:
\begin{equation}\label{Toric_Code_Model}
    \hat H_{{\rm TC}} = -\Delta_m\sum_{p} A_p - \Delta_e\sum_v B_v,
\end{equation}
where $A_p = \prod_{l\in p} Z_l$ for the product of all bonds associated to plaquette $p$ and $B_v = \prod_{l\in v} X_l$ for the product of all bonds associated to vertex $v$. Here $X = \sigma_x$, $Z = \sigma_z$ are Pauli matrices. The symmetry operator for translation of one unit cell along $x$ ($y$) direction is defined as $T_x$ ($T_y$). The symmetry fractionalization of translation symmetries $H=\mbz^{T_x}\times\mbz^{T_y}$ can be understood as follows~\cite{Essin2013}. Consider an eigenstate $\ket{\psi}$ of $\hat H_{\rm TC}$, where a single $e$ particle is created on top of the ground state (assuming the other $e$ particle is at infinity), if it carries the trivial element of the projective representation (i.e., the linear representation) of ${\mathcal H}^2_{\rm id}({\mathbb Z}^2,{\mathbb Z}_2^{\mathsf m}) $, we have $T_x^{-1} \circ T_y^{-1} \circ T_x \circ T_y \ket{\psi} = +\ket{\psi}$. This is e.g. the case when $\Delta_m>0$~\cite{Essin2013}. On the other hand, if it carries the non-trivial projective representation of the ${\mathcal H}^2_{\rm id}({\mathbb Z}^2,{\mathbb Z}_2^{\mathsf m}) $, we have $T_x^{-1} \circ T_y^{-1} \circ T_x \circ T_y \ket{\psi} = -\ket{\psi}$. The fact $T_x^{-1} \circ T_y^{-1} \circ T_x \circ T_y \ket{\psi} = -\ket{\psi}$ means that each plaquette has an ${\mathsf m}$ flux, or equivalently traps one ${\mathsf m}$ particle. This is the case when $\Delta_m<0$. Generally, the translation symmetry fractionalization class is given by
\bea\label{translation sym fractionalization}
\frac{\omega(T_x,T_y)}{\omega(T_y,T_x)}=a\in\mathcal{A}
\eea
where $a=1,m$ in the two scenarios with different sign of $\Delta_m$ (both with $\Delta_e>0$) as discussed above. As discussed in section \ref{Sec: Defect fractionalization}, if $\pi_1(G)=0$ as in this case, due to short exact sequence (\ref{eq:short sequence}), there is a one-to-one correspondence between symmetry defects as elements of $H$ and point defects as elements of $\pi_1(G/H)\simeq H$. Therefore, the above cohomology data of symmetry defects directly translate into cohomology data of point defects of order parameters:
\bea
{\omega(\tau_{\hat x},\tau_{\hat y})}{\omega(\tau_{-\hat x},\tau_{-\hat y})}=a\in\mathcal{A}
\eea
where we use $\tau_{\vec b}$ to label the dislocation (i.e. the point defect of translation symmetries) with Burgers vector $\vec b$. Physically, it means the four dislocations must fuse into an Abelian anyon $a$:
\bea\label{fusion rule:dislocations}
\omega_{\hat x}\times\omega_{\hat y}\times\omega_{-\hat x}\times\omega_{-\hat y}=a\in\mathcal{A}
\eea
where $a$ can be considered as the ``background anyon'' in each unit cell in a translational-invariant topological order\cite{Cheng2016}. Below, we demonstrate the fusion rule (\ref{fusion rule:dislocations}) for four dislocations, by an explicitly calculation of the anyon $a$ in the toric code on the square lattice.

We consider four pairs of dislocations $(A,A^\prime),(B,B^\prime),(C,C^\prime),(D,D^\prime)$, as shown in Fig.~\ref{Defect_Fusion}(b). As mentioned earlier, here we focus on distinguishing $a=1$ vs. $a=m$ (captured by $\mathcal{H}^2(H,Z_2^{\mathsf m})=\mbz_2$), which can be detected by a Wilson loop of $\mathsf e$ type. Such a Wilson loop (red lines) $W_p = \prod \sigma_z$ in Fig.~\ref{Defect_Fusion}(b) probes the parity of the number of ${\mathsf m}$ particles within it. A similar calculation for the $m$-type Wilson loop can fully determine the background anyon $a$ in fusion rule (\ref{fusion rule:dislocations}).

Our strategy is to use finite step local unitary operations to move the four inner dislocations $A,B,C,D$ into the Wilson loop, so that the Wilson loop $W_p$ in Fig.~\ref{Defect_Fusion}(b) can detect their fusion outcome $a=1$ vs. $a=m$. To do so, we first define the following unitary operations for $Z$ components:
\begin{subequations}
\begin{align}
    &U_{T_x^+}(v) Z_{v, v- \hat e_y} U_{T_x^+}^{-1}(v) = Z_{v - \hat e_x -\hat e_y, v } Z_{v -\hat e_y, v -\hat e_x - \hat e_y}, \label{Dislocation_Movement_Operator_1} \\
    &U_{T_x^-}(v) Z_{v,v+ \hat e_y} U_{T_x^-}^{-1}(v) = Z_{v, v + \hat e_x + \hat e_y} Z_{v + \hat e_y, v+ \hat e_x + \hat e_y}, \\
    &U_{T_y^+}(v) Z_{v, v - \hat e_x} U_{T_y^+}^{-1}(v) = Z_{v, v -\hat e_x - \hat e_y } Z_{v -\hat e_x, v -\hat e_x - \hat e_y}, \\
    &U_{T_y^-}(v) Z_{v, v +\hat e_x} U_{T_y^-}^{-1}(v) = Z_{v, v+ \hat e_x + \hat e_y} Z_{v + \hat e_x, v+ \hat e_x + \hat e_y},
\end{align}
\end{subequations}
where $v$ labels the vertex, and $\hat e_x$ ($\hat e_y$) stands for unit vector along the $\hat x$ ($\hat y$) direction. The symmetry operations have been shown in Fig.~\ref{Defect_Operation}(a). Note that these operations can be written in terms of two-qubit unitary gates. For example, the $U_{T_x^+}(v)$ operation in Eq.~\eqref{Dislocation_Movement_Operator_1} is nothing but the two-qubit swap gate:
\begin{equation}
    U_{T_x^+}(v) \!=\! U_{T_x^+}^{-1}(v) \!=\! \frac{1 + Z_{v, v- \hat e_x}}{2} + \frac{1- Z_{v,v-\hat e_x}}{2}X_{v,v+\hat e_y}.
\end{equation}
For the vertex $v=1$, $U_{T_x^+}(v)$ moves the dislocation $A$ in Fig.~\ref{Dislocation_Illustration}(c) rightward by one lattice constant $\hat x$, and arrives at the Fig.~\ref{Dislocation_Illustration}(d).

Similarly, for the $X$ components, we have:
\begin{subequations}
\begin{align}
    &U_{T_x^+}(v) X_{v, v- \hat e_y} U_{T_x^+}^{-1}(v) = X_{v, v - \hat e_x - \hat e_y} X_{v, v -\hat e_y}, \\
    &U_{T_x^-}(v) X_{v, v + \hat e_y} U_{T_x^-}^{-1}(v) = X_{v, v + \hat e_x + \hat e_y} X_{v, v+ \hat e_y}, \\
    &U_{T_y^+}(v) X_{v,v -\hat e_x} U_{T_y^+}^{-1}(v) = X_{v, v -\hat e_x - \hat e_y} X_{v, v-\hat e_x}, \\
    &U_{T_y^-}(v) X_{v, v +\hat e_x} U_{T_y^{-}}^{-1}(v) = X_{v, v + \hat e_x + \hat e_y} X_{v, v + \hat e_x}
\end{align}
\end{subequations}
Now, let us consider the following symmetry actions for the lattice defined in Fig.~\ref{Defect_Fusion}(b): Step 1, move the dislocation $A$ to the right, as shown in Fig.~\ref{Defect_Fusion}(c). The corresponding symmetry action reads: $U_1 = U_{T_x^+}(v_1)$.
Step 2, move the dislocation $B$ to the left, as shown in Fig.~\ref{Defect_Fusion}(d). The corresponding symmetry action will be applying $U_{T_x^-}$ three times: $U_2 = U_{T_x^-}(v_4)U_{T_x^-}(v_3)U_{T_x^-}(v_2)$. Step 3, move the dislocation $D$ upwards, as shown in Fig.~\ref{Defect_Fusion}(e). The corresponding symmetry action will be acting $U_{T_y^+}$ for three times: $U_3 = U_{T_y^+}(v_7)U_{T_y^+}(v_6)U_{T_y^+}(v_5)$. Step 4, move the dislocation $C$ downwards, as shown in Fig.~\ref{Defect_Fusion}(f). The corresponding symmetry action will be acting $U_{T_y^-}$ twice: $U_4 = U_{T_y^-}(v_9)U_{T_y^-}(v_8)$. The Hamiltonian for Fig.~\ref{Defect_Fusion}(b) and Fig.~\ref{Defect_Fusion}(f) are actually related by finite step unitary operations $U = U_4U_3U_2U_1$.

We note that not all plaquettes within the Wilson loop in Fig.~\ref{Defect_Fusion}(f) are squares. We have shown the part associated with non-square plaquettes and non-cross vertex in Fig.~\ref{Defect_Operation}(a). The Hamiltonian for Fig.~\ref{Defect_Operation}(a) reads:
\begin{equation}
\begin{aligned}
    H^\prime_a =& - Z_{l_1}Z_{l_2}Z_{l_3} Z_{l_{11}}Z_{l_{12}} Z_{l_{9}} Z_{l_{10}} -Z_{l_5} Z_{l_6} Z_{l_7} Z_{l_{12}} Z_{l_{11}} \\
    &- X_{l_3} X_{l_4} X_{l_5} X_{l_{11}} -\Delta_{11,12}X_{l_{11}} X_{l_{12}} - X_{l_{12}} X_{l_7} X_{l_8} X_{l_9} \\
    & - X_{l_2} X_{l_{13}} X_{l_3} - X_{l_9} X_{l_{14}} X_{l_{10}}.
\end{aligned}
\end{equation}
\begin{figure}[!h]
\centering
\includegraphics[width=1\columnwidth]{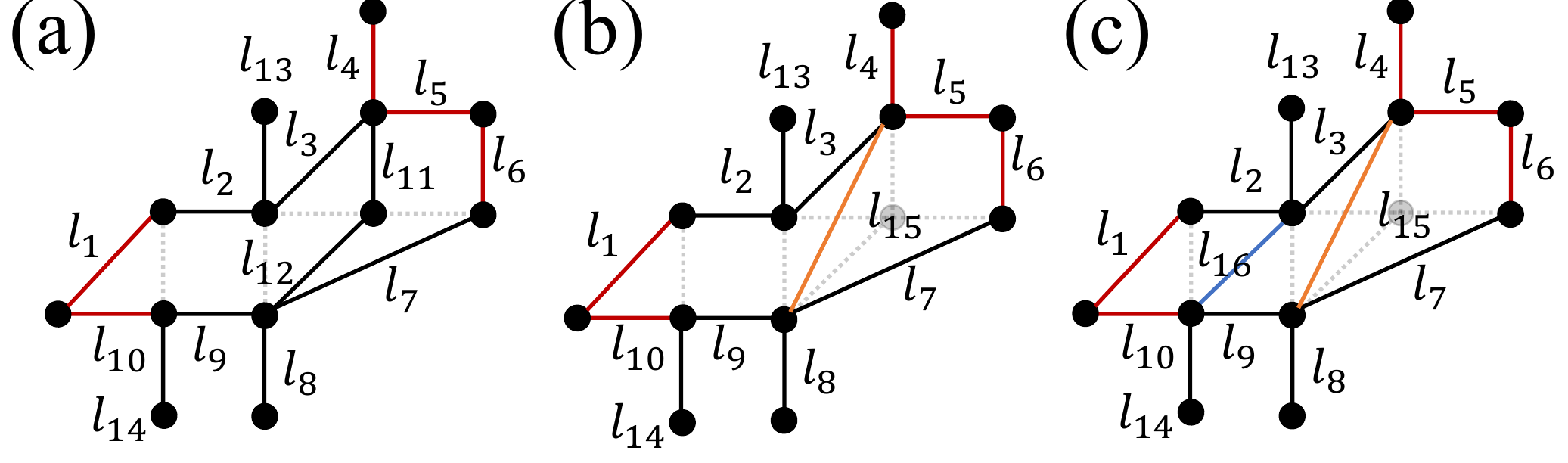}
\caption{\label{Defect_Operation}Merging line $l_{11}$ and $l_{12}$ in (a) to one line $l_{15}$ (orange) in (b) and adding a line $l_{16}$ (blue) in (c).}
\end{figure}
To fuse all the defects, we have to introduce two additional local operations. First, we need to merge line $l_{11}$ and $l_{12}$ to get the orange line $l_{15}$, as shown in Fig.~\ref{Defect_Operation}(b). This can be done by taking the coupling constant $\Delta_{11,12}$ of $X_{ll}X_{12}$ term to infinity, then redefining $X_{15}\equiv X_{11}=X_{12}$ and $Z_{15}=Z_{11}Z_{12}$ in the strong coupling limit. The Hamiltonian for Fig.~\ref{Defect_Operation}(b):
\begin{equation}
    \begin{aligned}
       &H^\prime_b=\lim_{\Delta_{11,12}\rightarrow\infty}H_a^\prime=\\ 
       & - Z_{l_1}Z_{l_2}Z_{l_3} Z_{l_{15}} Z_{l_{9}} Z_{l_{10}} -Z_{l_5} Z_{l_6} Z_{l_7} Z_{l_{15}}  \\
    &- X_{l_3} X_{l_4} X_{l_5} X_{l_{15}} - X_{l_{15}} X_{l_7} X_{l_8} X_{l_9} \\
    & - X_{l_2} X_{l_{13}} X_{l_3} - X_{l_9} X_{l_{14}} X_{l_{10}}.
    \end{aligned}
\end{equation}
The second step is to add the blue line $l_{16}$ in Fig.~\ref{Defect_Operation}(c). The Hamiltonian for Fig.~\ref{Defect_Operation}(c) reads:
\begin{equation}
    \begin{aligned}
       H^\prime_c = & - Z_{l_1}Z_{l_2} Z_{l_{16}} Z_{l_{10}} - Z_{l_3}  Z_{l_{15}} Z_{l_9} Z_{l_{16}}-Z_{l_5} Z_{l_6} Z_{l_7} Z_{l_{15} } \\
    &- X_{l_3} X_{l_4} X_{l_5} X_{l_{15}} - X_{l_{15}} X_{l_7} X_{l_8} X_{l_9} \\
    & - X_{l_2} X_{l_{13}} X_{l_3} X_{l_{16}} - X_{l_9} X_{l_{14}} X_{l_{10}}X_{l_{16}}.
    \end{aligned}
\end{equation}
One way to justify adding qubit $l_{16}$ in the last step is to see that in the ${h_{16}\rightarrow\infty}$ limit of a large Zeeman term $-h_{16}X_{16}$, the low-energy effective Hamiltonian of $H^\prime_c$ above reduces back to $H_b^\prime$.

Note that in all steps discussed above, the gap of the full Hamiltonian never closes. More precisely, we have constructed a smooth path of gapped Hamiltonians which connects the initial Hamiltonian $\hat H_0$ in Fig.~\ref{Defect_Fusion}(b), and the final Hamiltonian $H^\prime_c$ in Fig.~\ref{Defect_Fusion}(h). In other words, the movement of the 4 dislocations into Wilson loop $W_p$ are implemented by a local unitary quantum circuit of finite depth~\cite{Chen2010}. 

However, after moving the 4 dislocations into the Wilson loop $W_p$, one finds that the number of the plaquettes inside the Wilson loop changed from 9 to 10. In the case of a nontrivial translation fractionalization class, with $a=m$ in (\ref{translation sym fractionalization}) and realized by $\Delta_m<0$, there is one $m$ particle per unit cell (or plaquette), Therefore after moving the 4 dislocations into the Wilson loop, the Wilson loop now encloses one extra $m$ particle. This means the fusion outcome of the 4 dislocations is nothing but an $a=m$ anyon. On the other hand, if the translation symmetry fractionalization class is trivial with $a=1$ in (\ref{translation sym fractionalization}), e.g. realized in the original toric code with $\Delta_{e,m}>0$, the background anyon per unit cell (or plaquette) is trivial. Therefore moving the 4 dislocations into the Wilson loop does not bring in extra anyons inside the Wilson loop. As a result, the fusion outcome of the 4 dislocations in (\ref{fusion rule:dislocations}) is also trivial, with $a=1$. Therefore, we have proved in the context of the toric code model that the fusion rule of the dislocations in Eq. (\ref{fusion rule:dislocations}) is determined by the translation fractionalization class (\ref{translation sym fractionalization}).

\section{Skyrmions in intrinsic topological orders}\label{sec:texture}

\subsection{Field theory of the skyrmions in two-dimensional topological orders}

After discussing the point defects in topological orders with coexisting symmetry breaking, in this section we discuss smooth textures of symmetry-breaking order parameters in a topological order. We shall restrict ourselves to skyrmions, e.g. in a ferromagnet where $SO(3)$ spin rotation symmetry is spontaneously broken down to a $U(1)_{S^z}$ subgroup. We shall discuss in detail the universal physical properties of skyrmions following the group cohomology results in section \ref{cohomology:skyrmion:set}.

Consider a skyrmion in ferromagnetic topological orders, where $G=SO(3)$ spin rotational symmetry is spontaneously broken down to a $H=U(1)_{S^z}$ subgroup. As discussed in section \ref{cohomology:skyrmion:set}, due to the short exact sequence (\ref{five_term:texture:set}), a skyrmion of topological charge $\nu\in\mbz=\pi_2(G/H)$ is equivalent to $4\pi\nu$ flux of the remnant $U(1)_{S^z}$ symmetry. Therefore, we can write down a field theory for the ferromagnetic topological order. Note that the skyrmion 3-current~\cite{han2017skyrmions,Fradkinbook,Wilczek1983} should be treated in the same way as the $4\pi$ flux 3-current:
\begin{equation}
    J^\mu_{\rm skyrmion} \equiv\frac{1}{8\pi} \epsilon^{\mu \nu \lambda} \vec n \cdot{} (\partial_\nu \vec n \times \partial_\lambda \vec n)\simeq\epsilon^{\mu\nu\lambda}\frac{\partial_\nu A_\lambda}{4\pi}
\end{equation}
where unit vector $\vec n$ is the ferromagnetic order parameter, and $A_\mu$ is the vector potential for the conserved $U(1)_{S^z}$ symmetry. Now that the ferromagnetic topological order has a conserved spin $S^z$ associated with the remnant $U(1)_{S^z}$ symmetry, we can rewrite its conserved $S^z$ current using a dual gauge field $a_\mu$ such that
\bea
j^\mu_{S^z}=\frac{\epsilon^{\mu\nu\lambda}}{2\pi}\partial_\nu a_\lambda
\eea
Since the spin current couples with external vector potential $A_\mu$ minimally, the full Lagrangian density of the ferromagnetic topological order can be written as follows:
\bea\notag
&\mathcal{L}[\vec n,a_\mu]={\mathcal L}_{T.O.}[a_\mu]-\frac{\epsilon^{\mu\nu\lambda}}{2\pi}A_\mu\partial_\nu a_\lambda-2a_\mu J^\mu_{\rm skyrmion}+\cdots\\
\notag&={\mathcal{L}}_{T.O.}[a_\mu]-\frac{\epsilon^{\mu\nu\lambda}}{2\pi}A_\mu\partial_\nu a_\lambda\\
&-\frac{\epsilon^{\mu\nu\lambda}}{4\pi}a_\mu\big[\vec n \cdot{} (\partial_\nu \vec n \times \partial_\lambda \vec n)\big]+\cdots~~~\label{skyrmion field theory}
\eea
where $\mathcal{L}_{T.O.}[a_\mu]$ describes the intrinsic topological order using the dual gauge field. Integrating out the dual gauge field $a_\mu$ can yield an effective action $\mathcal{L}_{rm eff}[\vec n]$ for the ferromagnetic order parameter $\vec n$. In particular, as we will show below, for a generic topological order, this can induce a Hopf term for the ferromagnetic vector $\vec n$, giving rise to fractional statistics of the skyrmions~\cite{ABANOV2000,Sondhi1993}.

Finally we point out a universal  relation connecting the statistical angle (or topological spin) and $U(1)_{S^z}$ quantum number\footnote{Strictly speaking, since the $U(1)_{S^z}$ symmetry is broken by the skyrmion configuration, the $U(1)_{S^z}$ quantum number $Q$ is not well defined. More precisely, different local perturbations can change this quantum number by an integer. However, the fractional part of $Q$ is a universal number, and it shows up in the universal relation (\ref{skyrmion relation}).} of a skyrmion. The skyrmion as a spatial texture of order parameter $\vec{n}(x,y,t)$ is classified by the second homotopy group $\pi_2(G/H=S^2) = \mathbb{Z}$. A standard realization of a single skyrmion of topological charge $\nu\in\mbz$, of size $R$ and located at the origin, has the following form:
\bea
\vec n(\vec r ,t)\!=\!\big(\sin f(r)\cos(\nu\phi),\sin f(r)\sin(\nu\phi),\cos f(r)\big).~~~~
\eea
where we introduced the polar coordinate $\vec r=(x,y)=r(\cos\phi,\sin\phi)$ and a smooth function $f(r)$ satisfying $f(0)=0$ and $f(r\geq R)=\pi$. Clearly such a skyrmion texture is invariant under a combined spatial rotation (around origin) by angle $\theta$ and spin rotation (along $\hat z$ axis) by angle $\nu\theta$. Therefore, the statistical angle $\Theta_\nu\in U(1)$ of a charge-$\nu$ skyrmion must be related to its $U(1)_{S^z}$ quantum number $Q_\nu$ in the following manner:
\bea\label{skyrmion relation}
{\Theta_\nu}=e^{2\pi\nu\imth Q_\nu}
\eea

\subsection{Abelian topological orders}

To explicitly write down the field theory (\ref{skyrmion field theory}) and to obtain universal properties of skyrmions, in this section we focus on the case of an Abelian topological order, which are classified and described by Abelian Chern-Simons theory with a ${\bf K}$ matrix\cite{Wen1995}.

Suppose the underlying system has an Abelian topological order, characterized by the matrix ${\bf K}$ and a spin vector ${\bf q}$ associated with the remnant $U(1)_{S^z}$ symmetry. In particular, the dual gauge field for the conserved $U(1)_{S^z}$ 3-current is written as
\bea
a_\mu=\sum_I{\bf q}_Ia^I_\mu
\eea
From (\ref{skyrmion field theory}), the Lagrangian density for the system reads:
\begin{equation}\label{TO_Skyrmion}
\begin{aligned}
    {\mathcal L}  &\!=\! \frac{1}{4\pi}\epsilon^{\mu \nu \lambda} K_{IJ} a^I_\mu \partial_\nu a^J_\lambda \!-\! {\bf q}_I a_\mu^I \frac{\epsilon^{\mu \nu \lambda}}{2\pi} \partial_\nu A_\lambda \!-\! 2{\bf q}_I a_\mu^I  J^\mu_{\rm skyrmion},
\end{aligned}
\end{equation}
where the extra 2 in the coefficient of last term denotes the $4\pi$ flux carried by each skyrmion instead of ordinary $2\pi$, as we have discussed in Sec.~\ref{cohomology:skyrmion:set}.

Integrating out $a_\mu^I$, we obtain the spin Hall conductance associated with $A_\mu$-field~\cite{Wen1995}:
\begin{equation}\label{Hall_Conductance}
    \sigma_{xy}^{S_z} = {\bf q}^{T} {\bf K}^{-1} {\bf q}.
\end{equation}
Meanwhile, a skyrmion with topological charge $\nu\in\mbz$ corresponds to an Abelian anyon labeled by the vector ${\bf l}_\nu= 2\nu   {\bf q}$. The $U(1)_{S^z}$ (spin) quantum number carried by a texture of winding number $\nu$ reads:
\begin{equation}\label{Skyrmion_Charge}
    Q_\nu = {\bf l}^T {\bf K}^{-1} {\bf q} = 2\nu   {\bf q}^T {\bf K}^{-1} {\bf q}\mod1,
\end{equation}
and its self-statistics (unit of $a_\mu^I$ flux) reads:
\begin{equation}\label{Skyrmion_Statistics}
    \Theta_\nu = e^{\imth\pi {\bf l}^T {\bf K}^{-1} {\bf l}} = e^{4\pi\imth \nu^2  {\bf q}^T {\bf K}^{-1} {\bf q}}
\end{equation}
Indeed they obey the universal relation (\ref{skyrmion relation}).

\subsection{Half-integer spins with $SU(2)$ symmetry}\label{sec:SU(2)}

Lastly we clarify one subtle difference between $G=SO(3)$ in systems with integer spins, as discussed previously, and $G=SU(2)$ in systems with half-integer spins. 

For a bosonic or fermionic system with half-integer spins, the full symmetry group is $G=SU(2)$ rather than the previously discussed $SO(3)$ case. Since $\pi_1(SU(2))=0$, now the skyrmions as elements of $\pi_2(SU(2)/U(1))=\mbz$ and fluxes as elements of $\pi_1(U(1))=\mbz$ have a one-to-one correspondence realized by bijective map $f$ below:
\begin{equation}\label{eq:short exact sequence:texture:SU(2)/U(1)}
   \pi_2(SU(2))=0 \rightarrow
             \pi_2(S^2)  \xrightarrow{f}
                 \pi_1(U(1))   \rightarrow  \pi_1(SU(2))=0 .
\end{equation}
More precisely, a skyrmion of winding number $\nu$ is now mapped to a $2\pi\nu$ flux (i.e. $\nu$ flux quanta) of the unbroken $H=U(1)_{S^z}$ subgroup. The effective theory for an Abelian topological order with an odd ${\bf K}$ matrix reads:
\begin{equation}\label{TO_Skyrmion_Odd}
\begin{aligned}
    {\mathcal L}  &\!=\! \frac{1}{4\pi}\epsilon^{\mu \nu \lambda} K_{IJ} a^I_\mu \partial_\nu a^J_\lambda \!-\! {\bf q}_I a_\mu^I \frac{\epsilon^{\mu \nu \lambda}}{2\pi} \partial_\nu A_\lambda \!-\! {\bf q}_I a_\mu^I  J^\mu_{\rm skyrmion}.
\end{aligned}
\end{equation}
 Similar to the case for integer spins, a skyrmion of topological charge $\nu$, now labeled by vector ${\bf l}_{\nu}=\nu{\bf q}$, carries $U(1)_{S^z}$ quantum number $Q_\nu=\nu{\bf q}^T{\bf K}^{-1}{\bf q}\mod 1$ and self statistical angle $\Theta_\nu=e^{\imth\pi\nu^2{\bf q}^{T}{\bf K}^{-1}{\bf q}}$. 
 
For example, in the quantum Hall ferromagnet in the lowest Landau level~\cite{Sondhi1993} with ${\bf K}=1,~{\bf q}=1$, a $\nu=1$ skyrmion becomes a fermion with ${\Theta_{\nu=1}}=-1$~\cite{ABANOV2000}. An alternative way to understand this problem is to gauge the fermion parity symmetry $Z_2^F$ to map it to a bosonic topological order with $G=SO(3),~H=U(1)$ as discussed previously. In this case, the quantum Hall ferromagnet in the lowest Landau level~\cite{Sondhi1993} corresponds to a $U(1)_4$ Chern-Simons theory with ${\bf K}=4,~{\bf q}=1$ after gauging the fermion parity, where a $\nu=1$ skyrmion is again a fermion with $\Theta_{\nu=1}=-1$. \\

\section{Concluding remarks}\label{sec:conclusion}

In summary, we described a theoretical framework which classifies point defects and textures in two dimensional quantum phases, where the full symmetry $G$ is spontaneously broken down to a subgroup $H$ of remnant symmetries, so that each symmetry breaking ground state with a fixed (non-fluctuating) order parameter is an $H$-symmetry enriched topological ($H$-SET) state. Using the long exact sequence of homotopy groups that maps the point defects and textures of order parameters to symmetry defects in the $H$-SET phase (see section \ref{sec:theory framework}), we obtain a group cohomology classification for the point defects and textures (see section \ref{sec:group cohomology}), which is induced from the group cohomology classification of $H$-SET phases (including the $H$-SPT phases as a special case). 

Using this general framework and the group cohomology classification, we address their physical consequences focusing on three aspects. In section \ref{sec:spt+dqcp}, we studied point defects and textures of order parameters in $H$-SPT phases, and reveal their connection to deconfined quantum critical points (DQCPs). In section \ref{sec:point defect}, we studied point defects of order parameters in $H$-SET phases, showing that they can (1) permute anyons when braided around, and (2) fuse into Abelian anyons, a phenomenon we coined ``defect fractionalization''. In section \ref{sec:texture} we studied textures of order parameters in $H$-SET phases, establishing their field theory descriptions and the fractional statistics obeyed by the skyrmions. 

This work aims to understand the interplay between classical long-range orders of local order parameters, and quantum orders of long-range entanglement in the ground state \cite{Wen2017}. It serves as a first step towards a complete classification and characterization of quantum phases with both classical and quantum orders. While we focused on point defects and textures in two-dimensional bosonic systems in this work, three natural extensions are: (i) to understand domain walls of discrete symmetry breakings; (ii) to study fermionic systems with long-range orders; and (iii) to go beyond two spatial dimensions. For example, in three dimensions, the coexistence of long-range orders and quantum spin liquids (known as ``magnetic moment fragmentation'') has been proposed in quantum spin ice compounds~\cite{Savary2012,Brooks-bartlett2014}. 

In our study of the symmetry breaking in SET phases in Sec.~\ref{sec:point defect}, we focused on examples of Abelian topological orders, where we used an underlying toric code topological order throughout the analysis. This choice is made for simplicity and for the purpose of explicit lattice model construction in Sec.~\ref{sec4:c:lattice_models}. We believe the toric code suffices for illustrating the general principle that we laid out using an algebraic means. It is interesting to identify a non-Abelian system which harbors the nontrivial point defects and textures discussed in our formalism.

Now that we have studied point defects and smooth textures of order parameters associated with SSB from $G$ to $H$, condensing such point defects (i.e. vortices) or textures (skyrmions) is expected to restore the broken symmetries. However, as we have shown in this work, point defects can carry a projective representation of the remnant symmetry $H$, and textures can carry fractional statistics and fractional quantum numbers. As a result, the defect/texture condensation transition may spontaneously break other symmetries and/or alter the topological order in the ground state. This will lead to a family of quantum phase transitions beyond the Landau paradigm~\cite{Senthil2004,KONG2014,Bischoff2019}. For example, condensing a skyrmion obeying bosonic self statistics but nontrivial mutual statistics with other anyons can restore $SO(3)$ symmetry, leading to a paramagnetic ground state with a different topological order than the ferromagnetic phase. We leave these novel phase transitions as future projects.

Finally, note that in two spatial dimensions, point defects of the order parameters in our framework can be mapped to a one-parameter family of gapped $H$-SET Hamiltonians, while a smooth texture of order parameters in our framework can be mapped to a two-parameter family of gapped $H$-SET Hamiltonians~\cite{Kitaev2006}. Therefore, our classification of point defects and textures in $H$-SET phases also serves as a first step towards the classification and construction of adiabatic pumping cycles in SET phases~\cite{Kapustin2020,Aasen2022,Wen2021}. We also leave these developments to future works.

\section{Acknowledgment}
 We thank Joel E. Moore for helpful discussions, and Dominic Else for bringing to our attention a previous work~\cite{Else2021}. Y.-Q.W. was supported as part of the Center for Novel Pathways to Quantum Coherence in Materials, an Energy Frontier Research Center funded by the U.S. Department of Energy, Office of Science, Basic Energy Sciences. C.L. was supported by the EPiQS program of the Gordon and Betty Foundation. YML was supported by NSF under award number DMR-1653769.

\appendix

\noindent

\setcounter{section}{0}
\setcounter{equation}{0}
\setcounter{figure}{0}
\setcounter{table}{0}
\renewcommand\theequation{A\arabic{equation}}
\renewcommand\thefigure{A\arabic{figure}}
\renewcommand\thetable{A\arabic{table}}
%\renewcommand*{\bibnumfmt}[1]{[S#1]}
% \renewcommand\bibnumfmt[1]{[S#1]}

%\setcounter{enumiv}{0}

% \section*{Supplementary Material}

\begin{table*}[!htb]
\centering
\caption{\label{tab:realization} Table adapted and improved from Ref.~\cite{KOBAYASHI2011}. Phase, order parameter manifold, standard spinor, and the first ($\pi_1$), second ($\pi_2$) and third $(\pi_3)$ homotopy groups for scalr and spinor BEC. Here FM, P, UN, and BN stand for ferromagnetic, polar, uniaxial nematic and biaxial nematic, respectively. The $D_4$ and $T$ are the fourth dihedral group and the tetrahedral group, respectively. The subscripts $f,\phi$ and $(f,\phi)$ show that the symmetry is about spin, gauge and the combined spin and gauge. We denote a vortex as ${\mathbb Z} \times_h (K)_{{\bf F},\phi}$ in spinor BECs, where group $K$ is constructed based on the composite symmetry between the gauge symmetry $\phi$ and the hyperfine spin symmetry $F$. For any $n,m \in {\mathbb Z}$, $g,g^\prime \in K$, $(n,g) \in {\mathbb Z} \times_h (K)_{F,\phi}$ satisfies that  $(n,g) \cdot{} (m, g^\prime) = (n + m + h(g \cdot{} g^\prime), g \cdot{} g^\prime)$, where map $h$: $K \times K \rightarrow {\mathbb Z}$ is defined such that $h(g,g^\prime) =0$ when $\theta + \theta^\prime < 2\pi$ and $h(g,g^\prime) = 1$ when $\theta + \theta^\prime \geq 2\pi$. $K^*$ is defined as  $f(K):=K^*$ by the map $f:$ $U(1) \times SO(3) \rightarrow U(1) \times SU(2)$.}
\begin{tabular}{|c|c|c|c|c|c|c|}
\hline
System & Phase & $G$ & $H$ & ${\mathcal M}$ & $\pi_1({\mathcal M})$  & $\pi_2({\mathcal M})$ \\ \hline
Ions&Crystal&$R^d$&$\mbz^d$&$\big[U(1)\big]^d$&$\mbz^d$&0\\ \hline
Isotropic magnets&FM&$SO(3)$&$U(1)$&$S^2$&0&$\mbz$\\ \hline
Liquid crystal & UN & $SO(3)$ & $D_\infty$ & ${\mathbb R}P^2$&${\mathbb Z}_2$ & ${\mathbb Z}$ \\ \hline
Liquid crystal & BN & $SO(3)$ & $D_2$ & $SO(3)/D_2$ & $Q_8$ & 0  \\ \hline
Spin-0 bosons&$n$-boson BEC&$U(1)$&$\mbz_n$&$U(1)$&$\mbz$&0\\ \hline
Spin-1 bosons & FM & $U(1) \times SO(3)$ & $U(1)$ & $SO(3)$& ${\mathbb Z}_2$ & 0 \\ \hline
Spin-2 bosons &cyclic & $U(1) \times SO(3)$ & $T$ & $(U(1) \times SO(3))/T$ & ${\mathbb Z} \times T$ & 0 \\ \hline
Spin-2 bosons  & UN & $U(1) \times SO(3)$ & $({\mathbb Z}_2)\ltimes SO(2) $ & $U(1) \times S^2/({\mathbb Z}_2)$ & ${\mathbb Z} \times {\mathbb Z}_2$ & ${\mathbb Z}$ \\ \hline
Spin-2 bosons & BN & $U(1)\times SO(3)$ & $(D_4)$ & $(U(1) \times SO(3))/D_4)$  &${\mathbb Z} \times_h (D^*_4)$ & 0  \\ \hline
Spin-2 bosons & nematic & $U(1) \times SO(5)$ & $({\mathbb Z}_2)\ltimes SO(4)$ & $(U(1) \times S^4) / ({\mathbb Z}_2)$ & ${\mathbb Z} \times_h ({\mathbb Z}_2)$ & 0 \\ \hline
${}^3$He-A & dipole-free & $S^2 \times SO(3)$ & ${\mathbb Z}_2$ & $S^2 \times SO(3)/ ({\mathbb Z}_2)$ & ${\mathbb Z}_4$ & ${\mathbb Z}$ \\ \hline
${}^3$He-A & dipole-locked & $O(3) = SO(3) \times {\mathbb Z}_2$ & ${\mathbb Z}_2$ & $SO(3)$ & ${\mathbb Z}_2$ & 0 \\ \hline
\end{tabular}
\end{table*}

\section{Note on notation}

Many Abelian groups are defined in this work. When an Abelian group represents (the fusion of) Abelian anyons we will often denote by blackboard bold symbol, e.g. $\mathbb{Z}_2$. Quite often we use subscript to detail the anyon types, e.g. $\mathbb{Z}_2^{\mathsf{e}}$ denotes the (fusion) group of the trivial anyon $\mathsf{1}$ and the "electric" particle $\mathsf{e}$. We also denote the homotopy group and the cohomology group blackboard bold symbol. On the other hand, when an Abelian group appears as a group of symmetry operations we will often denote with the usual symbol, e.g. $Z_2$ and $Z_4$. Sometimes a group is indicated by its generators, for example, for the Abelian group generated by the two-fold element $h$ all the following notations are equivalent: $Z_2 = Z_2^h = \{1,h\} = \langle h\rangle$.

Several types of products appear in this work. Since quite often we are dealing with finite Abelian groups, we do not distinguish direct product and direct sum, for example, $\mathbb{Z}_2\times \mathbb{Z}_2$ and $\mathbb{Z}_2 \oplus \mathbb{Z}_2$, and $\mathbb{Z}^2_2$ all mean the same object. Note these are different from tensor product of groups. As an example, we have $\mathbb{Z}_2\otimes \mathbb{Z}_2 = \mathbb{Z}_2 \neq \mathbb{Z}^2_2$.

For the group cohomology $\mathcal{H}^*(G,\mathcal{A})$,  $G$ can either act trivially or nontrivially on the coefficient $\mathcal{A}$. Quite often when the action is nontrivial it is specified explicitly in some way (either stated in words or using symbols): for example in Eq.~\eqref{eqc3} the Abelian group $Z_2^h$ generated by an order-two element $h$ acts nontrivially on the anyons $\mathcal{A}$, and we remind this action by the subscript $h$ in the notation $\mathcal{H}^n_h$. As another example, in the five-term exact sequence, all the cohomology groups with possible nontrivial actions ($\rho\colon H\rightarrow \mathrm{Aut}(\mathcal{A}))$, $\widetilde{\rho}\colon \pi_1(G/H)\rightarrow \mathrm{Aut}(\mathcal{A}))$) are manifest by the subscripts, whereas the term $\mathcal{H}^1(\pi_1(G),\mathcal{A})$ without subscript (or with the subscript ``$\mathrm{id}$'') means $\pi_1(G)$ acts trivially on $\mathcal{A}$.

\section{A short introduction to homotopy and group cohomology}

\renewcommand{\theequation}{B\arabic{equation}}
% reset the counter
%\setcounter{equation}{1}

\subsection{Homotopy group and spontaneous symmetry breaking}
In this section we give a brief introduction to homotopy theory based on Ref.~\cite{Mermin1979}. Homotopy theory provides the natural language for the description and classification of defects in a large class of ordered system.
Consider the ground state of a  quantum many-body system, which exhibits a long-range order associated with spontaneous symmetry breaking. To be precise, we consider  the symmetry group $G$ of the Hamiltonian to spontaneously break down to a subgroup $H$ that is preserved by an ordered ground state.  Mathematically, the long-range order of spontaneous symmetry breaking is described by a local order parameter living on the manifold $\mathcal{M}$~\cite{Mermin1979}:
\bea
\hat O(\vec r)\in\mathcal{M}=G/H
\eea
which is the (left) coset space of $G$ modulo $H$, where the full symmetry $G$ of the Hamiltonian is spontaneously broken down to a subgroup $H$ in the ground state. In particular, the remnant symmetry $H$ is the subgroup of $G$ which keeps the order parameter $\{\hat O(\vec r)\}$ invariant:
\bea
H\equiv\{h\in G|h\hat O h^{-1}=\hat O\}.
\eea
The order parameter manifold is therefore given by the coset space $\mathcal{M}=G/H$. Topological defects of codimension $D+1$ in an ordered media is generally classified by the homotopy group $\pi_{D}(G/H)$. In particular, in two spatial dimensions, the different types of topological defects are classified by the following homotopy groups:

(i) Domain walls where order parameters vanish along a line, with codimension 1, classified by $\pi_0(G/H)$;

(ii) Point defects where order parameters vanish at a point, with codimension 2, classified by $\pi_1(G/H)$;

(iii) Smooth textures where order parameters are finite everywhere, classified by $\pi_2(G/H)$.

The following long exact sequence of fibration is a useful tool to compute these homotopy groups~\cite{Mermin1979}:
\begin{equation}
    \begin{aligned}
        \cdots&\longrightarrow\pi_n(H)\longrightarrow\pi_n(G)\longrightarrow\pi_n(G/H)\\
        &\longrightarrow\pi_{n-1}(H)\longrightarrow\pi_{n-1}(G)\longrightarrow\cdots.
    \end{aligned}
\end{equation}

\subsection{Theory of group cohomology}\label{App:B2}
Here we give a short introduction to group cohomology~\cite{Mesaros2013}. The concept of spectral sequence will be given in the next subsection.

\subsubsection{Cochain, cocycle and coboundary}

The input data for group cohomology is a group $G$ (Abelian or non-Abelian) and an Abelian group ${\mathcal A}$ equipped with an action of $G$: $G\times \mathcal{A}\rightarrow \mathcal{A}$, $g\times \mathsf{a} \mapsto g.\mathsf{a}$. Equivalently, the $G$ action on $\mathcal{A}$ defines a map $\rho\colon G\rightarrow \mathrm{Aut}(\mathcal{A})$ from $G$ to the automorphism group of $\mathcal{A}$. Consider $n$-argument functions $\omega(g_1,g_2,\cdots,g_n) \in {\mathcal A}$ that maps an $n$ tuple of group elements in $G$ to the Abelian group $\mathcal{A}$ \begin{equation}
    \omega: \underbrace{G\times G\times\cdots \times G}_{n \text{ times}} \rightarrow {\mathcal A}.
\end{equation}
Such a group function is called an {\it $n$-cochain}. The set of all $n$-cochains, which we denote by $C^n(G, {\mathcal A})$, forms an Abelian group under group multiplication in $\mathcal{A}$
\begin{equation}
    (\omega_1 \cdot \omega_2)(g_1,\cdots,g_n) \!=\! \omega_1(g_1,\cdots,g_n) \cdot{} \omega_2(g_1,\cdots, g_n).
\end{equation}
here we define the identity $n$-cochain to be the trivial group function whose value is always the identity in ${\mathcal A}$. One can define a map $\partial \colon C^n(G,{\mathcal A})\rightarrow C^{n+1}(G, {\mathcal A})$, $\omega \mapsto \partial \omega$ by
\begin{equation}\label{Boundary}
\begin{aligned}
    \partial \omega(&g_1,\cdots, g_{n+1}) \!=\!\\ &[g_1\,.\,\omega(g_2, \cdots, g_{n+1})] \cdot{} \omega^{(-1)^{n+1}}(g_1,\cdots, g_n) \\
    &\times \prod \omega^{(-1)^i} (g_1,\cdots,g_{i-1},g_i \cdot{} g_{i+1},g_{i+2},\cdots,g_{n+1}),
\end{aligned}
\end{equation}
where the symbol $g_1\,.\,\omega(g_2, \cdots, g_{n+1})$ denotes the action of the group element $g_1$ on the function $\omega$, inherited from the action of $G$ on $\mathcal{A}$.

One can check that (1) $\partial^2 \omega := \partial (\partial  \omega) = I$, where $I$ denotes the identity $(n+2)$-cochain, (2) for two $n$-cochains, $\omega_1,\omega_2$, $\partial(\omega_1\cdot \omega_2) = (\partial \omega_1) \cdot{} (\partial \omega_2)$.

An $n$-cochain $\omega(g_1,\cdots,g_n)$ is called an {\it $n$-cocycle} if and only if it is mapped to the trivial elemenet under the map $\partial$, i.e.  $\partial \omega = I$.
%When this condition is satisfied, we also say that $\omega(g_1,\cdots, g_n)$ is an $n$-cocycle of group $G$ with coefficients in ${\mathcal A}$.
The set of all $n$-cocycles, denoted by $Z^n(G,{\mathcal A})$ is a subgroup of $C^n(G, {\mathcal A})$.

Since $\partial^2$ maps every cochain to the trivial one, any $(n-1)$-cochain $c(g_1,\cdots, g_{n-1})$, defines an $n$-cocycle $\partial c$. If an $n$-cocycle $b$ can be represented as $b = \partial c$ for some $c \in C^{n-1}(G,{\mathcal A})$, $b$ is called an {\it $n$-coboundary}. The set of all $n$-coboundaries is a subgroup of $Z^n(G, {\mathcal A})$, which we denote by $B^n(G,{\mathcal A})$. Two $n$-cocycles $\omega_1$, $\omega_2$ are equivalent (denoted by $\omega_1 \sim \omega_2$) if and only if they differ by an $n$-coboundary: $\omega_1 = \omega_2 \cdot b$, where $b \in B^n(G, {\mathcal A})$. 

The {\it $n$th cohomology group} of a group $G$ with coefficients in ${\mathcal A}$, ${\mathcal H}^n(G, {\mathcal A})$, is formed by the equivalence classes in $Z^n(G, {\mathcal A})$ (i.e. up to $B^n(G,\mathcal{A})$). More precisely, we have:
\begin{equation}\label{defgc}
    {\mathcal H}^n_\rho(G, {\mathcal A}) = \frac{Z^n(G, {\mathcal A})}{B^n(G, {\mathcal A})}.
\end{equation}
Here the subscript $\rho$ in the cohomology group is a reminder of the group action $\rho\colon G\rightarrow \mathrm{Aut}(\mathcal{A})$.

For concreteness, we give the expression for the first and second group cohomology:

\begin{widetext}
\begin{equation}\label{Hupper1}
{\mathcal H}^1_\rho(G,\mathcal{A}) = \frac{Z^1(G,\mathcal{A})}{B^1(G,\mathcal{A})}
=
\frac{\{d\colon G\rightarrow \mathcal{A}| d(\text{id})=\mathsf{0},d(gh)=g.d(h)+d(g)\;\forall g,h \in G\}}{\{d_0\colon G\rightarrow \mathcal{A}|d_0(g) = g.\mathsf{a}-\mathsf{a}\text{ for some }\mathsf{a}\in \mathcal{A}\}}.
\end{equation}
\begin{equation}\label{Hupper2}
\begin{aligned}
{\mathcal H}^2_\rho(G,\mathcal{A}) &= \frac{Z^2(G,\mathcal{A})}{B^2(G,\mathcal{A})}\\
&=
\frac{\{\omega\colon G\!\times\! G\!\rightarrow\! \mathcal{A}| \omega(1,g_1)\!=\!\omega(g_1,1)\!=\!0,\omega(g_1,g_2)\!+\!\omega(g_1g_2,g_3)\!=\!g_1.\omega(g_2,g_3)\!+\!\omega(g_1,g_2g_3),\forall g_1,g_2,g_3\!\in\! G\}}{\{\omega\colon G\!\times \!G\!\rightarrow \!\mathcal{A}|\omega(g_1,g_2) \!=\! g_1.d(g_2)\!-\!d(g_1g_2)\!+\!d(g_1)\text{ for some }d\colon G\rightarrow  \mathcal{A}\text{ with } d(1)=0\}}.
\end{aligned}
\end{equation}
\end{widetext}

We note that the Abelian group $\mathcal{A}$ can be either finite (such as $\mathbb{Z}_2$), discrete infinite (such as $\mathbb{Z}$) or continuous (such as $U(1)$). The group $G$ can also in principle be finite, discrete or continuous (the continuous case may be treated with additional caution), and in this work we will mostly work with a discrete group $G$.

Finially, note that one can directly search for all the solutions to Eqs.~\eqref{Hupper1} and \eqref{Hupper2} by implementing them in a computer program, and this is the method we used for computing the cohomology of dihedral and quaternion groups. This method is quite elementary, brutal force in nature, but works well for any finite group $G$ with small order and finite Abelian group $\mathcal{A}$, and can provides the most complete data.

\subsubsection{K{\"u}nneth Formula}

When the group $G$ acts trivially on the coefficient $\mathcal{A}$,
a useful decomposition formula for $\mathcal{H}^n(G, \mathcal{A})$ exists for those $G$ that have a direct product form $G= G_1 \times G_2$, namely the K{\"u}nneth formula:
\begin{equation}\label{Kunneth_SPT}
    {\mathcal H}^n(G, \mathcal{A}) = \sum_{k = 0}^n {\mathcal H}^n (G_1, {\mathcal H}^{n-k}(G_2,\mathcal{A})),
\end{equation}
this says that the $n$th cohomology group of $G$ can be obtained from the cohomology groups of $G_1$ and $G_2$ in lower degree, but in a ``nested'' fashion. Note that here we allow the coefficient $\mathcal{A}$ to be either $U(1)$, which appears in the classification of bosonic SPT (in $d=n-1$ spatial dimensions), or a finite Abelian group, which appears in symmetry fractionalization of SET ($n=2$).

In the case of SPT phases with a $U(1)$ coefficient, this formula implies that SPT phases with $G$ symmetry in $d$ spatial dimension can be constructed from SPT phases with $G_1$ symmetry and $G_2$ symmetry in lower dimensions. Specifically, the $k =0$ term in the formula reads:
\begin{equation}
    {\mathcal H}^0(G_1, {\mathcal H}^{d+1}(G_2, U(1))) = {\mathcal H}^{d+1}(G_2, U(1)),
\end{equation}
physically, this means that some SPT phases with symmetry $G$ in $d$ dimension are identified with SPT phases with a subgroup symmetry $G_2$ in $d$ dimension. The $k=n$ term admits a similar meaning for the subgroup $G_1$.

In the special case of $\mathcal{A} = \mathbb{Z}_2$, the summands in the K{\"u}nneth formula can be further decomposed into tensor product of groups, and we have~\cite{rotman2009introduction}:
\begin{equation}\label{Kunneth_SET}
    {\mathcal H}^n_{\rm id}(G_1 \times G_2, \mathbb{Z}_2) = \bigoplus_{p+q = n} \bigg{(} {\mathcal H}^p_{\rm id}(G_1, \mathbb{Z}_2 ) \otimes {\mathcal H}^q_{\rm id} (G_2, \mathbb{Z}_2) \bigg{)}.
\end{equation}
The K{\"u}nneth formula \eqref{Kunneth_SPT} can be proved using the theory of spectral sequence, which we briefly introduce in App.~\ref{app:c}.

\subsection{Calculation of group cohomology: examples}

Using the definition of group cohomology, for $\mathcal{A}= (\mathbb{Z}_2)^m$ a product of $m$ copies of $\mathbb{Z}_2$ and $G=Z^h_2$ an Abelian group generated by an order-two element $h$, we have~\cite{rotman2009introduction}
\begin{equation}\label{Cyclic_Group}
{\mathcal H}^n(Z^h_2,\mathcal{A} ) =
 \mathcal{A}^h/(h.\mathcal{A}-\mathcal{A}),\quad \text{for }n\geq 1,
\end{equation}
The (generally nontrivial) action of $Z^h_2$ on $\mathcal{A}$ is reflected in the following symbols: $h.\mathcal{A}$ denotes the image of the action of $h$ on $\mathcal{A}$, and $\mathcal{A}^h$ denotes the elements of $\mathcal{A}$ that are left invariant under the action of $h$. Below we calculate the group cohomology that mentioned in the main text.

\subsubsection{${\mathcal H}^2(Z_2,\mathbb{Z}_2^2)$, trivial action}\label{Z2Trivial}
Let's first consider the case $m = 2$ in Eq.~\eqref{Cyclic_Group}. We can interpret $\mathcal{A} = \mathbb{Z}_2^{\mathsf{e}}\times \mathbb{Z}_2^{\mathsf{m}}$ as the toric code topological order, where ${\mathsf e}$ and ${\mathsf m}$ stand for charge and flux component, respectively. The elements in ${\mathcal A}$ are $\mathsf{1} = (0,0)$, ${\mathsf e} = (1 ,0)$, ${\mathsf m} = (0,1)$ and $ \varepsilon = (1,1)$. In the case of trivial action, $h$ does not permute $\mathsf{e}$ and $\mathsf{m}$, we have the invariant subgroup of ${\mathcal A}$ under $h$ as $\mathcal{A}^h = \mathcal{A}$ and the denominator above is trivial, so the right-hand side is just $\mathcal{A} =\mathbb{Z}_2^{\mathsf{e}}\times \mathbb{Z}_2^{\mathsf{m}}$.

\subsubsection{${\mathcal H}^2(Z_2,\mathbb{Z}_2^2)$, nontrivial action}\label{Z2Nontrivial}
Similar to the case in Sec.~\ref{Z2Trivial}, if $h$ permutes $\mathsf{e}$ and $\mathsf{m}$, then the invariant element of ${\mathcal A}$ reads  $\mathcal{A}^h = \mathbb{Z}_2^{\mathsf{\varepsilon}} = \{ 1, \varepsilon \}$. The denominator $h.{\mathcal A} - {\mathcal A} = \{ h. \mathsf{a} - \mathsf{a}| \mathsf{a} \in {\mathcal A} \} = \{1, \varepsilon \} \cong {\mathbb Z}_2$. Thus in this case the right-hand side of Eq.~\eqref{Cyclic_Group} is trivial.

\subsubsection{${\mathcal H}^2(Z_2^2,\mathbb{Z}_2^2)$, trivial action}\label{D2Trivbial}
As another example, we consider ${\mathcal H}^2(D_2,\mathcal{A})$ with $D_2=Z_2^g\times Z_2^h$ and $\mathcal{A}=\mathbb{Z}_2^{\mathsf{e}}\times \mathbb{Z}_2^{\mathsf{m}}$. In the case of trivial action, we have ${\mathcal H}^2(D_2,\mathcal{A}) = {\mathcal H}^2(D_2,\mathbb{Z}_2^{\mathsf{e}})\times {\mathcal H}^2(D_2,\mathbb{Z}_2^{\mathsf{m}}) $. The calculation of ${\mathcal H}^2_{\rm id}( D_2, \mathbb{Z}_2)$ for trivial action can be derived from the K\"unneth formula Eq.~\eqref{Kunneth_SET} with $G_1 = Z_2$ and $G_2 = Z_2$. With above, we shall see:
\begin{equation}
    {\mathcal H}^2_{\rm id} (D_2, {\mathbb Z}_2) =  {\mathbb Z}_2 \oplus ({\mathbb Z}_2\otimes {\mathbb Z}_2) \oplus {\mathbb Z}_2 \cong {\mathbb Z}_2^3.
\end{equation}
Thus we have ${\mathcal H}_{\rm id}^2(D_2, {\mathbb Z}_2^{\mathsf e} \times {\mathbb Z}_2^{\mathsf m})= \left(\mathbb{Z}_2^3\right)^{\mathsf{e}}\times \left(\mathbb{Z}_2^3\right)^{\mathsf{m}}$.

\subsubsection{${\mathcal H}^2(Z_2^2,\mathbb{Z}_2^2)$, nontrivial action}

Compared to the trivial action discussed in App.~\ref{D2Trivbial}, in the case of nontrivial action, $D_2 = Z_2^g\times Z_2^h$, where $h$ permutes $\mathsf{e}$ and $\mathsf{m}$
while $g$ does not permute $\mathsf{e}$ and $\mathsf{m}$.

A direct computation using definition \eqref{Hupper2} gives
\begin{equation}\label{eqc3}
{\mathcal H}^2_h(D_2=Z_2^g\times Z_2^h,\mathcal{A}=\mathbb{Z}_2^{\mathsf{e}}\times \mathbb{Z}_2^{\mathsf{m}})
=\mathbb{Z}_2.
\end{equation}
A more technical calculation using spectral sequence can be found in App.~\ref{app:c}.

\subsubsection{${\mathcal H}^2(Q_8,\mathbb{Z}_2^2)$, trivial action.}

Let us first define the elements of the quaternion group $Q_8$.
Denote $X := i \sigma^x$ and $Y = i\sigma^y$, then any element of $Q_8$ can be written in the standard form $X^x Y^y$, $x=0,1,2,3$, $y=0,1$. For $g_i = X^{x_i}Y^{y_i}$, $i=1,2$, the multiplication in standard form is
\begin{equation}
g_1g_2 = X^{x_1+x_2+2y_1 x_2+2y_1y_2}Y^{y_1+y_2-2y_1y_2},
\end{equation}
where the exponent of $X$, $x_1+x_2+2y_1 x_2+2y_1y_2$, is defined in the mod 4 sense.

Since the action of $Q_8$ on the coefficient $\mathcal{A}=\mathbb{Z}_2^{\mathsf{e}}\times \mathbb{Z}_2^{\mathsf{m}}$ is trivial, the cohomology $\mathcal{H}^2(Q_8,\mathbb{Z}_2^2)$ is the product of two copies of $\mathcal{H}^2(Q_8,\mathbb{Z}_2)$, labeled by $\mathsf{e}$ and $\mathsf{m}$. A direct computation using the definition \eqref{Hupper2} gives
\begin{equation}
\mathcal{H}^2(Q_8,\mathbb{Z}_2) = \mathbb{Z}_2^2,\qquad
\mathcal{H}^2(Q_8,\mathbb{Z}_2^2) = \mathbb{Z}_2^4.
\end{equation}

The explicit cocycles can be found in Table \ref{cocycle_table_trivial}. Importantly, one finds
\begin{equation}
\omega(X^2,X^2) = 0,
\end{equation}
which is independent of the coboundary values,
and that $\omega(X,X) = \Phi + \Omega +b_2$, $\omega(Y,Y)=\Omega+b_2$ and $\omega(XY,XY) = b_2$ where $b_2 \in \mathbb{Z}_2$ labels the different coboundaries and only $\Phi,\Omega \in \mathbb{Z}_2$ label different cocycles in $\mathcal{H}^2(D_2,\mathbb{Z}_2)$. This shows that the following are topological invariants of ${\mathcal H}^2(Q_8,\mathbb{Z}_2)$ (with trivial action):
\begin{equation}
    \frac{\omega(X,X)}{\omega(Y,Y)}=\Phi,\quad
    \frac{\omega(Y,Y)}{\omega(XY,XY)}=\Omega.
\end{equation}

A more technical calculation using spectral sequence is outlined in App.~\ref{app:c}.

 \subsubsection{${\mathcal H}^{1,2}(Q_8,\mathbb{Z}_2^2)$, nontrivial action.}

 We denote $\mathcal{A}=\mathbb{Z}^{\mathsf{e}}_2\times \mathbb{Z}^{\mathsf{m}}_2 = \{\mathsf{1},\mathsf{e},\mathsf{m},\mathsf{\varepsilon}\}$.  and $Q_8 = \{\pm 1, \pm i \sigma^x, \pm i \sigma^y,\pm i\sigma^z\}$. We assume that the nontrivial action comes from $\pm i\sigma^x, \pm i \sigma^z\colon \mathsf{e}\mapsto \mathsf{m}, \mathsf{m}\mapsto \mathsf{e}$, while $\pm 1$ and $\pm i\sigma^y$ have trivial action on ${\mathcal A}$. This is the only consistent way of having nontrivial action (any other nontrivial action is isomorphic to this one).

 First we calculate
 ${\mathcal H}^1(Q_8,\mathbb{Z}_2^2)$using the definition \eqref{Hupper1}.

We solve for $B^1(G,\mathcal{A})$: it is known that $B^1(G,\mathcal{A}) = \mathcal{A}/\mathcal{A}^{G}$, from which we get $B^1(Q_8,\mathcal{A}) = \mathcal{A}/\langle
\mathsf{\varepsilon}\rangle = \mathbb{Z}_2$. Then, we solve for $Z^1(G,\mathcal{A})$: the goal is to find all the crossed homomorphism  $f\colon (x,y)\rightarrow \mathcal{A}$ defined by  $f(x,y):=d(g)$ for $g=X^x Y^y$ subject to group , such that
\begin{equation}\label{conditioneq}
\begin{aligned}
&f(x_1+x_2+2y_1x_2+2 y_1y_2,y_1+y_2-2 y_1y_2)
\\
&+x_1.f(x_2,y_2) + f(x_1,y_1) = \mathsf{0},
\end{aligned}
\end{equation}
where we abused the notation $x_1.f(x_2,y_2) := g_1.f(x_2,y_2)$. The crossed homomorphism $f(x,y)$ is entirely determined by the group generator, i.e. the values of $f(1,0)$ and $f(0,1)$ in $\mathcal{A}$. Naively, there are 16 choices. However, it is easy to see that $f(0,1) = \mathsf{e}$ is forbidden: if $f(0,1)=\mathsf{e}$, then setting $(x_1,y_1,x_2,y_2) = (1,0,0,1)$, the condition \eqref{conditioneq} becomes
$ f(1,1)+\mathsf{m}+f(1,0) = \mathsf{1}$. Further using $f(1,1) = f(1,0)+f(0,1) = f(1,0) + \mathsf{e}$, we have $\mathsf{e}+\mathsf{m} + 2 f(1,0) = \mathsf{1}$, which is forbidden, so $f(0,1)=\mathsf{e}$ is forbidden. Similarly, using $(x_1,y_1,x_2,y_2) = (1,0,1,0)$ one can show that $f(1,0)=\mathsf{e}$ is forbidden. By symmetry, $f(1,0)=\mathsf{m}$ and $f(0,1)=\mathsf{m}$ are also forbidden, so the only possible crossed homomorphisms are $Z_1(G,\mathcal{A})
= \{f\colon G\rightarrow \mathcal{A}|\, (f(1,0),f(0,1)) = (\mathsf{0},\mathsf{0}),
(\mathsf{0},\mathsf{\varepsilon}),
(\mathsf{\varepsilon},\mathsf{0}),\text{ or }
(\mathsf{\varepsilon},\mathsf{\varepsilon})\}
\cong \mathbb{Z}_2^2$, so that
\begin{equation}\label{q8upper1nontrivial}
{\mathcal H}^1(Q_8,\mathcal{A}) = \mathbb{Z}_2.
\end{equation}

We next compute $\mathcal{H}^2(Q_8,\mathcal{A})$ using the definition \eqref{Hupper2}. Without providing detail we state the result
\begin{equation}\label{Q8Hupper2nontrivlal}
\mathcal{H}^2(Q_8,\mathcal{A}) = \mathbb{Z}_2^{\mathsf{\varepsilon}}.
\end{equation}
 We can write down explicit cocycles, see Table \ref{cocycle_table_nontrivial}. Importantly, we find a topological invariant
\begin{equation}\label{topoinvq8nontrivial}
 \omega(X^2,X^2) = (\Omega,\Omega)
 \end{equation}
 where $\Omega \in \mathbb{Z}^{\mathsf{\varepsilon}}_2$ labels the cohomology class.

 A more technical calculation for $\mathcal{H}^2(Q_8,\mathcal{A})$ using spectral sequence is given in App.~\ref{app:c}.

\begin{table*}[!thb]
    \centering
\caption{The cocycle element $\omega(g_1,g_2)$ for $\mathcal{H}^2(Q_8,\mathbb{Z}_2) = \mathbb{Z}_2^2$ with trivial action. The horizontal (vertical) group elements are for $g_2$ ($g_1$). The ordered pair denotes the charge in $\mathbb{Z}_2^{\mathsf{e}}\times \mathbb{Z}_2^{\mathsf{m}}$, where $e_1,e_2,...,e_7,\Omega, \Phi \in \mathbb{Z}_2$. Only $\Omega, \Phi \in \mathbb{Z}_2$ classify the cohomology, i.e. $\Omega=\Phi=0$ denotes the trivial  cocycle element of $\mathcal{H}^2(Q_8,\mathbb{Z}_2) = \mathbb{Z}_2^2$ while $(\Omega,\Phi)=(1,0),(0,1),(1,1)$ denote the nontrivial cocycle element of $\mathcal{H}^2(Q_8,\mathbb{Z}_2) = \mathbb{Z}_2^2$. The other 7 $\mathbb{Z}_2$ parameters, $b_1,b_2,...,b_7$, exhaust all the possible coboundary functions in $\mathcal{B}^2(Q_8,\mathbb{Z}_2)$, which we record here for the search of topological invariants.}\label{cocycle_table_trivial}    \begin{tabular}{c|cccccccc}
\hline\hline
$\omega(g_1,g_2)$&
$1$&$X$&$X^2$&$X^3$&$Y$&$XY$&$X^2Y$&$X^3Y$\\
\hline
$1$&$0$&$0$&$0$&$0$&$0$&$0$&$0$&$0$\\
$X$&$0$&$\Phi\!+\!\Omega\!+\!b_2$&$b_1\!+\!b_2\!+\!b_3$&$\Phi\!+\!\Omega\!+\!b_1\!+\!b_3$&$\Omega\!+\!b_1\!+\!b_4\!+\!b_5$&$\Phi\!+\!b_1\!+\!b_5\!+\!b_6$&$\Omega\!+\!b_1\!+\!b_6\!+\!b_7$&$\Phi\!+\!b_1\!+\!b_4\!+\!b_7$\\
$X^2$&$0$&$b_1\!+\!b_2\!+\!b_3$&$0$&$b_1\!+\!b_2\!+\!b_3$&$b_2\!+\!b_4\!+\!b_6$&$b_2\!+\!b_5\!+\!b_7$&$b_2\!+\!b_4\!+\!b_6$&$b_2\!+\!b_5\!+\!b_7$\\
$X^3$&$0$&$\Phi\!+\!\Omega\!+\!b_1\!+\!b_3$&$b_1\!+\!b_2\!+\!b_3$&$\Phi\!+\!\Omega\!+\!b_2$&$\Omega\!+\!b_3\!+\!b_4\!+\!b_7$&$\Phi\!+\!b_3\!+\!b_4\!+\!b_5$&$\Omega\!+\!b_3\!+\!b_5\!+\!b_6$&$\Phi\!+\!b_3\!+\!b_6\!+\!b_7$\\
$Y$&$0$&$\Phi\!+\!\Omega\!+\!b_1\!+\!b_4\!+\!b_7$&$b_2\!+\!b_4\!+\!b_6$&$\Phi\!+\!\Omega\!+\!b_3\!+\!b_4\!+\!b_5$&$\Omega\!+\!b_2$&$\Phi\!+\!b_1\!+\!b_4\!+\!b_5$&$\Omega\!+\!b_4\!+\!b_6$&$\Phi\!+\!b_3\!+\!b_4\!+\!b_7$\\
$XY$&$0$&$b_1\!+\!b_4\!+\!b_5$&$b_2\!+\!b_5\!+\!b_7$&$b_3\!+\!b_5\!+\!b_6$&$b_3\!+\!b_4\!+\!b_5$&$b_2$&$b_1\!+\!b_5\!+\!b_6$&$b_5\!+\!b_7$\\
$X^2Y$&$0$&$\Phi\!+\!\Omega\!+\!b_1\!+\!b_5\!+\!b_6$&$b_2\!+\!b_4\!+\!b_6$&$\Phi\!+\!\Omega\!+\!b_3\!+\!b_6\!+\!b_7$&$\Omega\!+\!b_4\!+\!b_6$&$\Phi\!+\!b_3\!+\!b_5\!+\!b_6$&$\Omega\!+\!b_2$&$\Phi\!+\!b_1\!+\!b_6\!+\!b_7$\\
$X^3$&$0$&$b_1\!+\!b_6\!+\!b_7$&$b_2\!+\!b_5\!+\!b_7$&$b_3\!+\!b_4\!+\!b_7$&$b_1\!+\!b_4\!+\!b_7$&$b_5\!+\!b_7$&$b_3\!+\!b_6\!+\!b_7$&$b_2$\\
\hline
\hline
\end{tabular}
\end{table*}

\begin{table*}[!htb]
\centering
\caption{The cocycle element $\omega(g_1,g_2)$ for $\mathcal{H}^2(Q_8,\mathbb{Z}_2^{\mathsf{e}}\times \mathbb{Z}_2^{\mathsf{m}}) = \mathbb{Z}_2^{\mathsf{\varepsilon}}$, with nontrivial action. The horizontal (vertical) group elements are for $g_2$ ($g_1$). The ordered pair denotes the charge in $\mathbb{Z}_2^{\mathsf{e}}\times \mathbb{Z}_2^{\mathsf{m}}$, where $e_1,e_2,...,e_7,m_1,m_2,...,m_7,\Omega \in \mathbb{Z}_2$. Only $\Omega \in \mathbb{Z}_2$ classifies the cohomology, i.e. $\Omega=0$ ($\Omega=1$) denotes the trivial (nontrivial) cocycle element of $\mathcal{H}^2(Q_8,\mathbb{Z}_2^{\mathsf{e}}\times \mathbb{Z}_2^{\mathsf{m}}) = \mathbb{Z}_2^{\mathsf{\varepsilon}}$. The other 14 $\mathbb{Z}_2$ parameters, $e_1,e_2,...,e_7,m_1,m_2,...,m_7$, exhaust all the possible coboundary functions in $B^2(Q_8,\mathbb{Z}_2^{\mathsf{e}}\times \mathbb{Z}_2^{\mathsf{m}})$, which we record here for the search of topological invariants.}\label{cocycle_table_nontrivial}
\resizebox{\textwidth}{!}{
\begin{tabular}{c|cccc}
\hline
\hline
$\omega(g_1,g_2)$&
$1$&$X$&$X^2$&$X^3$\\%&$Y$&$XY$&$X^2Y$&$X^3Y$\\
\hline
$1$ &$(0,0) $ & $(0,0) $ & $(0,0) $ & $(0,0) $
%& $(0,0) $ & $(0,0) $ & $(0,0) $ & $(0,0)$
\\
$X$&$(0,0) $ & $(\Omega\!+\!e_1\!+\!e_2\!+\!m_1,e_1\!+\!m_1\!+\!m_2) $ & $(\Omega\!+\!e_1\!+\!e_3\!+\!m_2,e_2\!+\!m_1\!+\!m_3) $ & $(e_1\!+\!m_3,e_3\!+\!m_1) $
%& $(e_1\!+\!e_5\!+\!m_4,\Omega\!+\!e_4\!+\!m_1\!+\!m_5) $ & $(e_1\!+\!e_6\!+\!m_5,e_5\!+\!m_1\!+\!m_6) $ & $(e_1\!+\!e_7\!+\!m_6,e_6\!+\!m_1\!+\!m_7) $ & $(e_1\!+\!e_4\!+\!m_7,\Omega\!+\!e_7\!+\!m_1\!+\!m_4)$
\\
$X^2$&$(0,0) $ & $(e_1\!+\!e_2\!+\!e_3,\Omega\!+\!m_1\!+\!m_2\!+\!m_3) $ & $(\Omega,\Omega) $ & $(\Omega\!+\!e_1\!+\!e_2\!+\!e_3,m_1\!+\!m_2\!+\!m_3) $
%& $(e_2\!+\!e_4\!+\!e_6,m_2\!+\!m_4\!+\!m_6) $ & $(\Omega\!+\!e_2\!+\!e_5\!+\!e_7,m_2\!+\!m_5\!+\!m_7) $ & $(\Omega\!+\!e_2\!+\!e_4\!+\!e_6,\Omega\!+\!m_2\!+\!m_4\!+\!m_6) $ & $(e_2\!+\!e_5\!+\!e_7,\Omega\!+\!m_2\!+\!m_5\!+\!m_7)$
\\
$X^3$&$(0,0) $ & $(e_3\!+\!m_1,e_1\!+\!m_3) $ & $(e_1\!+\!e_3\!+\!m_2,\Omega\!+\!e_2\!+\!m_1\!+\!m_3) $ & $(e_2\!+\!e_3\!+\!m_3,\Omega\!+\!e_3\!+\!m_2\!+\!m_3) $
%& $(\Omega\!+\!e_3\!+\!e_7\!+\!m_4,e_4\!+\!m_3\!+\!m_7) $ & $(\Omega\!+\!e_3\!+\!e_4\!+\!m_5,e_5\!+\!m_3\!+\!m_4) $ & $(e_3\!+\!e_5\!+\!m_6,e_6\!+\!m_3\!+\!m_5) $ & $(e_3\!+\!e_6\!+\!m_7,e_7\!+\!m_3\!+\!m_6)$
\\
$Y$&$(0,0) $ & $(e_1\!+\!e_4\!+\!e_7,m_1\!+\!m_4\!+\!m_7) $ & $(e_2\!+\!e_4\!+\!e_6,m_2\!+\!m_4\!+\!m_6) $ & $(e_3\!+\!e_4\!+\!e_5,m_3\!+\!m_4\!+\!m_5) $
%& $(\Omega\!+\!e_2,\Omega\!+\!m_2) $ & $(e_1\!+\!e_4\!+\!e_5,\Omega\!+\!m_1\!+\!m_4\!+\!m_5) $ & $(e_4\!+\!e_6,m_4\!+\!m_6) $ & $(\Omega\!+\!e_3\!+\!e_4\!+\!e_7,m_3\!+\!m_4\!+\!m_7)$
\\
$XY$&$(0,0) $ & $(e_4\!+\!e_5\!+\!m_1,e_1\!+\!m_4\!+\!m_5) $ & $(e_5\!+\!e_7\!+\!m_2,\Omega\!+\!e_2\!+\!m_5\!+\!m_7) $ & $(e_5\!+\!e_6\!+\!m_3,\Omega\!+\!e_3\!+\!m_5\!+\!m_6) $
%& $(e_3\!+\!e_5\!+\!m_4,e_4\!+\!m_3\!+\!m_5) $ & $(e_2\!+\!e_5\!+\!m_5,\Omega\!+\!e_5\!+\!m_2\!+\!m_5) $ & $(e_1\!+\!e_5\!+\!m_6,\Omega\!+\!e_6\!+\!m_1\!+\!m_5) $ & $(e_5\!+\!m_7,e_7\!+\!m_5)$
\\
$X^2Y$&$(0,0) $ & $(e_1\!+\!e_5\!+\!e_6,\Omega\!+\!m_1\!+\!m_5\!+\!m_6) $ & $(\Omega\!+\!e_2\!+\!e_4\!+\!e_6,\Omega\!+\!m_2\!+\!m_4\!+\!m_6) $ & $(\Omega\!+\!e_3\!+\!e_6\!+\!e_7,m_3\!+\!m_6\!+\!m_7) $
%& $(e_4\!+\!e_6,m_4\!+\!m_6) $ & $(e_3\!+\!e_5\!+\!e_6,m_3\!+\!m_5\!+\!m_6) $ & $(e_2,m_2) $ & $(e_1\!+\!e_6\!+\!e_7,m_1\!+\!m_6\!+\!m_7)$
\\
$X^3Y$&$(0,0) $ & $(\Omega\!+\!e_6\!+\!e_7\!+\!m_1,e_1\!+\!m_6\!+\!m_7) $ & $(\Omega\!+\!e_5\!+\!e_7\!+\!m_2,e_2\!+\!m_5\!+\!m_7) $ & $(e_4\!+\!e_7\!+\!m_3,e_3\!+\!m_4\!+\!m_7) $
%& $(e_1\!+\!e_7\!+\!m_4,e_4\!+\!m_1\!+\!m_7) $ & $(e_7\!+\!m_5,e_5\!+\!m_7) $ & $(\Omega\!+\!e_3\!+\!e_7\!+\!m_6,e_6\!+\!m_3\!+\!m_7) $ & $(\Omega\!+\!e_2\!+\!e_7\!+\!m_7,e_7\!+\!m_2\!+\!m_7)$
\\
\hline\hline
$\omega(g_1,g_2)$&
$Y$&$XY$&$X^2Y$&$X^3Y$\\
\hline
$1$ &%$(0,0) $ & $(0,0) $ & $(0,0) $ & $(0,0) $ &
$(0,0) $ & $(0,0) $ & $(0,0) $ & $(0,0)$\\
$X$&%$(0,0) $ & $(\Omega\!+\!e_1\!+\!e_2\!+\!m_1,e_1\!+\!m_1\!+\!m_2) $ & $(\Omega\!+\!e_1\!+\!e_3\!+\!m_2,e_2\!+\!m_1\!+\!m_3) $ & $(e_1\!+\!m_3,e_3\!+\!m_1) $ &
$(e_1\!+\!e_5\!+\!m_4,\Omega\!+\!e_4\!+\!m_1\!+\!m_5) $ & $(e_1\!+\!e_6\!+\!m_5,e_5\!+\!m_1\!+\!m_6) $ & $(e_1\!+\!e_7\!+\!m_6,e_6\!+\!m_1\!+\!m_7) $ & $(e_1\!+\!e_4\!+\!m_7,\Omega\!+\!e_7\!+\!m_1\!+\!m_4)$\\
$X^2$&%$(0,0) $ & $(e_1\!+\!e_2\!+\!e_3,\Omega\!+\!m_1\!+\!m_2\!+\!m_3) $ & $(\Omega,\Omega) $ & $(\Omega\!+\!e_1\!+\!e_2\!+\!e_3,m_1\!+\!m_2\!+\!m_3) $ &
$(e_2\!+\!e_4\!+\!e_6,m_2\!+\!m_4\!+\!m_6) $ & $(\Omega\!+\!e_2\!+\!e_5\!+\!e_7,m_2\!+\!m_5\!+\!m_7) $ & $(\Omega\!+\!e_2\!+\!e_4\!+\!e_6,\Omega\!+\!m_2\!+\!m_4\!+\!m_6) $ & $(e_2\!+\!e_5\!+\!e_7,\Omega\!+\!m_2\!+\!m_5\!+\!m_7)$\\
$X^3$&%$(0,0) $ & $(e_3\!+\!m_1,e_1\!+\!m_3) $ & $(e_1\!+\!e_3\!+\!m_2,\Omega\!+\!e_2\!+\!m_1\!+\!m_3) $ & $(e_2\!+\!e_3\!+\!m_3,\Omega\!+\!e_3\!+\!m_2\!+\!m_3) $ &
$(\Omega\!+\!e_3\!+\!e_7\!+\!m_4,e_4\!+\!m_3\!+\!m_7) $ & $(\Omega\!+\!e_3\!+\!e_4\!+\!m_5,e_5\!+\!m_3\!+\!m_4) $ & $(e_3\!+\!e_5\!+\!m_6,e_6\!+\!m_3\!+\!m_5) $ & $(e_3\!+\!e_6\!+\!m_7,e_7\!+\!m_3\!+\!m_6)$\\
$Y$&%$(0,0) $ & $(e_1\!+\!e_4\!+\!e_7,m_1\!+\!m_4\!+\!m_7) $ & $(e_2\!+\!e_4\!+\!e_6,m_2\!+\!m_4\!+\!m_6) $ & $(e_3\!+\!e_4\!+\!e_5,m_3\!+\!m_4\!+\!m_5) $ &
$(\Omega\!+\!e_2,\Omega\!+\!m_2) $ & $(e_1\!+\!e_4\!+\!e_5,\Omega\!+\!m_1\!+\!m_4\!+\!m_5) $ & $(e_4\!+\!e_6,m_4\!+\!m_6) $ & $(\Omega\!+\!e_3\!+\!e_4\!+\!e_7,m_3\!+\!m_4\!+\!m_7)$\\
$XY$&%$(0,0) $ & $(e_4\!+\!e_5\!+\!m_1,e_1\!+\!m_4\!+\!m_5) $ & $(e_5\!+\!e_7\!+\!m_2,\Omega\!+\!e_2\!+\!m_5\!+\!m_7) $ & $(e_5\!+\!e_6\!+\!m_3,\Omega\!+\!e_3\!+\!m_5\!+\!m_6) $ &
$(e_3\!+\!e_5\!+\!m_4,e_4\!+\!m_3\!+\!m_5) $ & $(e_2\!+\!e_5\!+\!m_5,\Omega\!+\!e_5\!+\!m_2\!+\!m_5) $ & $(e_1\!+\!e_5\!+\!m_6,\Omega\!+\!e_6\!+\!m_1\!+\!m_5) $ & $(e_5\!+\!m_7,e_7\!+\!m_5)$\\
$X^2Y$&%$(0,0) $ & $(e_1\!+\!e_5\!+\!e_6,\Omega\!+\!m_1\!+\!m_5\!+\!m_6) $ & $(\Omega\!+\!e_2\!+\!e_4\!+\!e_6,\Omega\!+\!m_2\!+\!m_4\!+\!m_6) $ & $(\Omega\!+\!e_3\!+\!e_6\!+\!e_7,m_3\!+\!m_6\!+\!m_7) $ &
$(e_4\!+\!e_6,m_4\!+\!m_6) $ & $(e_3\!+\!e_5\!+\!e_6,m_3\!+\!m_5\!+\!m_6) $ & $(e_2,m_2) $ & $(e_1\!+\!e_6\!+\!e_7,m_1\!+\!m_6\!+\!m_7)$\\
$X^3Y$&%$(0,0) $ & $(\Omega\!+\!e_6\!+\!e_7\!+\!m_1,e_1\!+\!m_6\!+\!m_7) $ & $(\Omega\!+\!e_5\!+\!e_7\!+\!m_2,e_2\!+\!m_5\!+\!m_7) $ & $(e_4\!+\!e_7\!+\!m_3,e_3\!+\!m_4\!+\!m_7) $ &
$(e_1\!+\!e_7\!+\!m_4,e_4\!+\!m_1\!+\!m_7) $ & $(e_7\!+\!m_5,e_5\!+\!m_7) $ & $(\Omega\!+\!e_3\!+\!e_7\!+\!m_6,e_6\!+\!m_3\!+\!m_7) $ & $(\Omega\!+\!e_2\!+\!e_7\!+\!m_7,e_7\!+\!m_2\!+\!m_7)$\\
\hline\hline
\end{tabular}}
\end{table*}

\section{Spectral sequence and group cohomology}\label{app:c}

\renewcommand{\theequation}{C\arabic{equation}}

\subsection{The LHS spectral sequence}

Consider the setup given in Eq.~\eqref{NGGN}, namely a group $G$, its normal subgroup $N$, the quotient group $Q:=G/N$, and an Abelian group $\mathcal{A}$ with $G$ action.

The Lyndon-Holchschild-Serre (LHS) spectral sequence is a computational tool to build the group cohomology of $G$ out of those of $N$ and $Q$. It associates with the group extension \eqref{NGGN} a three-dimensional array of Abelian groups $E^{p,q}_n$, $p,q,n \geq 0$. It is conventional to call the third argument $n$ a \emph{page}, and for each page $n$ we have a two-dimensional (semi-infinite) array
\begin{equation}\label{spectral_sequence_def}
\begin{array}{c|ccccc}
\vdots & \vdots & \vdots & \vdots &\vdots &  \\
q=3 & E^{0,3}_n & E^{1,3}_n & E^{2,3}_n& E^{3,3}_n  & \cdots\\
q=2 & E^{0,2}_n & E^{1,2}_n & E^{2,2}_n  & E^{3,2}_n & \cdots\\
q=1 & E^{0,1}_n & E^{1,1}_n & E^{2,1}_n  & E^{3,1}_n & \cdots\\
q=0 & E^{0,0}_n & E^{1,0}_n & E^{2,0}_n  & E^{3,0}_n & \cdots\\
\hline
E^{p,q}_n & p=0&p=1&p=2&p=3 &\cdots\end{array}
\end{equation}
the second page ($n=2$) has a concrete form:
\begin{equation}\label{SecondPageGeneric}
     E_2^{p,q} = {\mathcal H}^p(Q, {\mathcal H}^q(N,{\mathcal A})),
\end{equation}
where the cohomology of $Q$ has coefficients $\mathcal{H}^*(N,A)$; implicit in this is a predefined action of $Q$ on $\mathcal{H}^*(N,A)$, which is inherited from the action of $G$ on $\mathcal{A}$. Here we give the explicit form of this action: for $[\omega] \in \mathcal{H}^d(N,\mathcal{A})$ where $\omega$ is a $d$-cochain (i.e. a $d$-argument function $N\times \cdots N\rightarrow \mathcal{A}$), the action of $Q$ on $\omega$ is defined by
\begin{equation}\label{q_action_on_coho_of_n}
(q.\omega)(n_1,n_2,...,n_d)
:=q.(\omega(q^{-1}n_1q,q^{-1}n_2q,...,q^{-1}n_d q)),
\end{equation}
for any $q \in Q$, $n_1,...,n_d \in N$.
This explicit definition allows us to calculate the second page using elementary methods. The higher pages (whose definition will be given below)
 generally do not have a simple expression, but they are subgroups of the 2nd page: $E^{p,q}_n \subset E^{p,q}_2$, $n=3,4,...$. Note that here we only consider the case where $G$ is a discrete group, $Q$ is finite, and that all cohomology groups of $N$ and $Q$ considered here are also finite. In this case, the $n$-th cohomology group of $G$ is given by the direct sum on the $n=\infty$ page:
\begin{equation}\label{Sequence_Define}
    {\mathcal H}^n(G, {\mathcal A}) = \bigoplus_{p+q = n} E^{p,q}_\infty.
\end{equation}
note that the action of $G$ on the coefficient $\mathcal{A}$ can be either trivial or nontrivial.

The remaining task concerns how to go from the 2nd page to the $n=\infty$ page. This is done using the \emph{differential maps} $d^{p,q}_n$. These are predefined maps that comes with a LHS spectral sequence:
\begin{equation}
    d_m^{p,q}\colon E_m^{p,q} \rightarrow E_m^{p+m,q-m+1},\quad \forall p,q \in \mathbb{Z},
\end{equation}
with the understanding that $E^{p,q}_m = 0$ whenever $p<0$ or $q<0$.
The entries in the $(m+1)$-th page are defined as the homology of those in the $m$-th page:
\begin{equation}\label{ThirdPageGeneric}
    E_{m+1}^{p,q} = {\rm homology}(E_m^{p,q}) = \frac{{\rm ker}(d_m^{p,q})}{{\rm im}(d_m^{p-m,q+m-1})}.
\end{equation}
as special examples, we have
\begin{equation}
    \begin{aligned}
       E_3^{1,2} &= \frac{{\rm ker}(d_2^{1,2})}{{\rm im}(d_2^{-1,3})} = {\rm ker}(d_2^{1,2}), \\ E_3^{3,0} &= \frac{{\rm ker}(d_2^{3,0})}{{\rm im}(d_2^{1,1})}= \frac{E^{3,0}_2}{{\rm im}(d_2^{1,1})}.
    \end{aligned}
\end{equation}
the differential maps on a generic page $n$ are complicated. Here we write down the explicit formulas for two $d_2$ maps that already fulfills our purpose \footnote{We thank Bill Jacob for providing the explicit form of the $d_{0,1}$ and $d_{1,1}$ maps.}:
\begin{equation}\label{explicitdmaps}
\begin{aligned}
&(d_{0,1}(f))(q_1,q_2)
= f((l(q_1q_2))^{-1}l(q_1)l(q_2)),\\
&(d_{1,1}(\omega))(q_1,q_2,q_3)
= (\omega(q_3))((l(q_1q_2))^{-1}l(q_1)l(q_2)),
\end{aligned}
\end{equation}
here $q_1,q_2,q_3 \in Q$, $f \in Z^1(N,\mathcal{A})$, $\omega \in Z^1(Q,\mathcal{H}^1(N,\mathcal{A}))$,
and $l\colon Q\rightarrow G$ is a lifting, which is any function $Q\rightarrow G$ (not necessarily a homomorphism) that satisfies $p\circ l (q) = q$ for any $q \in Q$ and that $l(1_Q)=1_G$ for the identity elements of $1_Q \in Q$ and $1_G \in G$ (see \cite{rotman2009introduction}, P497). To make sense of Eq.~\eqref{explicitdmaps}, we must have that $d_{0,1}(f) \in \mathcal{H}^2(Q,\mathcal{A}^N)$ is a 2-cocycle and that $d_{1,1}(f) \in \mathcal{H}^3(Q,\mathcal{A}^N)$ is a 3-cocycle. These can be proved using the definition of group cohomology by elementary computation.

Due to the increasing span of the differential maps as one goes to higher pages, it is easy to see that the entries in the lower left corner of the spectral sequence (i.e. those with small $p$ and $q$) will be unmodified at some point when going to the next page. When this happens, we say that this entry is ``stabilized''. When all the entries in a page is stabilized, we say this page is ``stabilized'' or it ``collapses'' (to the infinity page). One can either check that
\begin{equation}
\begin{aligned}
E^{0,0}_2 &= E^{0,0}_\infty,~~
E^{1,0}_2 = E^{1,0}_\infty,\\
E^{0,1}_3 &= E^{0,1}_\infty,~~
E^{1,1}_3 = E^{1,1}_\infty,~~
E^{2,1}_3 = E^{2,1}_\infty,~~
E^{2,0}_3 = E^{2,0}_\infty,\\
E^{0,2}_4 &= E^{0,2}_\infty,\cdots\cdots
\end{aligned}
\end{equation}
And this shows that the calculation of the first and the second group cohomology stops at finite pages
\begin{equation}
\begin{aligned}
    \mathcal{H}^1(G,A) &= E^{0,1}_\infty \oplus E^{1,0}_\infty
    = E^{0,1}_3\oplus E^{1,0}_2,\\
    \mathcal{H}^2(G,{\mathcal A}) &= E^{0,2}_\infty \oplus E^{1,1}_\infty \oplus E^{2,0}_\infty\\
    &=E^{0,2}_4\oplus E^{1,1}_3\oplus E^{2,0}_3.
\end{aligned}
\end{equation}

The LHS spectral sequence is not only a powerful computational tool for group cohomology, but many results introduced in the previous sections are special cases of it. First of all, the K\"{u}nneth formula \eqref{Kunneth_SPT} is nothing but a special case of Eq.~\eqref{Sequence_Define} with $E^{p,q}_\infty = E^{p,q}_2$. This is indeed the case when $G = G_1\times G_2$ acts trivially on the coefficient $\mathcal{A}$, where one can show that all the differential maps $d^{p,q}_n$ for $n\geq 2$ vanishes. Second, a corollary of the LHS spectral sequence is the five-term exact sequence \cite{rotman2009introduction}
\begin{equation}
0\rightarrow E^{1,0}_2\rightarrow \mathcal{H}^1(G,\mathcal{A})
\rightarrow E^{0,1}_2\xrightarrow{d_2^{0,1}}E^{2,0}_2\rightarrow \mathcal{H}^2(G,\mathcal{A}),
\end{equation}
which is exactly Eq.~\eqref{five_term} in the main text.

\subsection{Application of the LHS spectral sequence}

In the following we apply the LHS spectral sequence to the calculation of certain cohomology groups that we referred to in the main text. While the $n$-th cohomology can always be obtained using the definition \eqref{defgc}, or more explicitly, \eqref{Hupper2} for the 2nd group cohomology case (see the comment below \eqref{Hupper2}), the spectral sequence calculation provides a detailed structure of the cohomology group.

\subsubsection{${\mathcal H}^2(Z_2^2,\mathbb{Z}_2^2)$, nontrivial action}

Recall that $D_2=Z_2^g\times Z_2^h$, where $h$ permutes $\mathsf{e}$ and $\mathsf{m}$
while $g$ does not permute $\mathsf{e}$ and $\mathsf{m}$. We define the normal subgroup to be $N=Z_2^h$, so we have $0\rightarrow Z_2^h\rightarrow Z_2^g\times Z_2^h\rightarrow \{Z_2^h,gZ_2^h\}\rightarrow 0$ and using the results from App.~\ref{Z2Nontrivial} for the nontrivial action case, we see that all the ${\mathcal H}^{n\geq 1}(Z_2^h,\mathcal{A})$ is trivial. This means that the only nontrivial places the spectral sequence Eq.~\eqref{spectral_sequence_def} is the $q=0$ row, where $\mathcal{A}^h =\{1, \varepsilon\} = \mathbb{Z}_2^{\mathsf{\varepsilon}}$. Then using the example for the trivial action case in App.~\ref{D2Trivbial}, we see that the $q=0$ row has $\mathbb{Z}^{\varepsilon}_2$ for all $p\geq 1$. From Eq.~\eqref{Sequence_Define}, we have:
\begin{equation}\label{eqc3}
{\mathcal H}^n_h(D_2=Z_2^g\times Z_2^h,\mathcal{A}=\mathbb{Z}_2^{\mathsf{e}}\times \mathbb{Z}_2^{\mathsf{m}})
=\mathbb{Z}^{\varepsilon}_2\quad \text{for }n\geq 1.
\end{equation}
If instead both $h$ and $g$ acts nontrivially by permuting the $\mathsf{e}$ and $\mathsf{m}$ particles, we would get the same result \eqref{eqc3} by using a redefined decomposition of $D_2$.

\subsubsection{${\mathcal H}^2(Q_8,\mathbb{Z}_2^2)$, trivial action}

The spectral sequence calculation is detailed in the textbook \cite{adem2013cohomology} (P128-129), which we outline here. The spectral sequence makes use of the fact that \begin{equation}\label{Q8_standard}
0\rightarrow Z_4 \rightarrow Q_8\rightarrow Z_2 \rightarrow 0.
\end{equation}
We have ${\mathcal H}^n(Z_4,\mathbb{Z}_2) = \mathbb{Z}_2$ for all $n$. The action of $Q:=Z_2$ on ${\mathcal H}^n(Z_4,\mathbb{Z}_2)$ is trivial (since the $\mathrm{Aut}(\mathbb{Z}_2)$ is trivial), meaning that $E^{p,q}_2 = \mathbb{Z}_2$ for all $p,q\geq 0$. The cohomology is not yet determined, and one must analyze the $E_3$ and $E_4$ pages of the spectral sequence using the differential maps. Without going into further detail, we give the final result: ${\mathcal H}^2(Q_8,\mathbb{Z}_2) = \mathbb{Z}_2^2$.

\subsubsection{${\mathcal H}^2(Q_8,\mathbb{Z}_2^2)$, nontrivial action}

Note that we have two short exact sequences for $Q_8$: one is Eq.~\eqref{Q8_standard}, and the other one is
\begin{equation}\label{Q8_homotopy}
0\rightarrow D_2 \rightarrow Q_8 \rightarrow Z_2 \rightarrow 0,
\end{equation}
note that the latter, Eq.~\eqref{Q8_homotopy}, is the one obtained from the short exact sequence of homotopy groups in Eq.~\eqref{eq:short sequence}. These two short exact sequences define two separate LHS spectral sequences from which ${\mathcal H}^2(Q_8,\mathbb{Z}_2^2)$ can be obtained. Now let us calculate these two spectral sequences.

\textbf{LHS spectral sequence associated with Eq.~\eqref{Q8_standard}}.

We use the notations given in the previous section. We further denote $N = \{\pm 1, \pm i \sigma^x\} \cong \mathbb{Z}_4$, and the quotient group $Q = Q_8/N = \{  N, i\sigma^y N\} := \{1,q\}$, where $q = N$ is the generator of $Q$. Using the standard notation above (see \cite{rotman2009introduction}), we now have
$${}_N\mathcal{A} = \mathcal{A},\quad N\mathcal{A} = \{0\},\quad D\mathcal{A} = \mathbb{Z}_2^{\mathsf{\varepsilon}}, \quad \mathcal{A}^{Q_8} = \mathbb{Z}_2^{\mathsf{\varepsilon}},
$$
so that we have ${\mathcal H}^{2n-1}(N,\mathcal{A}) = \mathcal{A}/\langle \mathsf{\varepsilon}\rangle$ and ${\mathcal H}^{2n}(N,\mathcal{A}) = \langle \mathsf{\varepsilon}\rangle$, where we use $\langle \mathsf{\varepsilon}\rangle $ and $\mathbb{Z}_2^{\mathsf{\varepsilon}}$ interchangeably.

Now, we need to examine the effect of $Q$ as in ${\mathcal H}^n(N,\mathcal{A})^Q$. By definition (see Eq.~\eqref{q_action_on_coho_of_n}), %(See \cite{rotman2009introduction}, P567-568 and \cite{Online1}),
for a cocycle $f(n) = \mathsf{a}$ in ${\mathcal H}^1(N,\mathcal{A})$, the restriction to ${\mathcal H}^1(N,\mathcal{A})^Q$ are the cocycles $f$ s.t. $q.f(q^{-1}nq) =f(n)$, up to coboundary, and for ${\mathcal H}^n$ with $n\geq 2$ a similar criterion can be defined. Here, we do have ${\mathcal H}^1(N,\mathcal{A})^Q = {\mathcal H}^1(N,\mathcal{A}) = \mathcal{A}/\langle\mathsf{\varepsilon}\rangle\cong\mathbb{Z}_2 $ and ${\mathcal H}^2(N,\mathcal{A})^Q = {\mathcal H}^2(N,\mathcal{A}) = \langle\mathsf{\varepsilon}\rangle\cong\mathbb{Z}_2$.

This allows us to write the $E_2$ page of the spectral sequence:
\begin{equation}\label{Q8E2}
\begin{tikzpicture}
  \matrix (m) [matrix of math nodes,
    nodes in empty cells,nodes={minimum width=8ex,
    minimum height=6ex,outer sep=0pt},
    column sep=2ex,row sep=6ex]{
q=2&   \mathbb{Z}_2   &     &     & & \\
q=1& \mathbb{Z}_2 & \mathbb{Z}_2& \mathbb{Z}_2 & \\
q=0&  \mathcal{A}  & \mathbb{Z}_2 &  \mathbb{Z}_2  & \mathbb{Z}_2 \\
\quad\strut &   p=0  &  p=1  &  p=2  & p=3& \strut \\};
  \draw[-stealth] (m-2-2.south east) -- (m-3-4.north west);
  \node[below = 0mm] at (m-2-3.south east) {$d^{0,1}_2\!\neq\!0~~~$};
    \draw[-stealth] (m-2-3.south east) -- (m-3-5.north west);
      \node[below=0mm] at (m-2-4.south east) {$d^{1,1}_2\!\neq\!0~$};
      \draw[-stealth] (m-1-2.south east) -- (m-2-4.north west) ;
      \node[below=0mm] at (m-1-3.south east) {$~d^{0,2}_2 = \!0$};
\draw[thick] (m-1-1.east) -- (m-4-1.east) ;
\draw[thick] (m-4-1.north) -- (m-4-5.north) ;
\end{tikzpicture}
\end{equation}

Similar to the case of $\mathcal{H}^2(Q_8,\mathbb{Z}_2^2)$ with trivial action, here we also have $E^{p,q}_2 \cong \mathbb{Z}_2$ for all $p,q \geq 0$; but we must now determine the content of the three differential maps, $d^{0,2}_2$, $d^{0,1}_2$, and $d^{1,1}_2$. First of all, using the property of the $\mathbb{Z} Q_8$ resolution we know that $d^{0,2}_2=0$ (see \cite{adem2013cohomology}, P129).

Next, using the result $\mathcal{H}^1(Q_8,\mathcal{A}) = \mathbb{Z}_2$ as given in Eq.~\eqref{q8upper1nontrivial}, we see that the map $d^{0,1}_2$ is surjective.
we can also explicitly compute the $d^{0,1}_2$ and $d^{1,1}_2$ maps using Eq.~\eqref{explicitdmaps}.
In our context, things are simplified because the only nontrivial value is when $q_1=q_2=q_3=q$. For any lifting $l\colon Q\rightarrow G$, we have $l(q_1)l(q_2) = (l(q))^2 = -1$ and $l(q_1q_2)=l(q^2) = l(1_Q) = 1_G$, meaning that
\begin{equation}
\begin{aligned}
   & d^{0,1}_2(f))(q,q) =f(-1) = \langle \mathsf{\varepsilon}\rangle
, \\
&(d^{1,1}_2(\omega))(q,q,q)
=(\omega(q))(-1)  = \langle \mathsf{\varepsilon}\rangle
\end{aligned}
\end{equation}
this means that $d^{0,1}_2(f)$ is a nontrivial cocycle in $Z^2(Q,\langle \mathsf{\varepsilon}\rangle)$ and $d^{1,1}_2(\omega)$ is a nontrivial cocycle in $Z^3(Q,\langle \mathsf{\varepsilon}\rangle)$, thus $d^{0,1}_2$ and $d^{1,1}_2$ are nontrivial. This means that, in going to the third page of the spectral sequence $E_3^{p,q}$, we get
\begin{equation}\label{Q8E3}
\begin{array}{c|cccc}
q=2 &  \mathbb{Z}^{\mathsf{\varepsilon}}_2  & \cdots &\cdots& \cdots \\
q=1 & 0& 0& \cdots & \cdots \\
q=0 & \mathcal{A} & \mathcal{H}^1(\langle q\rangle, \langle \mathsf{\varepsilon}\rangle)\cong \mathbb{Z}_2 & 0  & 0  \\
\hline
E_3^{p,q} & p=0 & p=1 & p=2 &p=3 \end{array}
\end{equation}
where all the entries explicitly written here stabilize to the $\infty$ page. From this we reproduce ${\mathcal H}^2_{\pm i \sigma^x, \pm i\sigma^z}(Q_8,\mathbb{Z}_2^2)=\mathbb{Z}_2^{\mathsf{\varepsilon}}$ which was given in Eq.~\eqref{Q8Hupper2nontrivlal}.

We see that the result differs from the trivial action case, for which ${\mathcal H}^2_{\text{id}}(Q_8,\mathbb{Z}_2^2) = \mathbb{Z}_2^2$. In terms of the spectral sequence, it is on the third-page $E_3^{p,q}$ that the difference emerges. Importantly, the spectral sequence also tells us that the result $\mathbb{Z}_2^{\mathsf{\varepsilon}}$ comes from $E_2^{0,2} = \mathcal{H}^2(N,\mathcal{A})^Q$, which in our context is $\mathcal{H}^2(\pi_1(G),\mathcal{A})^H$.

\textbf{LHS spectral sequence associated with Eq.~\eqref{Q8_homotopy}}.

Here we write the spectral sequence associated to Eq.~\eqref{Q8_homotopy}. In principle there is no reason to expect that this spectral sequence will be identical to the spectral sequence associated with  Eq.~\eqref{Q8_standard}. However, using the intermediate step ${\mathcal H}^1(H,{\mathcal H}^1_{\text{id}}(\pi_1(G),\mathcal{A})) = {\mathcal H}^1(H, \mathcal{A}) = \mathbb{Z}_2^{\mathsf{\varepsilon}} = \langle \alpha \colon H\rightarrow (\pi_1(G)\rightarrow \mathbb{Z}_2^{\mathsf{\varepsilon}})\rangle$ with $H:=D_2$, it turns out that the entries as explicitly shown in Eqs.~\eqref{Q8E2} and \eqref{Q8E3} are all isomorphic (and are stabilized), and for this spectral sequence (associated to Eq.~\eqref{Q8_homotopy}) we also have $d^{0,2}_2 = 0$, $d^{0,1}_2\neq0$ and $d^{1,1}_2\neq 0$, so we get the same result \eqref{Q8Hupper2nontrivlal}. 
This explicitly shows that $\mathcal{H}^2(Q_8,\mathcal{A}) = \mathbb{Z}_2^{\mathsf{\varepsilon}}$ is inherited from the cohomology of its subgroup $\pi_1(G)$: $\mathcal{H}^2(Q_8,\mathcal{A}) = E^{0,2}_2 =\mathcal{H}^2(\pi_1(G),\mathcal{A})^H$.

\bibliography{classical_quantum_topoloy_ref}

\begin{thebibliography}{116}
\expandafter\ifx\csname natexlab\endcsname\relax\def\natexlab#1{#1}\fi
\expandafter\ifx\csname bibnamefont\endcsname\relax
  \def\bibnamefont#1{#1}\fi
\expandafter\ifx\csname bibfnamefont\endcsname\relax
  \def\bibfnamefont#1{#1}\fi
\expandafter\ifx\csname citenamefont\endcsname\relax
  \def\citenamefont#1{#1}\fi
\expandafter\ifx\csname url\endcsname\relax
  \def\url#1{\texttt{#1}}\fi
\expandafter\ifx\csname urlprefix\endcsname\relax\def\urlprefix{URL }\fi
\providecommand{\bibinfo}[2]{#2}
\providecommand{\eprint}[2][]{\url{#2}}

\bibitem[{\citenamefont{Landau}(1937)}]{Landau1937}
\bibinfo{author}{\bibfnamefont{L.~D.} \bibnamefont{Landau}},
  \bibinfo{journal}{Zh. Eksp. Teor. Fiz.} \textbf{\bibinfo{volume}{11}},
  \bibinfo{pages}{19} (\bibinfo{year}{1937}).

\bibitem[{\citenamefont{Ginzburg and Landau}(2009)}]{Ginzburg2009}
\bibinfo{author}{\bibfnamefont{V.~L.} \bibnamefont{Ginzburg}} \bibnamefont{and}
  \bibinfo{author}{\bibfnamefont{L.~D.} \bibnamefont{Landau}}, in
  \emph{\bibinfo{booktitle}{On superconductivity and superfluidity}}
  (\bibinfo{publisher}{Springer}, \bibinfo{year}{2009}), pp.
  \bibinfo{pages}{113--137}.

\bibitem[{\citenamefont{Landau and
  Lifshitz}(2013{\natexlab{a}})}]{landau2013fluid}
\bibinfo{author}{\bibfnamefont{L.~D.} \bibnamefont{Landau}} \bibnamefont{and}
  \bibinfo{author}{\bibfnamefont{E.~M.} \bibnamefont{Lifshitz}},
  \emph{\bibinfo{title}{Fluid Mechanics: Landau and Lifshitz: Course of
  Theoretical Physics, Volume 6}}, vol.~\bibinfo{volume}{6}
  (\bibinfo{publisher}{Elsevier}, \bibinfo{year}{2013}{\natexlab{a}}).

\bibitem[{\citenamefont{Landau and
  Lifshitz}(2013{\natexlab{b}})}]{landau2013statistical}
\bibinfo{author}{\bibfnamefont{L.~D.} \bibnamefont{Landau}} \bibnamefont{and}
  \bibinfo{author}{\bibfnamefont{E.~M.} \bibnamefont{Lifshitz}},
  \emph{\bibinfo{title}{Statistical Physics: Volume 5}},
  vol.~\bibinfo{volume}{5} (\bibinfo{publisher}{Elsevier},
  \bibinfo{year}{2013}{\natexlab{b}}).

\bibitem[{\citenamefont{Nambu}(1960)}]{Naumbu1960}
\bibinfo{author}{\bibfnamefont{Y.}~\bibnamefont{Nambu}},
  \bibinfo{journal}{Phys. Rev.} \textbf{\bibinfo{volume}{117}},
  \bibinfo{pages}{648} (\bibinfo{year}{1960}),
  \urlprefix\url{https://link.aps.org/doi/10.1103/PhysRev.117.648}.

\bibitem[{\citenamefont{Goldstone}(1961)}]{Goldstone1961}
\bibinfo{author}{\bibfnamefont{J.}~\bibnamefont{Goldstone}},
  \bibinfo{journal}{Il Nuovo Cimento (1955-1965)}
  \textbf{\bibinfo{volume}{19}}, \bibinfo{pages}{154} (\bibinfo{year}{1961}),
  \urlprefix\url{https://doi.org/10.1007/BF02812722}.

\bibitem[{\citenamefont{Mermin}(1979)}]{Mermin1979}
\bibinfo{author}{\bibfnamefont{N.~D.} \bibnamefont{Mermin}},
  \bibinfo{journal}{Rev. Mod. Phys.} \textbf{\bibinfo{volume}{51}},
  \bibinfo{pages}{591} (\bibinfo{year}{1979}),
  \urlprefix\url{https://link.aps.org/doi/10.1103/RevModPhys.51.591}.

\bibitem[{\citenamefont{Berezinskii}(1971)}]{berezinskii1971destruction}
\bibinfo{author}{\bibfnamefont{V.}~\bibnamefont{Berezinskii}},
  \bibinfo{journal}{Sov. Phys. JETP} \textbf{\bibinfo{volume}{32}},
  \bibinfo{pages}{493} (\bibinfo{year}{1971}).

\bibitem[{\citenamefont{Berezinskii}(1972)}]{berezinskii1972destruction}
\bibinfo{author}{\bibfnamefont{V.}~\bibnamefont{Berezinskii}},
  \bibinfo{journal}{Sov. Phys. JETP} \textbf{\bibinfo{volume}{34}},
  \bibinfo{pages}{610} (\bibinfo{year}{1972}).

\bibitem[{\citenamefont{Kosterlitz and Thouless}(1973)}]{Kosterlitz1973}
\bibinfo{author}{\bibfnamefont{J.~M.} \bibnamefont{Kosterlitz}}
  \bibnamefont{and} \bibinfo{author}{\bibfnamefont{D.~J.}
  \bibnamefont{Thouless}}, \textbf{\bibinfo{volume}{6}}, \bibinfo{pages}{1181}
  (\bibinfo{year}{1973}),
  \urlprefix\url{https://dx.doi.org/10.1088/0022-3719/6/7/010}.

\bibitem[{\citenamefont{Kleman and Lavrentovich}(2003)}]{kleman2003soft}
\bibinfo{author}{\bibfnamefont{M.}~\bibnamefont{Kleman}} \bibnamefont{and}
  \bibinfo{author}{\bibfnamefont{O.~D.} \bibnamefont{Lavrentovich}},
  \emph{\bibinfo{title}{Soft matter physics: an introduction}}
  (\bibinfo{publisher}{Springer}, \bibinfo{year}{2003}).

\bibitem[{\citenamefont{Vafa}(2020)}]{vafa2022defect}
\bibinfo{author}{\bibfnamefont{F.}~\bibnamefont{Vafa}},
  \bibinfo{journal}{arXiv:2009.10723}  (\bibinfo{year}{2020}).

\bibitem[{\citenamefont{Lewenstein et~al.}(2012)\citenamefont{Lewenstein,
  Sanpera, and Ahufinger}}]{lewenstein2012ultracold}
\bibinfo{author}{\bibfnamefont{M.}~\bibnamefont{Lewenstein}},
  \bibinfo{author}{\bibfnamefont{A.}~\bibnamefont{Sanpera}}, \bibnamefont{and}
  \bibinfo{author}{\bibfnamefont{V.}~\bibnamefont{Ahufinger}},
  \emph{\bibinfo{title}{Ultracold Atoms in Optical Lattices: Simulating quantum
  many-body systems}} (\bibinfo{publisher}{OUP Oxford}, \bibinfo{year}{2012}).

\bibitem[{\citenamefont{Stamper-Kurn and Ueda}(2013)}]{stamper2013spinor}
\bibinfo{author}{\bibfnamefont{D.~M.} \bibnamefont{Stamper-Kurn}}
  \bibnamefont{and} \bibinfo{author}{\bibfnamefont{M.}~\bibnamefont{Ueda}},
  \bibinfo{journal}{Reviews of Modern Physics} \textbf{\bibinfo{volume}{85}},
  \bibinfo{pages}{1191} (\bibinfo{year}{2013}).

\bibitem[{\citenamefont{Klitzing et~al.}(1980)\citenamefont{Klitzing, Dorda,
  and Pepper}}]{Klitzing1980}
\bibinfo{author}{\bibfnamefont{K.~v.} \bibnamefont{Klitzing}},
  \bibinfo{author}{\bibfnamefont{G.}~\bibnamefont{Dorda}}, \bibnamefont{and}
  \bibinfo{author}{\bibfnamefont{M.}~\bibnamefont{Pepper}},
  \bibinfo{journal}{Phys. Rev. Lett.} \textbf{\bibinfo{volume}{45}},
  \bibinfo{pages}{494} (\bibinfo{year}{1980}),
  \urlprefix\url{https://link.aps.org/doi/10.1103/PhysRevLett.45.494}.

\bibitem[{\citenamefont{Tsui et~al.}(1982)\citenamefont{Tsui, Stormer, and
  Gossard}}]{Tsui1982}
\bibinfo{author}{\bibfnamefont{D.~C.} \bibnamefont{Tsui}},
  \bibinfo{author}{\bibfnamefont{H.~L.} \bibnamefont{Stormer}},
  \bibnamefont{and} \bibinfo{author}{\bibfnamefont{A.~C.}
  \bibnamefont{Gossard}}, \bibinfo{journal}{Phys. Rev. Lett.}
  \textbf{\bibinfo{volume}{48}}, \bibinfo{pages}{1559} (\bibinfo{year}{1982}),
  \urlprefix\url{https://link.aps.org/doi/10.1103/PhysRevLett.48.1559}.

\bibitem[{\citenamefont{Laughlin}(1983)}]{Laughlin1983}
\bibinfo{author}{\bibfnamefont{R.~B.} \bibnamefont{Laughlin}},
  \bibinfo{journal}{Phys. Rev. Lett.} \textbf{\bibinfo{volume}{50}},
  \bibinfo{pages}{1395} (\bibinfo{year}{1983}),
  \urlprefix\url{https://link.aps.org/doi/10.1103/PhysRevLett.50.1395}.

\bibitem[{\citenamefont{Wen}(2017)}]{Wen2017}
\bibinfo{author}{\bibfnamefont{X.-G.} \bibnamefont{Wen}},
  \bibinfo{journal}{Rev. Mod. Phys.} \textbf{\bibinfo{volume}{89}},
  \bibinfo{pages}{041004} (\bibinfo{year}{2017}),
  \urlprefix\url{https://link.aps.org/doi/10.1103/RevModPhys.89.041004}.

\bibitem[{\citenamefont{WEN}(1990)}]{Wen1990}
\bibinfo{author}{\bibfnamefont{X.~G.} \bibnamefont{WEN}},
  \bibinfo{journal}{International Journal of Modern Physics B}
  \textbf{\bibinfo{volume}{04}}, \bibinfo{pages}{239} (\bibinfo{year}{1990}),
  \urlprefix\url{https://doi.org/10.1142/S0217979290000139}.

\bibitem[{\citenamefont{Gu and Wen}(2009)}]{Gu2009}
\bibinfo{author}{\bibfnamefont{Z.-C.} \bibnamefont{Gu}} \bibnamefont{and}
  \bibinfo{author}{\bibfnamefont{X.-G.} \bibnamefont{Wen}},
  \bibinfo{journal}{Phys. Rev. B} \textbf{\bibinfo{volume}{80}},
  \bibinfo{pages}{155131} (\bibinfo{year}{2009}),
  \urlprefix\url{https://link.aps.org/doi/10.1103/PhysRevB.80.155131}.

\bibitem[{\citenamefont{Chen et~al.}(2010)\citenamefont{Chen, Gu, and
  Wen}}]{Chen2010}
\bibinfo{author}{\bibfnamefont{X.}~\bibnamefont{Chen}},
  \bibinfo{author}{\bibfnamefont{Z.-C.} \bibnamefont{Gu}}, \bibnamefont{and}
  \bibinfo{author}{\bibfnamefont{X.-G.} \bibnamefont{Wen}},
  \bibinfo{journal}{Phys. Rev. B} \textbf{\bibinfo{volume}{82}},
  \bibinfo{pages}{155138} (\bibinfo{year}{2010}),
  \urlprefix\url{https://link.aps.org/doi/10.1103/PhysRevB.82.155138}.

\bibitem[{\citenamefont{Chen et~al.}(2011{\natexlab{a}})\citenamefont{Chen, Gu,
  and Wen}}]{Chen2011a}
\bibinfo{author}{\bibfnamefont{X.}~\bibnamefont{Chen}},
  \bibinfo{author}{\bibfnamefont{Z.-C.} \bibnamefont{Gu}}, \bibnamefont{and}
  \bibinfo{author}{\bibfnamefont{X.-G.} \bibnamefont{Wen}},
  \bibinfo{journal}{Phys. Rev. B} \textbf{\bibinfo{volume}{83}},
  \bibinfo{pages}{035107} (\bibinfo{year}{2011}{\natexlab{a}}),
  \urlprefix\url{https://link.aps.org/doi/10.1103/PhysRevB.83.035107}.

\bibitem[{\citenamefont{Chen et~al.}(2011{\natexlab{b}})\citenamefont{Chen, Gu,
  and Wen}}]{Chen2011b}
\bibinfo{author}{\bibfnamefont{X.}~\bibnamefont{Chen}},
  \bibinfo{author}{\bibfnamefont{Z.-C.} \bibnamefont{Gu}}, \bibnamefont{and}
  \bibinfo{author}{\bibfnamefont{X.-G.} \bibnamefont{Wen}},
  \bibinfo{journal}{Phys. Rev. B} \textbf{\bibinfo{volume}{84}},
  \bibinfo{pages}{235128} (\bibinfo{year}{2011}{\natexlab{b}}),
  \urlprefix\url{https://link.aps.org/doi/10.1103/PhysRevB.84.235128}.

\bibitem[{\citenamefont{Chen et~al.}(2011{\natexlab{c}})\citenamefont{Chen,
  Liu, and Wen}}]{Chen2011c}
\bibinfo{author}{\bibfnamefont{X.}~\bibnamefont{Chen}},
  \bibinfo{author}{\bibfnamefont{Z.-X.} \bibnamefont{Liu}}, \bibnamefont{and}
  \bibinfo{author}{\bibfnamefont{X.-G.} \bibnamefont{Wen}},
  \bibinfo{journal}{Phys. Rev. B} \textbf{\bibinfo{volume}{84}},
  \bibinfo{pages}{235141} (\bibinfo{year}{2011}{\natexlab{c}}),
  \urlprefix\url{https://link.aps.org/doi/10.1103/PhysRevB.84.235141}.

\bibitem[{\citenamefont{Chen et~al.}(2012)\citenamefont{Chen, Gu, Liu, and
  Wen}}]{Chen2012}
\bibinfo{author}{\bibfnamefont{X.}~\bibnamefont{Chen}},
  \bibinfo{author}{\bibfnamefont{Z.-C.} \bibnamefont{Gu}},
  \bibinfo{author}{\bibfnamefont{Z.-X.} \bibnamefont{Liu}}, \bibnamefont{and}
  \bibinfo{author}{\bibfnamefont{X.-G.} \bibnamefont{Wen}},
  \bibinfo{journal}{Science} \textbf{\bibinfo{volume}{338}},
  \bibinfo{pages}{1604} (\bibinfo{year}{2012}),
  \urlprefix\url{https://doi.org/10.1126/science.1227224}.

\bibitem[{\citenamefont{Chen et~al.}(2013)\citenamefont{Chen, Gu, Liu, and
  Wen}}]{Chen2013}
\bibinfo{author}{\bibfnamefont{X.}~\bibnamefont{Chen}},
  \bibinfo{author}{\bibfnamefont{Z.-C.} \bibnamefont{Gu}},
  \bibinfo{author}{\bibfnamefont{Z.-X.} \bibnamefont{Liu}}, \bibnamefont{and}
  \bibinfo{author}{\bibfnamefont{X.-G.} \bibnamefont{Wen}},
  \bibinfo{journal}{Phys. Rev. B} \textbf{\bibinfo{volume}{87}},
  \bibinfo{pages}{155114} (\bibinfo{year}{2013}),
  \urlprefix\url{https://link.aps.org/doi/10.1103/PhysRevB.87.155114}.

\bibitem[{\citenamefont{Chen et~al.}(2014)\citenamefont{Chen, Lu, and
  Vishwanath}}]{Chen2014}
\bibinfo{author}{\bibfnamefont{X.}~\bibnamefont{Chen}},
  \bibinfo{author}{\bibfnamefont{Y.-M.} \bibnamefont{Lu}}, \bibnamefont{and}
  \bibinfo{author}{\bibfnamefont{A.}~\bibnamefont{Vishwanath}},
  \bibinfo{journal}{Nature Communications} \textbf{\bibinfo{volume}{5}},
  \bibinfo{pages}{3507} (\bibinfo{year}{2014}),
  \urlprefix\url{https://doi.org/10.1038/ncomms4507}.

\bibitem[{\citenamefont{Pollmann et~al.}(2012)\citenamefont{Pollmann, Berg,
  Turner, and Oshikawa}}]{Pollmann2012}
\bibinfo{author}{\bibfnamefont{F.}~\bibnamefont{Pollmann}},
  \bibinfo{author}{\bibfnamefont{E.}~\bibnamefont{Berg}},
  \bibinfo{author}{\bibfnamefont{A.~M.} \bibnamefont{Turner}},
  \bibnamefont{and} \bibinfo{author}{\bibfnamefont{M.}~\bibnamefont{Oshikawa}},
  \bibinfo{journal}{Phys. Rev. B} \textbf{\bibinfo{volume}{85}},
  \bibinfo{pages}{075125} (\bibinfo{year}{2012}),
  \urlprefix\url{https://link.aps.org/doi/10.1103/PhysRevB.85.075125}.

\bibitem[{\citenamefont{Lu and Vishwanath}(2012)}]{Lu2012}
\bibinfo{author}{\bibfnamefont{Y.-M.} \bibnamefont{Lu}} \bibnamefont{and}
  \bibinfo{author}{\bibfnamefont{A.}~\bibnamefont{Vishwanath}},
  \bibinfo{journal}{Phys. Rev. B} \textbf{\bibinfo{volume}{86}},
  \bibinfo{pages}{125119} (\bibinfo{year}{2012}),
  \urlprefix\url{https://link.aps.org/doi/10.1103/PhysRevB.86.125119}.

\bibitem[{\citenamefont{Wen}(2013)}]{Wen2013}
\bibinfo{author}{\bibfnamefont{X.-G.} \bibnamefont{Wen}},
  \bibinfo{journal}{Phys. Rev. D} \textbf{\bibinfo{volume}{88}},
  \bibinfo{pages}{045013} (\bibinfo{year}{2013}),
  \urlprefix\url{https://link.aps.org/doi/10.1103/PhysRevD.88.045013}.

\bibitem[{\citenamefont{Levin and Gu}(2012)}]{Levin2012}
\bibinfo{author}{\bibfnamefont{M.}~\bibnamefont{Levin}} \bibnamefont{and}
  \bibinfo{author}{\bibfnamefont{Z.-C.} \bibnamefont{Gu}},
  \bibinfo{journal}{Phys. Rev. B} \textbf{\bibinfo{volume}{86}},
  \bibinfo{pages}{115109} (\bibinfo{year}{2012}),
  \urlprefix\url{https://link.aps.org/doi/10.1103/PhysRevB.86.115109}.

\bibitem[{\citenamefont{Wen}(2014)}]{Wen2014}
\bibinfo{author}{\bibfnamefont{X.-G.} \bibnamefont{Wen}},
  \bibinfo{journal}{Phys. Rev. B} \textbf{\bibinfo{volume}{89}},
  \bibinfo{pages}{035147} (\bibinfo{year}{2014}),
  \urlprefix\url{https://link.aps.org/doi/10.1103/PhysRevB.89.035147}.

\bibitem[{\citenamefont{Essin and Hermele}(2013)}]{Essin2013}
\bibinfo{author}{\bibfnamefont{A.~M.} \bibnamefont{Essin}} \bibnamefont{and}
  \bibinfo{author}{\bibfnamefont{M.}~\bibnamefont{Hermele}},
  \bibinfo{journal}{Phys. Rev. B} \textbf{\bibinfo{volume}{87}},
  \bibinfo{pages}{104406} (\bibinfo{year}{2013}),
  \urlprefix\url{https://link.aps.org/doi/10.1103/PhysRevB.87.104406}.

\bibitem[{\citenamefont{Mesaros and Ran}(2013)}]{Mesaros2013}
\bibinfo{author}{\bibfnamefont{A.}~\bibnamefont{Mesaros}} \bibnamefont{and}
  \bibinfo{author}{\bibfnamefont{Y.}~\bibnamefont{Ran}},
  \bibinfo{journal}{Phys. Rev. B} \textbf{\bibinfo{volume}{87}},
  \bibinfo{pages}{155115} (\bibinfo{year}{2013}),
  \urlprefix\url{https://link.aps.org/doi/10.1103/PhysRevB.87.155115}.

\bibitem[{\citenamefont{Lu and Vishwanath}(2016)}]{Lu2016}
\bibinfo{author}{\bibfnamefont{Y.-M.} \bibnamefont{Lu}} \bibnamefont{and}
  \bibinfo{author}{\bibfnamefont{A.}~\bibnamefont{Vishwanath}},
  \bibinfo{journal}{Phys. Rev. B} \textbf{\bibinfo{volume}{93}},
  \bibinfo{pages}{155121} (\bibinfo{year}{2016}),
  \urlprefix\url{https://link.aps.org/doi/10.1103/PhysRevB.93.155121}.

\bibitem[{\citenamefont{Tarantino et~al.}(2016)\citenamefont{Tarantino,
  Lindner, and Fidkowski}}]{Tarantino2016}
\bibinfo{author}{\bibfnamefont{N.}~\bibnamefont{Tarantino}},
  \bibinfo{author}{\bibfnamefont{N.~H.} \bibnamefont{Lindner}},
  \bibnamefont{and}
  \bibinfo{author}{\bibfnamefont{L.}~\bibnamefont{Fidkowski}},
  \bibinfo{journal}{New Journal of Physics} \textbf{\bibinfo{volume}{18}},
  \bibinfo{pages}{035006} (\bibinfo{year}{2016}),
  \urlprefix\url{http://dx.doi.org/10.1088/1367-2630/18/3/035006}.

\bibitem[{\citenamefont{Barkeshli et~al.}(2019)\citenamefont{Barkeshli,
  Bonderson, Cheng, and Wang}}]{Barkeshli2019}
\bibinfo{author}{\bibfnamefont{M.}~\bibnamefont{Barkeshli}},
  \bibinfo{author}{\bibfnamefont{P.}~\bibnamefont{Bonderson}},
  \bibinfo{author}{\bibfnamefont{M.}~\bibnamefont{Cheng}}, \bibnamefont{and}
  \bibinfo{author}{\bibfnamefont{Z.}~\bibnamefont{Wang}},
  \bibinfo{journal}{Phys. Rev. B} \textbf{\bibinfo{volume}{100}},
  \bibinfo{pages}{115147} (\bibinfo{year}{2019}),
  \urlprefix\url{https://link.aps.org/doi/10.1103/PhysRevB.100.115147}.

\bibitem[{\citenamefont{Feldman et~al.}(2016)\citenamefont{Feldman, Randeria,
  Gyenis, Wu, Ji, Cava, MacDonald, and Yazdani}}]{Feldman2016}
\bibinfo{author}{\bibfnamefont{B.~E.} \bibnamefont{Feldman}},
  \bibinfo{author}{\bibfnamefont{M.~T.} \bibnamefont{Randeria}},
  \bibinfo{author}{\bibfnamefont{A.}~\bibnamefont{Gyenis}},
  \bibinfo{author}{\bibfnamefont{F.}~\bibnamefont{Wu}},
  \bibinfo{author}{\bibfnamefont{H.}~\bibnamefont{Ji}},
  \bibinfo{author}{\bibfnamefont{R.~J.} \bibnamefont{Cava}},
  \bibinfo{author}{\bibfnamefont{A.~H.} \bibnamefont{MacDonald}},
  \bibnamefont{and} \bibinfo{author}{\bibfnamefont{A.}~\bibnamefont{Yazdani}},
  \bibinfo{journal}{Science} \textbf{\bibinfo{volume}{354}},
  \bibinfo{pages}{316} (\bibinfo{year}{2016}),
  \urlprefix\url{https://doi.org/10.1126/science.aag1715}.

\bibitem[{\citenamefont{Liu et~al.}(2019)\citenamefont{Liu, Wang, Sato,
  Hohenadler, Wang, Guo, and Assaad}}]{Liu2019}
\bibinfo{author}{\bibfnamefont{Y.}~\bibnamefont{Liu}},
  \bibinfo{author}{\bibfnamefont{Z.}~\bibnamefont{Wang}},
  \bibinfo{author}{\bibfnamefont{T.}~\bibnamefont{Sato}},
  \bibinfo{author}{\bibfnamefont{M.}~\bibnamefont{Hohenadler}},
  \bibinfo{author}{\bibfnamefont{C.}~\bibnamefont{Wang}},
  \bibinfo{author}{\bibfnamefont{W.}~\bibnamefont{Guo}}, \bibnamefont{and}
  \bibinfo{author}{\bibfnamefont{F.~F.} \bibnamefont{Assaad}},
  \bibinfo{journal}{Nature Communications} \textbf{\bibinfo{volume}{10}},
  \bibinfo{pages}{2658} (\bibinfo{year}{2019}),
  \urlprefix\url{https://doi.org/10.1038/s41467-019-10372-0}.

\bibitem[{\citenamefont{Petit et~al.}(2016)\citenamefont{Petit, Lhotel, Canals,
  Ciomaga~Hatnean, Ollivier, Mutka, Ressouche, Wildes, Lees, and
  Balakrishnan}}]{Petit2016}
\bibinfo{author}{\bibfnamefont{S.}~\bibnamefont{Petit}},
  \bibinfo{author}{\bibfnamefont{E.}~\bibnamefont{Lhotel}},
  \bibinfo{author}{\bibfnamefont{B.}~\bibnamefont{Canals}},
  \bibinfo{author}{\bibfnamefont{M.}~\bibnamefont{Ciomaga~Hatnean}},
  \bibinfo{author}{\bibfnamefont{J.}~\bibnamefont{Ollivier}},
  \bibinfo{author}{\bibfnamefont{H.}~\bibnamefont{Mutka}},
  \bibinfo{author}{\bibfnamefont{E.}~\bibnamefont{Ressouche}},
  \bibinfo{author}{\bibfnamefont{A.~R.} \bibnamefont{Wildes}},
  \bibinfo{author}{\bibfnamefont{M.~R.} \bibnamefont{Lees}}, \bibnamefont{and}
  \bibinfo{author}{\bibfnamefont{G.}~\bibnamefont{Balakrishnan}},
  \bibinfo{journal}{Nature Physics} \textbf{\bibinfo{volume}{12}},
  \bibinfo{pages}{746} (\bibinfo{year}{2016}),
  \urlprefix\url{https://doi.org/10.1038/nphys3710}.

\bibitem[{\citenamefont{Sondhi et~al.}(1993)\citenamefont{Sondhi, Karlhede,
  Kivelson, and Rezayi}}]{Sondhi1993}
\bibinfo{author}{\bibfnamefont{S.~L.} \bibnamefont{Sondhi}},
  \bibinfo{author}{\bibfnamefont{A.}~\bibnamefont{Karlhede}},
  \bibinfo{author}{\bibfnamefont{S.~A.} \bibnamefont{Kivelson}},
  \bibnamefont{and} \bibinfo{author}{\bibfnamefont{E.~H.}
  \bibnamefont{Rezayi}}, \bibinfo{journal}{Phys. Rev. B}
  \textbf{\bibinfo{volume}{47}}, \bibinfo{pages}{16419} (\bibinfo{year}{1993}),
  \urlprefix\url{https://link.aps.org/doi/10.1103/PhysRevB.47.16419}.

\bibitem[{\citenamefont{Volovik}(1999)}]{Volovik1999}
\bibinfo{author}{\bibfnamefont{G.~E.} \bibnamefont{Volovik}},
  \bibinfo{journal}{Journal of Experimental and Theoretical Physics Letters}
  \textbf{\bibinfo{volume}{70}}, \bibinfo{pages}{609} (\bibinfo{year}{1999}),
  \urlprefix\url{https://doi.org/10.1134/1.568223}.

\bibitem[{\citenamefont{Ivanov}(2001)}]{Ivanov2001}
\bibinfo{author}{\bibfnamefont{D.~A.} \bibnamefont{Ivanov}},
  \bibinfo{journal}{Phys. Rev. Lett.} \textbf{\bibinfo{volume}{86}},
  \bibinfo{pages}{268} (\bibinfo{year}{2001}),
  \urlprefix\url{https://link.aps.org/doi/10.1103/PhysRevLett.86.268}.

\bibitem[{\citenamefont{Read and Green}(2000)}]{Read2000}
\bibinfo{author}{\bibfnamefont{N.}~\bibnamefont{Read}} \bibnamefont{and}
  \bibinfo{author}{\bibfnamefont{D.}~\bibnamefont{Green}},
  \bibinfo{journal}{Phys. Rev. B} \textbf{\bibinfo{volume}{61}},
  \bibinfo{pages}{10267} (\bibinfo{year}{2000}),
  \urlprefix\url{https://link.aps.org/doi/10.1103/PhysRevB.61.10267}.

\bibitem[{\citenamefont{Zhang et~al.}(2009)\citenamefont{Zhang, Ran, and
  Vishwanath}}]{Zhang2009}
\bibinfo{author}{\bibfnamefont{Y.}~\bibnamefont{Zhang}},
  \bibinfo{author}{\bibfnamefont{Y.}~\bibnamefont{Ran}}, \bibnamefont{and}
  \bibinfo{author}{\bibfnamefont{A.}~\bibnamefont{Vishwanath}},
  \bibinfo{journal}{Phys. Rev. B} \textbf{\bibinfo{volume}{79}},
  \bibinfo{pages}{245331} (\bibinfo{year}{2009}),
  \urlprefix\url{https://link.aps.org/doi/10.1103/PhysRevB.79.245331}.

\bibitem[{\citenamefont{Ran et~al.}(2009)\citenamefont{Ran, Zhang, and
  Vishwanath}}]{Ran2009}
\bibinfo{author}{\bibfnamefont{Y.}~\bibnamefont{Ran}},
  \bibinfo{author}{\bibfnamefont{Y.}~\bibnamefont{Zhang}}, \bibnamefont{and}
  \bibinfo{author}{\bibfnamefont{A.}~\bibnamefont{Vishwanath}},
  \bibinfo{journal}{Nature Physics} \textbf{\bibinfo{volume}{5}},
  \bibinfo{pages}{298} (\bibinfo{year}{2009}).

\bibitem[{\citenamefont{Teo and Kane}(2010)}]{Teo2010}
\bibinfo{author}{\bibfnamefont{J.~C.~Y.} \bibnamefont{Teo}} \bibnamefont{and}
  \bibinfo{author}{\bibfnamefont{C.~L.} \bibnamefont{Kane}},
  \bibinfo{journal}{Phys. Rev. B} \textbf{\bibinfo{volume}{82}},
  \bibinfo{pages}{115120} (\bibinfo{year}{2010}),
  \urlprefix\url{https://link.aps.org/doi/10.1103/PhysRevB.82.115120}.

\bibitem[{\citenamefont{Barkeshli et~al.}(2013)\citenamefont{Barkeshli, Jian,
  and Qi}}]{Barkeshli2013}
\bibinfo{author}{\bibfnamefont{M.}~\bibnamefont{Barkeshli}},
  \bibinfo{author}{\bibfnamefont{C.-M.} \bibnamefont{Jian}}, \bibnamefont{and}
  \bibinfo{author}{\bibfnamefont{X.-L.} \bibnamefont{Qi}},
  \bibinfo{journal}{Phys. Rev. B} \textbf{\bibinfo{volume}{87}},
  \bibinfo{pages}{045130} (\bibinfo{year}{2013}),
  \urlprefix\url{https://link.aps.org/doi/10.1103/PhysRevB.87.045130}.

\bibitem[{\citenamefont{Barkeshli and Qi}(2014)}]{Barkeshili2014}
\bibinfo{author}{\bibfnamefont{M.}~\bibnamefont{Barkeshli}} \bibnamefont{and}
  \bibinfo{author}{\bibfnamefont{X.-L.} \bibnamefont{Qi}},
  \bibinfo{journal}{Phys. Rev. X} \textbf{\bibinfo{volume}{4}},
  \bibinfo{pages}{041035} (\bibinfo{year}{2014}),
  \urlprefix\url{https://link.aps.org/doi/10.1103/PhysRevX.4.041035}.

\bibitem[{\citenamefont{Balram et~al.}(2015)\citenamefont{Balram, Wurstbauer,
  Wójs, Pinczuk, and Jain}}]{Balram2015}
\bibinfo{author}{\bibfnamefont{A.~C.} \bibnamefont{Balram}},
  \bibinfo{author}{\bibfnamefont{U.}~\bibnamefont{Wurstbauer}},
  \bibinfo{author}{\bibfnamefont{A.}~\bibnamefont{Wójs}},
  \bibinfo{author}{\bibfnamefont{A.}~\bibnamefont{Pinczuk}}, \bibnamefont{and}
  \bibinfo{author}{\bibfnamefont{J.~K.} \bibnamefont{Jain}},
  \bibinfo{journal}{Nature Communications} \textbf{\bibinfo{volume}{6}},
  \bibinfo{pages}{8981} (\bibinfo{year}{2015}).

\bibitem[{\citenamefont{Wang et~al.}(2022)\citenamefont{Wang, Liu, and
  Moore}}]{Wang2022}
\bibinfo{author}{\bibfnamefont{Y.-Q.} \bibnamefont{Wang}},
  \bibinfo{author}{\bibfnamefont{C.}~\bibnamefont{Liu}}, \bibnamefont{and}
  \bibinfo{author}{\bibfnamefont{J.~E.} \bibnamefont{Moore}},
  \bibinfo{journal}{arXiv:2208.14056}  (\bibinfo{year}{2022}).

\bibitem[{\citenamefont{Else}(2021)}]{Else2021}
\bibinfo{author}{\bibfnamefont{D.~V.} \bibnamefont{Else}},
  \bibinfo{journal}{Phys. Rev. B} \textbf{\bibinfo{volume}{104}},
  \bibinfo{pages}{115129} (\bibinfo{year}{2021}),
  \urlprefix\url{https://link.aps.org/doi/10.1103/PhysRevB.104.115129}.

\bibitem[{\citenamefont{Teo}(2016)}]{Teo2016}
\bibinfo{author}{\bibfnamefont{J.~C.~Y.} \bibnamefont{Teo}},
  \bibinfo{journal}{Journal of Physics: Condensed Matter}
  \textbf{\bibinfo{volume}{28}}, \bibinfo{pages}{143001}
  (\bibinfo{year}{2016}),
  \urlprefix\url{https://doi.org/10.1088/0953-8984/28/14/143001}.

\bibitem[{\citenamefont{Senthil
  et~al.}(2004{\natexlab{a}})\citenamefont{Senthil, Vishwanath, Balents,
  Sachdev, and Fisher}}]{Senthil20041}
\bibinfo{author}{\bibfnamefont{T.}~\bibnamefont{Senthil}},
  \bibinfo{author}{\bibfnamefont{A.}~\bibnamefont{Vishwanath}},
  \bibinfo{author}{\bibfnamefont{L.}~\bibnamefont{Balents}},
  \bibinfo{author}{\bibfnamefont{S.}~\bibnamefont{Sachdev}}, \bibnamefont{and}
  \bibinfo{author}{\bibfnamefont{M.~P.~A.} \bibnamefont{Fisher}},
  \bibinfo{journal}{Science} \textbf{\bibinfo{volume}{303}},
  \bibinfo{pages}{1490} (\bibinfo{year}{2004}{\natexlab{a}}),
  \urlprefix\url{https://doi.org/10.1126/science.1091806}.

\bibitem[{\citenamefont{Senthil
  et~al.}(2004{\natexlab{b}})\citenamefont{Senthil, Balents, Sachdev,
  Vishwanath, and Fisher}}]{Senthil2004}
\bibinfo{author}{\bibfnamefont{T.}~\bibnamefont{Senthil}},
  \bibinfo{author}{\bibfnamefont{L.}~\bibnamefont{Balents}},
  \bibinfo{author}{\bibfnamefont{S.}~\bibnamefont{Sachdev}},
  \bibinfo{author}{\bibfnamefont{A.}~\bibnamefont{Vishwanath}},
  \bibnamefont{and} \bibinfo{author}{\bibfnamefont{M.~P.~A.}
  \bibnamefont{Fisher}}, \bibinfo{journal}{Phys. Rev. B}
  \textbf{\bibinfo{volume}{70}}, \bibinfo{pages}{144407}
  (\bibinfo{year}{2004}{\natexlab{b}}),
  \urlprefix\url{https://link.aps.org/doi/10.1103/PhysRevB.70.144407}.

\bibitem[{\citenamefont{Wang et~al.}(2017)\citenamefont{Wang, Nahum, Metlitski,
  Xu, and Senthil}}]{Wang2017}
\bibinfo{author}{\bibfnamefont{C.}~\bibnamefont{Wang}},
  \bibinfo{author}{\bibfnamefont{A.}~\bibnamefont{Nahum}},
  \bibinfo{author}{\bibfnamefont{M.~A.} \bibnamefont{Metlitski}},
  \bibinfo{author}{\bibfnamefont{C.}~\bibnamefont{Xu}}, \bibnamefont{and}
  \bibinfo{author}{\bibfnamefont{T.}~\bibnamefont{Senthil}},
  \bibinfo{journal}{Phys. Rev. X} \textbf{\bibinfo{volume}{7}},
  \bibinfo{pages}{031051} (\bibinfo{year}{2017}),
  \urlprefix\url{https://link.aps.org/doi/10.1103/PhysRevX.7.031051}.

\bibitem[{\citenamefont{Hatcher}(2002)}]{hatcher2002algebraic}
\bibinfo{author}{\bibfnamefont{A.}~\bibnamefont{Hatcher}},
  \emph{\bibinfo{title}{Algebraic Topology}} (\bibinfo{publisher}{Cambridge
  University Press}, \bibinfo{year}{2002}).

\bibitem[{\citenamefont{Rotman and Rotman}(2009)}]{rotman2009introduction}
\bibinfo{author}{\bibfnamefont{J.~J.} \bibnamefont{Rotman}} \bibnamefont{and}
  \bibinfo{author}{\bibfnamefont{J.~J.} \bibnamefont{Rotman}},
  \emph{\bibinfo{title}{An introduction to homological algebra}},
  vol.~\bibinfo{volume}{2} (\bibinfo{publisher}{Springer},
  \bibinfo{year}{2009}).

\bibitem[{\citenamefont{Wen}(2002)}]{Wen2002}
\bibinfo{author}{\bibfnamefont{X.-G.} \bibnamefont{Wen}},
  \bibinfo{journal}{Phys. Rev. B} \textbf{\bibinfo{volume}{65}},
  \bibinfo{pages}{165113} (\bibinfo{year}{2002}),
  \urlprefix\url{https://link.aps.org/doi/10.1103/PhysRevB.65.165113}.

\bibitem[{\citenamefont{Propitius}(1995)}]{propitius1995topological}
\bibinfo{author}{\bibfnamefont{M.~d.~W.} \bibnamefont{Propitius}},
  \bibinfo{journal}{arXiv preprint hep-th/9511195}  (\bibinfo{year}{1995}).

\bibitem[{\citenamefont{Kitaev}(2006)}]{Kitaev2006}
\bibinfo{author}{\bibfnamefont{A.}~\bibnamefont{Kitaev}},
  \bibinfo{journal}{Annals of Physics} \textbf{\bibinfo{volume}{321}},
  \bibinfo{pages}{2} (\bibinfo{year}{2006}),
  \urlprefix\url{https://www.sciencedirect.com/science/article/pii/S0003491605002381}.

\bibitem[{\citenamefont{Kitaev and Kong}(2012)}]{Kitaev2012}
\bibinfo{author}{\bibfnamefont{A.}~\bibnamefont{Kitaev}} \bibnamefont{and}
  \bibinfo{author}{\bibfnamefont{L.}~\bibnamefont{Kong}},
  \bibinfo{journal}{Communications in Mathematical Physics}
  \textbf{\bibinfo{volume}{313}}, \bibinfo{pages}{351} (\bibinfo{year}{2012}).

\bibitem[{\citenamefont{Bridgeman and Barter}(2020)}]{Bridgeman2020}
\bibinfo{author}{\bibfnamefont{J.~C.} \bibnamefont{Bridgeman}}
  \bibnamefont{and} \bibinfo{author}{\bibfnamefont{D.}~\bibnamefont{Barter}},
  \bibinfo{journal}{{Quantum}} \textbf{\bibinfo{volume}{4}},
  \bibinfo{pages}{277} (\bibinfo{year}{2020}),
  \urlprefix\url{https://doi.org/10.22331/q-2020-06-04-277}.

\bibitem[{\citenamefont{Hung and Wan}(2015)}]{Hung2015}
\bibinfo{author}{\bibfnamefont{L.-Y.} \bibnamefont{Hung}} \bibnamefont{and}
  \bibinfo{author}{\bibfnamefont{Y.}~\bibnamefont{Wan}},
  \bibinfo{journal}{Journal of High Energy Physics}
  \textbf{\bibinfo{volume}{2015}}, \bibinfo{pages}{120} (\bibinfo{year}{2015}).

\bibitem[{\citenamefont{Lan et~al.}(2015)\citenamefont{Lan, Wang, and
  Wen}}]{Lan2015}
\bibinfo{author}{\bibfnamefont{T.}~\bibnamefont{Lan}},
  \bibinfo{author}{\bibfnamefont{J.~C.} \bibnamefont{Wang}}, \bibnamefont{and}
  \bibinfo{author}{\bibfnamefont{X.-G.} \bibnamefont{Wen}},
  \bibinfo{journal}{Phys. Rev. Lett.} \textbf{\bibinfo{volume}{114}},
  \bibinfo{pages}{076402} (\bibinfo{year}{2015}),
  \urlprefix\url{https://link.aps.org/doi/10.1103/PhysRevLett.114.076402}.

\bibitem[{\citenamefont{Hu et~al.}(2017)\citenamefont{Hu, Wan, and
  Wu}}]{Hu2017}
\bibinfo{author}{\bibfnamefont{Y.}~\bibnamefont{Hu}},
  \bibinfo{author}{\bibfnamefont{Y.}~\bibnamefont{Wan}}, \bibnamefont{and}
  \bibinfo{author}{\bibfnamefont{Y.-S.} \bibnamefont{Wu}},
  \bibinfo{journal}{Chinese Physics Letters} \textbf{\bibinfo{volume}{34}},
  \bibinfo{eid}{077103} (\bibinfo{year}{2017}),
  \urlprefix\url{http://cpl.iphy.ac.cn/EN/abstract/article_70810.shtml}.

\bibitem[{\citenamefont{Bischoff et~al.}(2019)\citenamefont{Bischoff, Jones,
  Lu, and Penneys}}]{Bischoff2019}
\bibinfo{author}{\bibfnamefont{M.}~\bibnamefont{Bischoff}},
  \bibinfo{author}{\bibfnamefont{C.}~\bibnamefont{Jones}},
  \bibinfo{author}{\bibfnamefont{Y.-M.} \bibnamefont{Lu}}, \bibnamefont{and}
  \bibinfo{author}{\bibfnamefont{D.}~\bibnamefont{Penneys}},
  \bibinfo{journal}{Journal of High Energy Physics}
  \textbf{\bibinfo{volume}{2019}}, \bibinfo{pages}{62} (\bibinfo{year}{2019}).

\bibitem[{\citenamefont{Lan et~al.}(2020)\citenamefont{Lan, Wen, Kong, and
  Wen}}]{Lan2020}
\bibinfo{author}{\bibfnamefont{T.}~\bibnamefont{Lan}},
  \bibinfo{author}{\bibfnamefont{X.}~\bibnamefont{Wen}},
  \bibinfo{author}{\bibfnamefont{L.}~\bibnamefont{Kong}}, \bibnamefont{and}
  \bibinfo{author}{\bibfnamefont{X.-G.} \bibnamefont{Wen}},
  \bibinfo{journal}{Phys. Rev. Research} \textbf{\bibinfo{volume}{2}},
  \bibinfo{pages}{023331} (\bibinfo{year}{2020}),
  \urlprefix\url{https://link.aps.org/doi/10.1103/PhysRevResearch.2.023331}.

\bibitem[{KON(2018)}]{KONG2018}
\bibinfo{journal}{Nuclear Physics B} \textbf{\bibinfo{volume}{927}},
  \bibinfo{pages}{140} (\bibinfo{year}{2018}).

\bibitem[{KON(2021)}]{KONG2021}
\bibinfo{journal}{Nuclear Physics B} \textbf{\bibinfo{volume}{966}},
  \bibinfo{pages}{115384} (\bibinfo{year}{2021}).

\bibitem[{\citenamefont{Kawamura and Miyashita}(1984)}]{Kawamura1984}
\bibinfo{author}{\bibfnamefont{H.}~\bibnamefont{Kawamura}} \bibnamefont{and}
  \bibinfo{author}{\bibfnamefont{S.}~\bibnamefont{Miyashita}},
  \bibinfo{journal}{Journal of the Physical Society of Japan}
  \textbf{\bibinfo{volume}{53}}, \bibinfo{pages}{4138} (\bibinfo{year}{1984}),
  \urlprefix\url{https://doi.org/10.1143/JPSJ.53.4138}.

\bibitem[{\citenamefont{Dombre and Read}(1989)}]{Dombre1989}
\bibinfo{author}{\bibfnamefont{T.}~\bibnamefont{Dombre}} \bibnamefont{and}
  \bibinfo{author}{\bibfnamefont{N.}~\bibnamefont{Read}},
  \bibinfo{journal}{Phys. Rev. B} \textbf{\bibinfo{volume}{39}},
  \bibinfo{pages}{6797} (\bibinfo{year}{1989}),
  \urlprefix\url{https://link.aps.org/doi/10.1103/PhysRevB.39.6797}.

\bibitem[{\citenamefont{Grover and Senthil}(2008)}]{Grover2008}
\bibinfo{author}{\bibfnamefont{T.}~\bibnamefont{Grover}} \bibnamefont{and}
  \bibinfo{author}{\bibfnamefont{T.}~\bibnamefont{Senthil}},
  \bibinfo{journal}{Phys. Rev. Lett.} \textbf{\bibinfo{volume}{100}},
  \bibinfo{pages}{156804} (\bibinfo{year}{2008}),
  \urlprefix\url{https://link.aps.org/doi/10.1103/PhysRevLett.100.156804}.

\bibitem[{\citenamefont{Kitaev}(2009)}]{Kitaev2009}
\bibinfo{author}{\bibfnamefont{A.}~\bibnamefont{Kitaev}}, \bibinfo{journal}{AIP
  Conference Proceedings} \textbf{\bibinfo{volume}{1134}}, \bibinfo{pages}{22}
  (\bibinfo{year}{2009}),
  \urlprefix\url{https://aip.scitation.org/doi/abs/10.1063/1.3149495}.

\bibitem[{LU2(2020)}]{LU2020}
\bibinfo{journal}{Annals of Physics} \textbf{\bibinfo{volume}{413}},
  \bibinfo{pages}{168060} (\bibinfo{year}{2020}).

\bibitem[{\citenamefont{Levin and Senthil}(2004)}]{Levin2004}
\bibinfo{author}{\bibfnamefont{M.}~\bibnamefont{Levin}} \bibnamefont{and}
  \bibinfo{author}{\bibfnamefont{T.}~\bibnamefont{Senthil}},
  \bibinfo{journal}{Phys. Rev. B} \textbf{\bibinfo{volume}{70}},
  \bibinfo{pages}{220403} (\bibinfo{year}{2004}),
  \urlprefix\url{https://link.aps.org/doi/10.1103/PhysRevB.70.220403}.

\bibitem[{\citenamefont{Haldane}(1988)}]{Haldane1988}
\bibinfo{author}{\bibfnamefont{F.~D.~M.} \bibnamefont{Haldane}},
  \bibinfo{journal}{Phys. Rev. Lett.} \textbf{\bibinfo{volume}{61}},
  \bibinfo{pages}{1029} (\bibinfo{year}{1988}),
  \urlprefix\url{https://link.aps.org/doi/10.1103/PhysRevLett.61.1029}.

\bibitem[{\citenamefont{Tanaka and Hu}(2005)}]{Tanaka2005}
\bibinfo{author}{\bibfnamefont{A.}~\bibnamefont{Tanaka}} \bibnamefont{and}
  \bibinfo{author}{\bibfnamefont{X.}~\bibnamefont{Hu}}, \bibinfo{journal}{Phys.
  Rev. Lett.} \textbf{\bibinfo{volume}{95}}, \bibinfo{pages}{036402}
  (\bibinfo{year}{2005}),
  \urlprefix\url{https://link.aps.org/doi/10.1103/PhysRevLett.95.036402}.

\bibitem[{\citenamefont{Senthil and Fisher}(2006)}]{Senthil2006}
\bibinfo{author}{\bibfnamefont{T.}~\bibnamefont{Senthil}} \bibnamefont{and}
  \bibinfo{author}{\bibfnamefont{M.~P.~A.} \bibnamefont{Fisher}},
  \bibinfo{journal}{Phys. Rev. B} \textbf{\bibinfo{volume}{74}},
  \bibinfo{pages}{064405} (\bibinfo{year}{2006}),
  \urlprefix\url{https://link.aps.org/doi/10.1103/PhysRevB.74.064405}.

\bibitem[{\citenamefont{Wess and Zumino}(1974)}]{Wess1974}
\bibinfo{author}{\bibfnamefont{J.}~\bibnamefont{Wess}} \bibnamefont{and}
  \bibinfo{author}{\bibfnamefont{B.}~\bibnamefont{Zumino}},
  \bibinfo{journal}{Nuclear Physics B} \textbf{\bibinfo{volume}{70}},
  \bibinfo{pages}{39} (\bibinfo{year}{1974}),
  \urlprefix\url{https://www.sciencedirect.com/science/article/pii/0550321374903551}.

\bibitem[{\citenamefont{Witten}(1983)}]{Witten1983}
\bibinfo{author}{\bibfnamefont{E.}~\bibnamefont{Witten}},
  \bibinfo{journal}{Nuclear Physics B} \textbf{\bibinfo{volume}{223}},
  \bibinfo{pages}{422} (\bibinfo{year}{1983}),
  \urlprefix\url{https://www.sciencedirect.com/science/article/pii/0550321383900639}.

\bibitem[{\citenamefont{Abanov}(2000)}]{ABANOV2000}
\bibinfo{author}{\bibfnamefont{A.~G.} \bibnamefont{Abanov}},
  \bibinfo{journal}{Physics Letters B} \textbf{\bibinfo{volume}{492}},
  \bibinfo{pages}{321} (\bibinfo{year}{2000}),
  \urlprefix\url{https://www.sciencedirect.com/science/article/pii/S0370269300011187}.

\bibitem[{\citenamefont{{Lu}}(2017)}]{Lu2017}
\bibinfo{author}{\bibfnamefont{Y.-M.} \bibnamefont{{Lu}}},
  \bibinfo{journal}{arXiv e-prints} \bibinfo{eid}{arXiv:1705.04691}
  (\bibinfo{year}{2017}).

\bibitem[{\citenamefont{Yang et~al.}(2018)\citenamefont{Yang, Jiang,
  Vishwanath, and Ran}}]{Yang2018}
\bibinfo{author}{\bibfnamefont{X.}~\bibnamefont{Yang}},
  \bibinfo{author}{\bibfnamefont{S.}~\bibnamefont{Jiang}},
  \bibinfo{author}{\bibfnamefont{A.}~\bibnamefont{Vishwanath}},
  \bibnamefont{and} \bibinfo{author}{\bibfnamefont{Y.}~\bibnamefont{Ran}},
  \bibinfo{journal}{Phys. Rev. B} \textbf{\bibinfo{volume}{98}},
  \bibinfo{pages}{125120} (\bibinfo{year}{2018}),
  \urlprefix\url{https://link.aps.org/doi/10.1103/PhysRevB.98.125120}.

\bibitem[{\citenamefont{Kane and Mele}(2005)}]{Kane2005}
\bibinfo{author}{\bibfnamefont{C.~L.} \bibnamefont{Kane}} \bibnamefont{and}
  \bibinfo{author}{\bibfnamefont{E.~J.} \bibnamefont{Mele}},
  \bibinfo{journal}{Phys. Rev. Lett.} \textbf{\bibinfo{volume}{95}},
  \bibinfo{pages}{226801} (\bibinfo{year}{2005}),
  \urlprefix\url{https://link.aps.org/doi/10.1103/PhysRevLett.95.226801}.

\bibitem[{\citenamefont{Bernevig et~al.}(2006)\citenamefont{Bernevig, Hughes,
  and Zhang}}]{Bernevig2006}
\bibinfo{author}{\bibfnamefont{B.~A.} \bibnamefont{Bernevig}},
  \bibinfo{author}{\bibfnamefont{T.~L.} \bibnamefont{Hughes}},
  \bibnamefont{and} \bibinfo{author}{\bibfnamefont{S.-C.} \bibnamefont{Zhang}},
  \bibinfo{journal}{Science} \textbf{\bibinfo{volume}{314}},
  \bibinfo{pages}{1757} (\bibinfo{year}{2006}),
  \urlprefix\url{https://www.science.org/doi/abs/10.1126/science.1133734}.

\bibitem[{\citenamefont{König et~al.}(2007)\citenamefont{König, Wiedmann,
  Brüne, Roth, Buhmann, Molenkamp, Qi, and Zhang}}]{Konig2007}
\bibinfo{author}{\bibfnamefont{M.}~\bibnamefont{König}},
  \bibinfo{author}{\bibfnamefont{S.}~\bibnamefont{Wiedmann}},
  \bibinfo{author}{\bibfnamefont{C.}~\bibnamefont{Brüne}},
  \bibinfo{author}{\bibfnamefont{A.}~\bibnamefont{Roth}},
  \bibinfo{author}{\bibfnamefont{H.}~\bibnamefont{Buhmann}},
  \bibinfo{author}{\bibfnamefont{L.~W.} \bibnamefont{Molenkamp}},
  \bibinfo{author}{\bibfnamefont{X.-L.} \bibnamefont{Qi}}, \bibnamefont{and}
  \bibinfo{author}{\bibfnamefont{S.-C.} \bibnamefont{Zhang}},
  \bibinfo{journal}{Science} \textbf{\bibinfo{volume}{318}},
  \bibinfo{pages}{766} (\bibinfo{year}{2007}),
  \urlprefix\url{https://www.science.org/doi/abs/10.1126/science.1148047}.

\bibitem[{\citenamefont{Bernevig and Zhang}(2006)}]{Bernevig2006a}
\bibinfo{author}{\bibfnamefont{B.~A.} \bibnamefont{Bernevig}} \bibnamefont{and}
  \bibinfo{author}{\bibfnamefont{S.-C.} \bibnamefont{Zhang}},
  \bibinfo{journal}{Phys. Rev. Lett.} \textbf{\bibinfo{volume}{96}},
  \bibinfo{pages}{106802} (\bibinfo{year}{2006}),
  \urlprefix\url{https://link.aps.org/doi/10.1103/PhysRevLett.96.106802}.

\bibitem[{\citenamefont{Khalaf et~al.}(2021)\citenamefont{Khalaf, Chatterjee,
  Bultinck, Zaletel, and Vishwanath}}]{Khalaf2021}
\bibinfo{author}{\bibfnamefont{E.}~\bibnamefont{Khalaf}},
  \bibinfo{author}{\bibfnamefont{S.}~\bibnamefont{Chatterjee}},
  \bibinfo{author}{\bibfnamefont{N.}~\bibnamefont{Bultinck}},
  \bibinfo{author}{\bibfnamefont{M.~P.} \bibnamefont{Zaletel}},
  \bibnamefont{and}
  \bibinfo{author}{\bibfnamefont{A.}~\bibnamefont{Vishwanath}},
  \bibinfo{journal}{Science Advances} \textbf{\bibinfo{volume}{7}},
  \bibinfo{pages}{eabf5299} (\bibinfo{year}{2021}),
  \urlprefix\url{https://www.science.org/doi/abs/10.1126/sciadv.abf5299}.

\bibitem[{\citenamefont{Wen and Zee}(1992)}]{Wen1992}
\bibinfo{author}{\bibfnamefont{X.~G.} \bibnamefont{Wen}} \bibnamefont{and}
  \bibinfo{author}{\bibfnamefont{A.}~\bibnamefont{Zee}},
  \bibinfo{journal}{Phys. Rev. Lett.} \textbf{\bibinfo{volume}{69}},
  \bibinfo{pages}{953} (\bibinfo{year}{1992}),
  \urlprefix\url{https://link.aps.org/doi/10.1103/PhysRevLett.69.953}.

\bibitem[{\citenamefont{Han et~al.}(2019)\citenamefont{Han, Wang, and
  Ye}}]{Han2019}
\bibinfo{author}{\bibfnamefont{B.}~\bibnamefont{Han}},
  \bibinfo{author}{\bibfnamefont{H.}~\bibnamefont{Wang}}, \bibnamefont{and}
  \bibinfo{author}{\bibfnamefont{P.}~\bibnamefont{Ye}}, \bibinfo{journal}{Phys.
  Rev. B} \textbf{\bibinfo{volume}{99}}, \bibinfo{pages}{205120}
  (\bibinfo{year}{2019}),
  \urlprefix\url{https://link.aps.org/doi/10.1103/PhysRevB.99.205120}.

\bibitem[{\citenamefont{Litvin}(2008)}]{litvin2008tables}
\bibinfo{author}{\bibfnamefont{D.}~\bibnamefont{Litvin}},
  \bibinfo{journal}{Acta Crystallographica Section A: Foundations of
  Crystallography} \textbf{\bibinfo{volume}{64}}, \bibinfo{pages}{419}
  (\bibinfo{year}{2008}).

\bibitem[{\citenamefont{Metlitski and Thorngren}(2018)}]{PhysRevB.98.085140}
\bibinfo{author}{\bibfnamefont{M.~A.} \bibnamefont{Metlitski}}
  \bibnamefont{and}
  \bibinfo{author}{\bibfnamefont{R.}~\bibnamefont{Thorngren}},
  \bibinfo{journal}{Phys. Rev. B} \textbf{\bibinfo{volume}{98}},
  \bibinfo{pages}{085140} (\bibinfo{year}{2018}),
  \urlprefix\url{https://link.aps.org/doi/10.1103/PhysRevB.98.085140}.

\bibitem[{\citenamefont{Ye et~al.}(2021)\citenamefont{Ye, Guo, He, Wang, and
  Zou}}]{dqcp2021topological}
\bibinfo{author}{\bibfnamefont{W.}~\bibnamefont{Ye}},
  \bibinfo{author}{\bibfnamefont{M.}~\bibnamefont{Guo}},
  \bibinfo{author}{\bibfnamefont{Y.-C.} \bibnamefont{He}},
  \bibinfo{author}{\bibfnamefont{C.}~\bibnamefont{Wang}}, \bibnamefont{and}
  \bibinfo{author}{\bibfnamefont{L.}~\bibnamefont{Zou}} (\bibinfo{year}{2021}).

\bibitem[{\citenamefont{Barkeshli and Qi}(2012)}]{Barkeshli2012}
\bibinfo{author}{\bibfnamefont{M.}~\bibnamefont{Barkeshli}} \bibnamefont{and}
  \bibinfo{author}{\bibfnamefont{X.-L.} \bibnamefont{Qi}},
  \bibinfo{journal}{Phys. Rev. X} \textbf{\bibinfo{volume}{2}},
  \bibinfo{pages}{031013} (\bibinfo{year}{2012}),
  \urlprefix\url{https://link.aps.org/doi/10.1103/PhysRevX.2.031013}.

\bibitem[{\citenamefont{Teo et~al.}(2014)\citenamefont{Teo, Roy, and
  Chen}}]{Teo2014}
\bibinfo{author}{\bibfnamefont{J.~C.~Y.} \bibnamefont{Teo}},
  \bibinfo{author}{\bibfnamefont{A.}~\bibnamefont{Roy}}, \bibnamefont{and}
  \bibinfo{author}{\bibfnamefont{X.}~\bibnamefont{Chen}},
  \bibinfo{journal}{Phys. Rev. B} \textbf{\bibinfo{volume}{90}},
  \bibinfo{pages}{115118} (\bibinfo{year}{2014}),
  \urlprefix\url{https://link.aps.org/doi/10.1103/PhysRevB.90.115118}.

\bibitem[{\citenamefont{Qi}(2013)}]{Qi2013}
\bibinfo{author}{\bibfnamefont{X.-L.} \bibnamefont{Qi}}, \bibinfo{journal}{New
  Journal of Physics} \textbf{\bibinfo{volume}{15}}, \bibinfo{pages}{065002}
  (\bibinfo{year}{2013}),
  \urlprefix\url{https://doi.org/10.1088/1367-2630/15/6/065002}.

\bibitem[{\citenamefont{Gu and Levin}(2014)}]{Gu2014}
\bibinfo{author}{\bibfnamefont{Z.-C.} \bibnamefont{Gu}} \bibnamefont{and}
  \bibinfo{author}{\bibfnamefont{M.}~\bibnamefont{Levin}},
  \bibinfo{journal}{Phys. Rev. B} \textbf{\bibinfo{volume}{89}},
  \bibinfo{pages}{201113} (\bibinfo{year}{2014}),
  \urlprefix\url{https://link.aps.org/doi/10.1103/PhysRevB.89.201113}.

\bibitem[{\citenamefont{Lieb}(1994)}]{Lieb1994}
\bibinfo{author}{\bibfnamefont{E.~H.} \bibnamefont{Lieb}},
  \bibinfo{journal}{Phys. Rev. Lett.} \textbf{\bibinfo{volume}{73}},
  \bibinfo{pages}{2158} (\bibinfo{year}{1994}),
  \urlprefix\url{https://link.aps.org/doi/10.1103/PhysRevLett.73.2158}.

\bibitem[{\citenamefont{Zhang et~al.}(2013)\citenamefont{Zhang, Yin, and
  Wang}}]{Zhang2013}
\bibinfo{author}{\bibfnamefont{W.}~\bibnamefont{Zhang}},
  \bibinfo{author}{\bibfnamefont{R.}~\bibnamefont{Yin}}, \bibnamefont{and}
  \bibinfo{author}{\bibfnamefont{Y.}~\bibnamefont{Wang}},
  \bibinfo{journal}{Phys. Rev. B} \textbf{\bibinfo{volume}{88}},
  \bibinfo{pages}{174515} (\bibinfo{year}{2013}),
  \urlprefix\url{https://link.aps.org/doi/10.1103/PhysRevB.88.174515}.

\bibitem[{\citenamefont{Diener and Ho}(2006)}]{Diener2006}
\bibinfo{author}{\bibfnamefont{R.~B.} \bibnamefont{Diener}} \bibnamefont{and}
  \bibinfo{author}{\bibfnamefont{T.-L.} \bibnamefont{Ho}},
  \bibinfo{journal}{Phys. Rev. Lett.} \textbf{\bibinfo{volume}{96}},
  \bibinfo{pages}{190405} (\bibinfo{year}{2006}),
  \urlprefix\url{https://link.aps.org/doi/10.1103/PhysRevLett.96.190405}.

\bibitem[{\citenamefont{Bernier et~al.}(2006)\citenamefont{Bernier, Sengupta,
  and Kim}}]{Bernier2006}
\bibinfo{author}{\bibfnamefont{J.-S.} \bibnamefont{Bernier}},
  \bibinfo{author}{\bibfnamefont{K.}~\bibnamefont{Sengupta}}, \bibnamefont{and}
  \bibinfo{author}{\bibfnamefont{Y.~B.} \bibnamefont{Kim}},
  \bibinfo{journal}{Phys. Rev. B} \textbf{\bibinfo{volume}{74}},
  \bibinfo{pages}{155124} (\bibinfo{year}{2006}),
  \urlprefix\url{https://link.aps.org/doi/10.1103/PhysRevB.74.155124}.

\bibitem[{\citenamefont{Song et~al.}(2007)\citenamefont{Song, Semenoff, and
  Zhou}}]{Song2007}
\bibinfo{author}{\bibfnamefont{J.~L.} \bibnamefont{Song}},
  \bibinfo{author}{\bibfnamefont{G.~W.} \bibnamefont{Semenoff}},
  \bibnamefont{and} \bibinfo{author}{\bibfnamefont{F.}~\bibnamefont{Zhou}},
  \bibinfo{journal}{Phys. Rev. Lett.} \textbf{\bibinfo{volume}{98}},
  \bibinfo{pages}{160408} (\bibinfo{year}{2007}),
  \urlprefix\url{https://link.aps.org/doi/10.1103/PhysRevLett.98.160408}.

\bibitem[{\citenamefont{Wen}(1995)}]{Wen1995}
\bibinfo{author}{\bibfnamefont{X.-G.} \bibnamefont{Wen}},
  \bibinfo{journal}{Advances in Physics} \textbf{\bibinfo{volume}{44}},
  \bibinfo{pages}{405} (\bibinfo{year}{1995}),
  \urlprefix\url{https://doi.org/10.1080/00018739500101566}.

\bibitem[{\citenamefont{Cheng et~al.}(2016)\citenamefont{Cheng, Zaletel,
  Barkeshli, Vishwanath, and Bonderson}}]{Cheng2016}
\bibinfo{author}{\bibfnamefont{M.}~\bibnamefont{Cheng}},
  \bibinfo{author}{\bibfnamefont{M.}~\bibnamefont{Zaletel}},
  \bibinfo{author}{\bibfnamefont{M.}~\bibnamefont{Barkeshli}},
  \bibinfo{author}{\bibfnamefont{A.}~\bibnamefont{Vishwanath}},
  \bibnamefont{and}
  \bibinfo{author}{\bibfnamefont{P.}~\bibnamefont{Bonderson}},
  \bibinfo{journal}{Phys. Rev. X} \textbf{\bibinfo{volume}{6}},
  \bibinfo{pages}{041068} (\bibinfo{year}{2016}),
  \urlprefix\url{https://link.aps.org/doi/10.1103/PhysRevX.6.041068}.

\bibitem[{\citenamefont{Han}(2017)}]{han2017skyrmions}
\bibinfo{author}{\bibfnamefont{J.~H.} \bibnamefont{Han}},
  \emph{\bibinfo{title}{Skyrmions in condensed matter}}, vol.
  \bibinfo{volume}{278} (\bibinfo{publisher}{Springer}, \bibinfo{year}{2017}).

\bibitem[{\citenamefont{Fradkin}(2013)}]{Fradkinbook}
\bibinfo{author}{\bibfnamefont{E.}~\bibnamefont{Fradkin}},
  \emph{\bibinfo{title}{Field Theories of Condensed Matter Physics, 2nd ed}}
  (\bibinfo{publisher}{Cambridge University Press, Cambridge, England},
  \bibinfo{year}{2013}).

\bibitem[{\citenamefont{Wilczek and Zee}(1983)}]{Wilczek1983}
\bibinfo{author}{\bibfnamefont{F.}~\bibnamefont{Wilczek}} \bibnamefont{and}
  \bibinfo{author}{\bibfnamefont{A.}~\bibnamefont{Zee}},
  \bibinfo{journal}{Phys. Rev. Lett.} \textbf{\bibinfo{volume}{51}},
  \bibinfo{pages}{2250} (\bibinfo{year}{1983}),
  \urlprefix\url{https://link.aps.org/doi/10.1103/PhysRevLett.51.2250}.

\bibitem[{\citenamefont{Savary and Balents}(2012)}]{Savary2012}
\bibinfo{author}{\bibfnamefont{L.}~\bibnamefont{Savary}} \bibnamefont{and}
  \bibinfo{author}{\bibfnamefont{L.}~\bibnamefont{Balents}},
  \bibinfo{journal}{Phys. Rev. Lett.} \textbf{\bibinfo{volume}{108}},
  \bibinfo{pages}{037202} (\bibinfo{year}{2012}),
  \urlprefix\url{https://link.aps.org/doi/10.1103/PhysRevLett.108.037202}.

\bibitem[{\citenamefont{Brooks-Bartlett
  et~al.}(2014)\citenamefont{Brooks-Bartlett, Banks, Jaubert, Harman-Clarke,
  and Holdsworth}}]{Brooks-bartlett2014}
\bibinfo{author}{\bibfnamefont{M.~E.} \bibnamefont{Brooks-Bartlett}},
  \bibinfo{author}{\bibfnamefont{S.~T.} \bibnamefont{Banks}},
  \bibinfo{author}{\bibfnamefont{L.~D.~C.} \bibnamefont{Jaubert}},
  \bibinfo{author}{\bibfnamefont{A.}~\bibnamefont{Harman-Clarke}},
  \bibnamefont{and} \bibinfo{author}{\bibfnamefont{P.~C.~W.}
  \bibnamefont{Holdsworth}}, \bibinfo{journal}{Phys. Rev. X}
  \textbf{\bibinfo{volume}{4}}, \bibinfo{pages}{011007} (\bibinfo{year}{2014}),
  \urlprefix\url{https://link.aps.org/doi/10.1103/PhysRevX.4.011007}.

\bibitem[{\citenamefont{Kong}(2014)}]{KONG2014}
\bibinfo{author}{\bibfnamefont{L.}~\bibnamefont{Kong}},
  \bibinfo{journal}{Nuclear Physics B} \textbf{\bibinfo{volume}{886}},
  \bibinfo{pages}{436} (\bibinfo{year}{2014}),
  \urlprefix\url{https://www.sciencedirect.com/science/article/pii/S0550321314002223}.

\bibitem[{\citenamefont{Kapustin and Spodyneiko}(2020)}]{Kapustin2020}
\bibinfo{author}{\bibfnamefont{A.}~\bibnamefont{Kapustin}} \bibnamefont{and}
  \bibinfo{author}{\bibfnamefont{L.}~\bibnamefont{Spodyneiko}},
  \bibinfo{journal}{Phys. Rev. B} \textbf{\bibinfo{volume}{101}},
  \bibinfo{pages}{235130} (\bibinfo{year}{2020}),
  \urlprefix\url{https://link.aps.org/doi/10.1103/PhysRevB.101.235130}.

\bibitem[{\citenamefont{Aasen et~al.}(2022)\citenamefont{Aasen, Wang, and
  Hastings}}]{Aasen2022}
\bibinfo{author}{\bibfnamefont{D.}~\bibnamefont{Aasen}},
  \bibinfo{author}{\bibfnamefont{Z.}~\bibnamefont{Wang}}, \bibnamefont{and}
  \bibinfo{author}{\bibfnamefont{M.~B.} \bibnamefont{Hastings}},
  \bibinfo{journal}{Phys. Rev. B} \textbf{\bibinfo{volume}{106}},
  \bibinfo{pages}{085122} (\bibinfo{year}{2022}),
  \urlprefix\url{https://link.aps.org/doi/10.1103/PhysRevB.106.085122}.

\bibitem[{\citenamefont{{Wen} et~al.}(2021)\citenamefont{{Wen}, {Qi},
  {Beaudry}, {Moreno}, {Pflaum}, {Spiegel}, {Vishwanath}, and
  {Hermele}}}]{Wen2021}
\bibinfo{author}{\bibfnamefont{X.}~\bibnamefont{{Wen}}},
  \bibinfo{author}{\bibfnamefont{M.}~\bibnamefont{{Qi}}},
  \bibinfo{author}{\bibfnamefont{A.}~\bibnamefont{{Beaudry}}},
  \bibinfo{author}{\bibfnamefont{J.}~\bibnamefont{{Moreno}}},
  \bibinfo{author}{\bibfnamefont{M.~J.} \bibnamefont{{Pflaum}}},
  \bibinfo{author}{\bibfnamefont{D.}~\bibnamefont{{Spiegel}}},
  \bibinfo{author}{\bibfnamefont{A.}~\bibnamefont{{Vishwanath}}},
  \bibnamefont{and}
  \bibinfo{author}{\bibfnamefont{M.}~\bibnamefont{{Hermele}}},
  \bibinfo{journal}{arXiv e-prints} \bibinfo{eid}{arXiv:2112.07748}
  (\bibinfo{year}{2021}).

\bibitem[{\citenamefont{Kobayashi et~al.}(2012)\citenamefont{Kobayashi,
  Kobayashi, Kawaguchi, Nitta, and Ueda}}]{KOBAYASHI2011}
\bibinfo{author}{\bibfnamefont{S.}~\bibnamefont{Kobayashi}},
  \bibinfo{author}{\bibfnamefont{M.}~\bibnamefont{Kobayashi}},
  \bibinfo{author}{\bibfnamefont{Y.}~\bibnamefont{Kawaguchi}},
  \bibinfo{author}{\bibfnamefont{M.}~\bibnamefont{Nitta}}, \bibnamefont{and}
  \bibinfo{author}{\bibfnamefont{M.}~\bibnamefont{Ueda}},
  \bibinfo{journal}{Nuclear Physics B} \textbf{\bibinfo{volume}{856}},
  \bibinfo{pages}{577} (\bibinfo{year}{2012}),
  \urlprefix\url{https://www.sciencedirect.com/science/article/pii/S0550321311006316}.

\bibitem[{\citenamefont{Adem and Milgram}(2013)}]{adem2013cohomology}
\bibinfo{author}{\bibfnamefont{A.}~\bibnamefont{Adem}} \bibnamefont{and}
  \bibinfo{author}{\bibfnamefont{R.~J.} \bibnamefont{Milgram}},
  \emph{\bibinfo{title}{Cohomology of finite groups}}, vol.
  \bibinfo{volume}{309} (\bibinfo{publisher}{Springer Science \& Business
  Media}, \bibinfo{year}{2013}).

\end{thebibliography}
\bibliographystyle{apsrev}

\end{document}